\documentclass[%
 reprint,
superscriptaddress,
 amsmath,amssymb,
 aps,
]{revtex4-1}

\usepackage{graphicx}
\usepackage{dcolumn}
\usepackage{bm}
\usepackage{caption}
\usepackage{subcaption}
\usepackage[usenames, dvipsnames]{color}

\usepackage{amsmath}	
\usepackage{amssymb}	

\begin{document}

\title{Constraining strangeness in dense matter with GW170817}
\date{\today}
\author{R.O. Gomes}
\affiliation{Frankfurt Institute for Advanced Studies,
Frankfurt am Main, Germany.}
\affiliation{Astronomy Department, Universidade Federal do Rio Grande do Sul (UFRGS), Porto Alegre, Brazil}
\email{rosana.gomes@ufrgs.br}
\author{P. Char}
\affiliation{Inter-University Centre for Astronomy and Astrophysics, Post Bag 4, Ganeshkhind, Pune - 411 007, India}

\author{S. Schramm}
\affiliation{Frankfurt Institute for Advanced Studies,
Frankfurt am Main, Germany.}

\begin{abstract}

Particles with strangeness content are predicted to populate dense matter, modifying the equation of state of matter inside neutron stars as well as their structure and evolution. In this work, we show how the modeling of strangeness content in dense matter affects the properties of isolated neutrons stars and the tidal deformation in binary systems. For describing nucleonic and hyperonic stars we use the many-body forces model (MBF) at zero temperature, including the $\phi$ mesons for the description of repulsive hyperon-hyperon interactions. Hybrid stars are modeled using the MIT Bag Model with vector interaction (vMIT) in both Gibbs and Maxwell constructions, for different values of bag constant and vector interaction couplings. A parametrization with  a Maxwell construction, which gives rise to third family of compact stars (twin stars), is also investigated. 
We calculate the tidal contribution that adds to the post-Newtonian point-particle corrections, the  associated love number for sequences of stars of different composition (nucleonic, hyperonic, hybrid and twin stars), and determine signatures of the phase transition on the gravitational waves in the accumulated phase correction during the inspirals among different scenarios for binary systems. On the light of the recent results from GW170817 and the implications for the radius of $\sim1.4\,\mathrm{M_{\odot}}$ stars, our results show that hybrid stars can only exist if a phase transition takes place at low densities close to saturation.
\end{abstract}

\maketitle

\section{Introduction}


At the core of neutron stars, densities many orders of magnitude higher than the ones measured in experiments on Earth can be reached. It is predicted that exotic degrees of freedom can also populate these objects, and that even the dissolution of baryons into their quark constituents through a phase transition to quark matter could be facilitated in such extreme environments. The impact of exotic compositions on the structure of isolated neutron stars has been studied in the past for stars composed of hyperons \cite{Dexheimer:2008ax,Gomes:2014aka,Oertel:2014qza,Chatterjee:2015pua,Vidana:2015rsa,Yamamoto:2015lwa, Torres:2016ydl,Tolos:2016hhl,Mishra:2016qhw}, Delta baryon resonances \cite{Schurhoff:2010ph,Drago:2013fsa,Drago:2015cea,Cai:2015hya,Zhu:2016mtc}, meson condensates \cite{Menezes:2005ic,Mishra:2009bp,Alford:2009jm,Fernandez:2010zzc,Mesquita:2010zzb,Lim:2013tqa,Muto:2015sgx}, quarks or even color superconducting quark matter \cite{Buballa:2003et,Alford:2004pf,Bombaci:2006cs,Bonanno:2011ch,Alford:2017qgh}.
Such degrees of freedom are usually associated with a softening of the equation of state (EoS), impacting the maximum mass and stability of stars \cite{Hempel:2009vp,Klahn:2013kga}.


When hyperons are taken into account in relativistic mean field models, 
especial attention must be given to the hyperon-hyperon interaction modeling due to their 
internal strangeness degree of freedom. 
This is done through the introduction of $\phi$ and $\sigma^*$ mesons, which mediate the interaction among these particles by describing repulsion and attraction features, respectively \cite{Schaffner:1995th}.
The competition between softness and stiffness of the EoS due to the presence of hyperons   
has been widely discussed in the literature under the name of \emph{hyperon puzzle}  \cite{Bednarek:2011gd,Oertel:2014qza,Gomes:2014aka,Chatterjee:2015pua,Zhao:2015ncr,Haidenbauer:2016vfq,Bombaci:2016xzl,Torres:2016ydl,Fortin:2016hny,Tolos:2016hhl,Tolos:2017lgv,deOliveira:2017gty}.
So far, the scarce information regarding the hyperon-meson coupling, largely obtained by reproducing hyperon potentials deduced from experiment \cite{Weissenborn:2011kb,Gomes:2014aka,Sun:2017fnf,Fortin:2017cvt}, makes this topic still unresolved. 


With respect to a possible quark phase in astrophysical environments, the indirect observation of the quark-gluon plasma in the regime of high temperatures and low densities in ultra-relativistic heavy-ion collisions is in accordance with the existence of quark matter at the beginning of the universe \cite{Shuryak:2008eq}. 
The general study of the QCD phase structure relates the fields of heavy-ion collisions and neutron stars, where both approaches cover complementary regimes of density and temperature. Here, the experimental
efforts towards lower heavy-ion beam energies with rather large baryon densities at the future Facility for Antiproton and Ion Research (FAIR) and  Nuclotron-based Ion Collider fAcility (NICA), and the conditions of the recently observed neutron star mergers might lead to overlapping conditions 
of density and temperature in heavy-ion collisions and merger events. 
However, the fact that the regime of (star-relevant) high densities and low temperatures cannot be accessed by lattice or perturbative calculations makes the quark matter EoS even less constrained than the hadronic one, where at least there is some information on ground state matter properties. For this reason, effective models are used, rendering the determination of the phase transition point extremely model dependent.


Hybrid stars have been applied to investigate a broad range of topics regarding compact stars, such as nucleation in hadronic matter \cite{Bombaci:2006cs, Bombaci:2009jt,Bombaci:2016xuj}, color superconductivity in quark matter \cite{Buballa:2003et,Alford:2004pf,Bombaci:2006cs,Bonanno:2011ch,Alford:2017qgh}, stability \cite{Buballa:2003et,Ippolito:2007hn,Pereira:2017rmp}, rotation of neutron stars 
\cite{Glendenning:1997fy,Chubarian:1999yn,Glendenning:2000zz,Zdunik:2005kh, Ippolito:2007hn,Dimmelmeier:2009vw,Ayvazyan:2013cva,Bejger:2016emu}, 
magnetic neutrons stars \cite{Rabhi:2009ih,Sotani:2014rva,Franzon:2015sya,Franzon:2016urz}, thermal evolution \cite{Page:2005fq,Dexheimer:2012eu,Dexheimer:2014pea,deCarvalho:2015lpa}, 
proto-neutron stars \cite{Pons:2001ar,Yasutake:2009kj,Mariani:2017fqr}, supernovae \cite{Hempel:2009vp, Heinimann:2016zbx}, radial oscillations \cite{Brillante:2014lwa,Alford:2015gna,Pereira:2017rmp}, etc. 
The deconfinement phase transition has been extensively studied with different models such as Nambu-Jona-Lasinio \cite{Yasutake:2009kj,Lenzi:2012xz,Shao:2013toa,Klahn:2013kga,Benic:2014jia,Ranea-Sandoval:2015ldr,Alvarez-Castillo:2016wqj,Pereira:2016dfg,Miyatsu:2017teh}, MIT bag model \cite{Burgio:2002sn,Alford:2004pf,Bhattacharyya:2009fg,Yasutake:2010eq,Shao:2012tu,Franzon:2016urz,Pereira:2017rmp}, quark-meson coupling models \cite{Mishra:2016qhw,Whittenbury:2015ziz,Zacchi:2015oma,Miyatsu:2017teh}, and other approaches 
\cite{Dexheimer:2009hi,Chen:2011my,Dexheimer:2014pea,Li:2015ida,Schramm:2015hba,Burgio:2015zka,Alvarez-Castillo:2016oln}, showing that the features of the models used to describe both phases have implications for the determination of the macroscopic properties and composition of stars.  

Phase transitions taking place in neutron stars are usually described either in sharp or smooth scenarios:  
in the case of a sharp transition, a local charge neutrality is imposed in what is named a Maxwell construction; 
while in a Gibbs construction scenario a global charge neutrality takes place, generating a mixed phase structure \cite{Glendenning:1992vb}. However, which approach might be more realistic depends on poorly known strong interaction quantities like the QCD surface tension (w.g. \cite{Heiselberg:1992dx,Voskresensky:2002hu,Alford:2001zr,Pinto:2012aq,Stiele:2016cfs}). 
Both scenarios have been applied to describe hybrid stars \cite{Bhattacharyya:2009fg,Hempel:2009vp,Yasutake:2009kj,Yasutake:2010eq,Alaverdyan:2010zz,Contrera:2016phj}, indicating that the internal composition of stars can change substantially. 
In particular, the occurrence of a sharp phase transition to quark matter can in some cases allow for the appearance of a third family of compact stars and, consequently, for twin star configurations in which two stars have the same gravitational mass but different radii and composition
\cite{Alford:2015gna,Dexheimer:2014pea,Ayriyan:2017nhp}. 
 


In this sense, observational data from neutron stars can better constrain the modeling of strangeness in dense matter and works as a powerful tool to address the hyperon puzzle, as well as the interaction among quarks in hybrid stars. Almost a decade ago, the observation of massive neutron stars imposed constraints on the EoS of nuclear and quark matter, indicating that repulsion must be strong enough to reach $\sim 2\,M_{\odot}$ \cite{Demorest:2010bx,Antoniadis:2013pzd}. 

A further stellar property that is affected by possible exotic phases is the neutron star radius. The accuracy of radius determinations, which come from energy flux measurements including assumptions about the stellar atmosphere, is still quite limited. One of the current problems for acquiring such data is due to the impossibility of accurately measuring luminosity and distance of the stars separately. 
Furthermore, the temperature determination of these objects is complicated due to non-thermal emissions on their spectra arising from cyclotron radiation. For these reasons, values of neutron stars radii have 
large error bars and substantial efforts are being made to obtain much improved results in the future, like the missions Neutron Star Interior Composition Explorer (NICER)
\footnote{http://heasarc.gsfc.nasa.gov/docs/nicer/index.html}, 
Nuclear Spectroscopic Telescope Array (NUSTAR) \footnote{https://www.nustar.caltech.edu} and the Square Kilometer Array (SKA)
\footnote{http://www.skatelescope.org}.


The recent detection of gravitational waves (GW) events has started the multi-messenger era in astrophysics, and due to the extremely compact nature of neutron stars, it has proven to be a superb tool for constraining the nuclear matter equation of state \cite{TheLIGOScientific:2017qsa,Banik:2017zia}. It is expected that more GW events will be detected with the new generation and forthcoming ground-based detectors such as aLIGO \footnote{LIGO: http://www.ligo.caltech.edu.}, aVIRGO \footnote{VIRGO: http://virgo.infn.it}, KAGRA \footnote{Kagra: http://gwcenter.icrr.u-tokyo.ac.jp/en/} and the proposed Einstein Telescope \footnote{ET: http://www.et-gw.eu/}. 

In a coalescing binary neutron star system, stars exert a tidal force on each other and, when they are close enough, this tidal perturbation leads to a deformation on them. For this system, the imprint of this deformation can be clearly seen in the early inspiral stages of the GW signal. It modifies the waveform from the point-particle structure and results in a lag in the waveform phase. 
In particular, analysis from the first direct BNS GW event, the GW170817 event, detected from Advanced LIGO in 2017 \cite{TheLIGOScientific:2017qsa}, pose new constraints on the tidal deformability features of neutron stars. From preliminar results for low spin systems, it is estimated that the dimensionless tidal deformability $\tilde{\Lambda}$ has upper and lower limits of $\tilde{\Lambda} < 800-1000$ \cite{TheLIGOScientific:2017qsa,De:2018uhw} and $\tilde{\Lambda} > 400$ \cite{Radice:2017lry}, respectively.

The phase correction due to tidal deformation is characterized by a single EoS-dependent parameter, the Love number $k_2$ of the NS \cite{love}, also associated to the tidal deformation parameter $\lambda$. Tidal effects in this part of the signal will result in a phase shift accumulated over the cycles, which comes from the deviation from the point particle post-Newtonian waveforms. This is defined as $\bar{\lambda}$, the weighted average of the deformations of individual stars, and also depends on the chirp mass of the system. Therefore, a sufficiently large signal-to-noise ratio detection can be used to distinguish such corrections to the waveform to extract the EoS informations. Since the detection of GWs from a double black hole system, excitement about a double neutron stars system took place, and recent research has been performed on tidal deformation in neutron stars \cite{Hinderer:2009ca, Damour:2009vw, Postnikov:2010yn,Drago:2017bnf,Nandi:2017rhy,Kumar:2016dks,Alford:2017qgh,Marques:2017zju,Banik:2017zia,Ayriyan:2017nby,Paschalidis:2017qmb,Paschalidis:2017qmb,Most:2018hfd,Burgio:2018yix,Alvarez-Castillo:2018pve,Rueda:2018fky} 
and also on constraining the nuclear matter EoS  
\cite{Banik:2017zia,Alford:2017qgh,Annala:2017llu,Zhou:2017pha,Drago:2018nzf,Krastev:2018nwr,Zhu:2018ona,Raithel:2018ncd,Malik:2018zcf}.


In  this work we aim for constraining the composition of neutron stars through the analysis of isolated and binary systems and the comparison within the new constraints from observational data. The hadronic phase (with nucleons, hyperons and leptons) is described by the many-body forces model (MBF model), which simulates the effects of many-body forces via non-linear scalar fields contributions in the effective coupling \cite{Gomes:2014aka}. 
For the quark phase,we use the MIT bag model with vector interaction contributions (vMIT). When hybrid stars are modeled, the two phases are connected through a first-order phase transition and, in order to search for signals for the deconfinement, we investigate Maxwell and Gibbs construction scenarios. 

In the case of isolated stars, we explore the effects of different compositions on the properties of stars, focusing on how the modeling of interaction among particles with strangeness content (hyperons and quarks) impacts the phase transition scenario and, consequently, the radius of stars. We also identify the existence of a third family of compact stars when a sharp phase transition takes place. For binary systems, we use a post-newtonian approximation to calculate the structure of tidally deformed stars, in order to constrain the composition of neutron stars, as the comparatively cleaner signal from the early inspiral stage of the merger can present a detectable tidal signature \cite{Flanagan:2007ix}. Furthermore, due to the difference in radius for twin stars, we use the accumulated phase (sensitive to the fifth power of the radius) in order to identify a signal to distinguish these stars.

The article is organized as follows: in Section II we present the two models used for describing the EoS of matter inside stars, as well as the Maxwell and Gibbs construction formalisms; Section III is dedicated to the formalism used for describing binary systems in a post-newtonian approximation and in Section IV, we discuss the results for isolated and binary systems; finally, in Section V, discussion and perspectives are presented.

\section{Equation of state}\label{Section2}
In this session, we present the many-body forces (MBF) model and the MIT bag model with vector interaction (vMIT) used to describe the hadronic and quark matter in our analysis, respectively. The two phases are connected by a first order phase transition, which are modeled both in a Maxwell and a Gibbs construction.

\subsection{HADRONIC PHASE}\label{Hphase} 

The many-body forces (MBF) model \cite{Gomes:2014aka} is a relativistic mean field model in which meson field dependences are introduced in the couplings of baryons to mesons. 
The Lagrangian density of the MBF model reads:
\footnotesize
\begin{equation}\begin{split}\label{lagrangian}
\mathcal{L}&= \underset{b}{\sum}\overline{\psi}_{b}\left[\gamma_{\mu}\left(i\partial^{\mu}-g_{\omega b}\omega^{\mu} 
-g_{\varrho b}\mathbf{\textrm{\ensuremath{I_{3b}}\ensuremath{\varrho_3^{\mu}}}}\right)
-m^*_{b \zeta}\right]\psi_{b}
\\& +\left(\frac{1}{2}\partial_{\mu}\sigma\partial^{\mu}\sigma-m_{\sigma}^{2}\sigma^{2}\right)
+\frac{1}{2}\left(-\frac{1}{2}\omega_{\mu\nu}\omega^{\mu\nu}+m_{\omega}^{2}\omega_{\mu}\omega^{\mu}\right)
\\&+\frac{1}{2}\left(-\frac{1}{2}\boldsymbol{\varrho_{\mu\nu}.\varrho^{\mu\nu}}+m_{\varrho}^{2}\boldsymbol{\varrho_{\mu}.\varrho^{\mu}}\right)
+\left(\frac{1}{2}\partial_{\mu}\boldsymbol{\delta.}\partial^{\mu}\boldsymbol{\delta}-m_{\delta}^{2}\boldsymbol{\delta}^{2}\right)
\\& +\frac{1}{2}\left(-\frac{1}{2}\phi_{\mu\nu}\phi^{\mu\nu}+m_{\phi}^{2}\phi_{\mu}\phi^{\mu}\right)
+\underset{l}{\sum}\overline{\psi}_{l}\gamma_{\mu}\left(i\partial^{\mu}-m_{l}\right)\psi_{l},
\end{split}\end{equation}
\normalsize
where $b$ and $l$ correspond to the degrees of freedom of baryons ($n$, $p$, $\Lambda$, $\Sigma^+$, $\Sigma^0$, $\Sigma^-$, $\Xi^0$, $\Xi^-$) and leptons ($e^-$, $\mu^-$), respectively. The scalar mesons $\sigma$ and $\delta$ account for the description of attraction among baryons, while repulsion is described by the vector mesons $\omega$, $\varrho$ and $\phi$. Note that the $\phi$ mesons mediates interaction among hyperons, providing extra repulsion, which plays an important role in the description of massive stars with hyperon content \cite{Schaffner:1995th}. 
For more about the properties of baryons, leptons and mesons in this work, see Ref. \cite{Gomes:2014aka}.

The many-body forces contribution is introduced in the coupling of scalar fields as:
\begin{equation}
g_{i b}^*  = \left(1+    \frac{g_{\sigma b}\sigma+ g_{\delta b}I_{3b}\delta_{3}}{\zeta m_{b}}  \right)^{-\zeta} g_{i b},
\label{gss}
\end{equation}
for $i= \sigma,\delta$, directly affecting the effective masses of baryons and, consequently, their chemical potentials.
The field dependence in the couplings is, hence, introduced as a medium effect and is controlled by the $\zeta$ parameter. This parameter can be associated to non-linear terms that appear if one expands the term in Eq. \ref{gss}.

As strangeness is not conserved inside neutron stars due to the fact that typical time scales are much longer than the one for weak interaction, 
we allow for the appearance of hyperon degrees of freedom in our calculations. 
We model the coupling of hyperons using the SU(6) spin-flavor symmetry \cite{Dover:1985ba,Schaffner:1993qj}, and also by fixing the hyperon potentials as $U^{N}_{\Lambda}= −28\,\mathrm{MeV}$, 
$U^{N}_{\Sigma}= +30\,\mathrm{MeV}$, $U^{N}_{\Xi}= −18\,\mathrm{MeV}$ 
(see Ref. \cite{Gomes:2014aka} for other choices of hyperon potentials). 

The MBF model reproduces both nuclear matter properties at saturation and the observational properties of neutron stars with hyperons \cite{Gomes:2014aka} and magnetic hybrid stars \cite{Franzon:2016urz,Dexheimer:2017fhy}. 
For this work, the following nuclear properties at saturation were used: binding energy per nucleon $B/A= -15.75\,\mathrm{MeV} $, saturation density $\rho_0= 0.15\,\mathrm{fm^{-3}} $, symmetry energy and its slope $a^0_{sym}= 32\,\mathrm{MeV}$ and $L_0= 97\,\mathrm{MeV}$, respectively.

\subsection{QUARK PHASE}\label{Qphase} 
The quark matter is described by the MIT bag model with vector interactions contributions (vMIT). The Lagrangian density of the model reads:
\small
\begin{align}
\mathcal{L} = & \underset{q}{\sum}\left[ \overline{\psi}_{q}  (i\gamma_{\mu}\partial^{\mu}-g_{V q}\gamma_{\mu}V^{\mu} -m_q - B)\psi_{q}\right]\theta_H  \nonumber\\
+ & \underset{l}{\sum}\overline{\psi}_{l}\gamma_{\mu}\left(i\partial^{\mu}-m_{l}\right)\psi_{l} ,
\label{lagrangianQ}
\end{align}
\normalsize
where we consider the $u$, $d$ and $s$ quarks and $e^-$ and $\mu^-$ as leptons, labeled by the subscripts $q$ and $l$, respectively.
Also in this expression $B$ denotes the bag constant and $\theta_H$ is the Heaviside step function which allows for a confinement/deconfinement feature of the model ($\theta_H = 1$ inside the bag; $\theta_H = 0$ outside) \cite{Farhi:1984qu}. 

The vector interaction is introduced via the coupling of a vector-isoscalar meson $V^\mu$ to the quarks (with coupling constant $g_V$). Such a field is analogous to the $\omega$ field in hadronic models \cite{Shao:2013toa} and has direct impact on the chemical potential of quarks, i.e., on the equilibrium conditions as:
\begin{align}
\mu^*_q = \sqrt{(k_{F,q})^2+(m_q)^2}+g_{Vq} V,
\label{muq}
\end{align}
where $k_{F,q}$ and $(m_q)$ are respectively the Fermi momenta and bare masses of quarks.
Also, due to the beta equilibrium condition, the chemical potentials of quarks are related as $\mu_s=\mu_d=\mu_u+\mu_e$ and $\mu_u= \frac{1}{3}\mu_n-\frac{2}{3}\mu_e$, meaning that both hadronic and quark phases can be determined from the chemical potentials of neutrons and electrons.

It has been shown that vector interaction contributions in the description of quark matter plays a crucial role in the description of massive hybrid stars in agreement with observational data  \cite{Bonanno:2011ch,Shao:2012tu,Masuda:2012kf,denke2013influence,contrera2014hadron,menezes2014repulsive,deCarvalho:2015lpa,klahn2015vector,ranea2015constant,Contrera:2016phj}. These contributions have been investigated for different quark models \cite{klahn2015vector,contrera2014hadron,klahn2015vector, ranea2015constant}), and applied to neutron stars studies \cite{Shao:2012tu, contrera2014hadron,denke2013influence,menezes2014repulsive,deCarvalho:2015lpa}. 

Although attempts to constrain the value of the vector interaction coupling have been made (for example, by incorporating higher orders of perturbation theory and radiative corrections \cite{Fraga:2001id,Fraga:2013qra,Restrepo:2014fna}), this quantity remains widely uncertain. 
In order to investigate the impact of vector interactions in hybrid stars, we vary the coupling constant value and adopt the following set of parameters: $m_u=m_d=1\,\mathrm{MeV}$, $m_s=100\,\mathrm{MeV}$, $a_0= (g_V/m_{\omega})=1.8,\,2.0,\,2.2\,\mathrm{fm^{-2}}$, $m_{\omega}=782\,\mathrm{MeV}$, $B^{1/4}=170\,\mathrm{MeV}$.

\subsection{PHASE TRANSITION}\label{maxwell_section} 

In the zero temperature regime present inside neutron stars, it is predicted that a first order phase transition from hadronic to a deconfined quark matter takes place at high chemical potentials (baryon densities). 
In order to cover a wide range of scenarios, we investigate hybrid stars both with Gibbs and Maxwell  constructions.

In a Gibbs construction the interface between the two phases is smoother, allowing for a mixed phase to appear. The mixed phase is constructed under the assumption of  global charge neutrality, which leads to a charge rearrangement in both phases. 
On the other hand, a first order phase transition described in a Maxwell construction has a sharp interface and, in order to guarantee the electric charge neutrality in the system, local charge neutrality is imposed in both phases separately. 

The Maxwell criteria for zero temperature system reads (for both constructions):
\begin{equation}\label{maxwell}
P_H = P_Q , \qquad \mu_n^ {H}= \mu_n^ {Q}.
\end{equation}
where $P_H$ ($P_Q$) and $\mu_n^ {H}\,(\mu_n^ {Q})$ stand for the pressures and chemical potentials in the hadronic (quark) phase, respectively. 

The chemical equilibrium condition determines the coexistence phase and the way that the charge conservation is introduced in the system has direct impact on the pressure behavior at the transition point. When a local charge neutrality is imposed, the pressure depends only on the baryon chemical potential ($P_H(\mu_n),\,P_Q(\mu_n)$) and is constant along the phase transition as a consequence of the Maxwell criteria. However, in a Gibbs construction scenario in which charge neutrality is global, the pressure depends both on the electrical and baryon chemical potentials ($P_H(\mu_n,\mu_e),\,P_Q(\mu_n,\mu_e)$). This extra degree of freedom allows the pressure to vary as a function of the baryon chemical potential and still obey the pressure equality condition in both phases.

The comparison of both scenarios in the investigation of hybrid stars has been studied in several works, see e.g. \cite{Bhattacharyya:2009fg,Hempel:2009vp,Yasutake:2009kj,Yasutake:2010eq,Alaverdyan:2010zz}, where essentially the surface tension between the two phases determines which scenario is more appropriate. 
The threshold value for the surface tension has been calculated in several works \cite{Alford:2001zr,Voskresensky:2002hu,Maruyama:2007ey,Maruyama:2007ss,Palhares:2010be,Pinto:2012aq,Yasutake:2013sza,Lugones:2013ema,Garcia:2013eaa,Lugones:2016ytl}, indicating that values are high and more compatible with a Maxwell construction.

In particular, a minimal threshold of $\sigma \sim  40\,\mathrm{MeV/fm^2}$ \cite{Endo:2005zt,Yasutake:2014oxa} for this construction was estimated.
However, such estimates depend strongly on the equation of state (see Ref. \cite{Voskresensky:2002hu}), leaving the correct treatment of the phase transition as still an open question.

\section{Love number and tidal deformation}\label{Section3}
 We follow the perturbation scheme developed in Refs. \cite{Flanagan:2007ix,Hinderer:2007mb,Hinderer:2009ca} to compute the tidal deformability and the associated Love number.
 If a static spherically symmetric star of mass $M$ and radius $R$ is placed in a time-independent external tidal field $\mathcal{E}_{ij}$, 
 a quadrupole moment  $Q_{ij}$ is induced onto the star and to linear order, they satisfy the relation,
 \begin{equation}
  Q_{ij} = - \lambda \mathcal{E}_{ij},
 \end{equation}
 where $\lambda$ is defined as the tidal deformation parameter. It is related to the $l=2$ dimensionless Love number $k_2$, which is associated
 with the most dominant contribution to the stellar deformation as,
 \begin{equation}
 \lambda = \frac{2}{3} k_2 R^{5}.
 \end{equation}
 We consider only the leading order ($l=2$), static perturbation, which is axisymmetric ($m=0$) around the line joining the stars. The perturbed metric
 in Regge-Wheeler gauge \cite{Regge:1957td} can be written as \cite{Hinderer:2009ca}, 
\begin{eqnarray}
ds^2 &=& - e^{2\nu(r)} \left[1 + H(r)
Y_{20}(\theta,\phi)\right]dt^2
\nonumber\\
& & + e^{2\kappa(r)} \left[1 - H(r)
Y_{20}(\theta,\phi)\right]dr^2
\nonumber \\
& & + r^2 \left[1-K(r) Y_{20}(\theta,\phi)\right] \left( d\theta^2+ \sin^2\theta
d\phi^2 \right),
\nonumber\\
& &
\end{eqnarray}
where, $K(r)$ and $H(r)$ are the perturbed metric functions related by $K'(r)=H'(r)+2 H(r) \Phi'(r)$. Substituting the fluid perturbations,
 $\delta T_0^{\,0}=-\delta\epsilon(r) Y_{20} (\theta,\phi)$ and $\delta T_i^{\,i}=\delta p(r) Y_{20}(\theta,\phi)$, one obtains the equation for
the function $H(r)$ \cite{Hinderer:2009ca},
\begin{eqnarray}
\left(-\frac{6e^{2\kappa}}{r^2}-2(\nu ')^2+2\nu
  ''+\frac{3}{r}\kappa '\right.\nonumber\\
\left.+\frac{7}{r}\nu '- 2 \nu'
  \kappa '+ \frac{f}{r}(\nu '+\kappa ')\right)H\nonumber\\
+ \left(\frac{2}{r}+\nu '-\kappa'\right)H' + H'' =0
.\label{eq:h}
\end{eqnarray}
where, $f=d\epsilon/d p$. 
To calculate the tidal deformation, equation (\ref{eq:h}) is to be solved simultaneously with the TOV equations,
\begin{eqnarray}
e^{2 \kappa} &=& \left(1-\frac{2m}{r}\right)^{-1},\label{eq:lambda}\\
\frac{d\nu}{dr} &=& - \frac{1}{\epsilon+p}\frac{dp}{dr}, \label{eq:dphidr}\\
\frac{dp}{dr} &=& -(\epsilon+p)\frac{m+4\pi r^3 p}{r(r-2m)}, \label{eq:dpdr}\\
\frac{dm}{dr} &=& 4 \pi r^2 \epsilon.  \label{eq:dmdr}
\end{eqnarray}
The second-order differential equation for $H(r)$ can be rewritten as two coupled first-order differential equations and 
substituting the equlibrium metric functions using the TOV equations, taking the form, 
\begin{eqnarray}
\frac{dH}{dr}&=& Z\\
\frac{dZ}{dr}&=&2 \left(1 - 2\frac{m}{r}\right)^{-1} H\left\{-2\pi
  \left[5\epsilon+9
    p+f(\epsilon+p)\right]\phantom{\frac{3}{r^2}} \right. \nonumber\\
&& \quad \left. +\frac{3}{r^2}+2\left(1 - 2\frac{m}{r}\right)^{-1}
  \left(\frac{m}{r^2}+4\pi r p\right)^2\right\}\nonumber\\
&&+\frac{2Z}{r}\left(1 -
  2\frac{m}{r}\right)^{-1}\left\{-1+\frac{m}{r}+2\pi r^2
  (\epsilon-p)\right\}.\nonumber\\
\end{eqnarray}
The system is then solved radially outward from the center to the surface and matched with the exterior solution at the surface.
The love number $k_2$ is given by,
\begin{eqnarray}
k_2 &=& \frac{8C^5}{5}(1-2C)^2[2+2C(y-1)-y]\nonumber\\
      & & \times\bigg\{2C[6-3y+3C(5y-8)]\nonumber\\
      & & ~~~+4C^3[13-11y+C(3y-2)+2C^2(1+y)]\nonumber\\
      & & ~~~+3(1-2C)^2[2-y+2C(y-1)] \ln(1-2C)\bigg\}^{-1},\nonumber\\
\label{eq:k2}
\end{eqnarray}
where a quantity $y = \frac{ R\, Z(R)} {H(R)}$ is introduced and $C=\frac{M}{R}$ is the compactness of the star.

The observable signature of relativistic tidal deformation will have an imprint of the phase evolution of the GW spectrum of a binary NS system.
To calculate the phase evolution for different EoS's, we again follow the formalism used by Hinderer et al. \cite{Hinderer:2009ca}. , 
which shows in the secular limit, where the orbital period is much less than the gravitational radiation reaction time scale, 
that the tidal correction term adds linearly to the leading order post-Newtonian point particle corrections. Also, the signal will have cumulative effects of deformation coming from both the stars. Therefore, it is calculated from the weighted average of quadrupolar responses for the stars in a quasi-circular orbit, given by
\begin{equation}
\tilde\lambda=\frac{1}{26}\left[\frac{m_1 + 12 m_2}{m_1}\lambda_1
+ \frac{m_2 + 12 m_1}{m_2}\lambda_2 \right], 
\label{eq:lambda}
\end{equation}
where, $\lambda_1$ and $\lambda_2$ are the tidal deformations of stars $m_1$ and $m_2$.
Finally, the tidal correction term to the phase evolution is given by,
\begin{equation}
 \delta\Psi^{\rm tidal}=-\frac{117
\tilde \lambda x^{5/2}}{8 \eta M^5},
\end{equation}
where $M=m_1+m_2$ is the total mass of the binary,  $x = (\omega M )^{2/3}$, is a dimensionless post-Newtonian parameter and
$\eta = m_1 m_2 / M^2$, is the symmetric mass ratio. We also use dimensionless tidal deformability $\Lambda$ in our calculations, defined as $\Lambda = \lambda/M^5$. Similarly, eq. \ref{eq:lambda} can be recast as,
\begin{equation}
\tilde{\Lambda} = \frac{16}{13}\frac{\left(M_1 + 12 M_2\right)M_1^4 \Lambda_1 + \left(M_2 + 12 M_1\right)M_2^4 \Lambda_2}{\left(M_1 + M_2\right)^5}.
\label{eq:Lambda}
\end{equation}

\section{Analysis and discussion}\label{Section4}

In this section we discuss the impact of different neutron star compositions on the properties of isolated and binary stars systems. The analysis was carried out using the the models presented in Section \ref{Section2}, for nucleonic (Set 1), hyperonic (Set 2) and hybrid stars in Gibbs (Sets 3-5) and Maxwell (Sets 6-9) constructions. 

For hadronic matter, we use two parametrizations of the MBF model,  $\zeta=0.040, 0.085$, which are able to describe nuclear properties at saturation and, in particular, have a compressibility $K_0= 297,\,225\,\mathrm{MeV}$, respectively \cite{Gomes:2014aka}. When hyperons are allowed, $\phi$ mesons are introduced in the calculations. We have chosen not to consider the minor contributions of the scalar meson $\sigma^*$ in our calculations, as we are interested in massive neutron stars.

For both Gibbs and Maxwell constructions, we vary the values of the bag constant and vector coupling strengths in the ranges $B^{1/4}=160,170$ MeV and $a_0= 1.8, 2.2$ fm$^2$, respectively. For a Maxwell construction, we also identify a third family configuration of stars (twin stars) for the parametrization $B^{1/4}=171$ MeV and $a = 1.7$ fm$^2$, which is present in the sets of our analysis. The choice of parameters for both phases allows for a description of symmetric nuclear matter in terms of a pure 
hadronic phase at low densities, and the regime of high densities to be described by a quark phase.
All sets of parameters used in this work are summarized in Table \ref{Table1_sets}.

\begin{table}
  \caption{\label{Table1_sets} Sets of parameters used in the analysis, where \emph{Y} stands for hyperons ($\Lambda,\,\Sigma,\,\Xi$) and \emph{Q} stands for quarks ($u,\,d,\,s$). The columns are (when applicable): name of the set, particles population, strangeness modeling, Bag constant $B$ and vector interaction parameter $a_0$.}
\begin{center}
\begin{tabular}{cccccc}
 \hline \hline 
Set & Composition  & Strangeness  & $B^{1/4}$ &  $a_0$ & $\zeta$  \\
 &   & Modeling& $(\mathrm{MeV})$ & $(\mathrm{fm^2})$ &     \\
  \hline \hline
1 & $n,\,p,\,e,\,\mu$ & n.a. & n.a. & n.a. & $0.040$ \tabularnewline
2 & $n,\,p,\,Y,\,e,\,\mu$ & Y-Y int. ($\phi$) & n.a. & n.a.  & $0.040$ \tabularnewline
3 & $n,\,p,\,Y,\,e,\,\mu,\,Q$ & Gibbs & $160$ & $1.8$   & $0.085$ 
\tabularnewline
4 & $n,\,p,\,Y,\,e,\,\mu,\,Q$ & Gibbs & $170$ & $1.8$   & $0.040$
\tabularnewline
5 & $n,\,p,\,Y,\,e,\,\mu,\,Q$ & Gibbs & $170$ & $2.2$ & $0.040$\tabularnewline
6 & $n,\,p,\,Y,\,e,\,\mu,\,Q$ & Maxwell & $160$ & $1.8$ &  $0.085$  \tabularnewline
7 & $n,\,p,\,Y,\,e,\,\mu,\,Q$ & Maxwell & $170$ & $1.8$ &  $0.040$ \tabularnewline
8 & $n,\,p,\,Y,\,e,\,\mu,\,Q$ & Maxwell & $170$ & $2.2$ &  $0.040$ \tabularnewline
9 & $n,\,p,\,e,\,\mu,\,Q$ & Maxwell & $171$ & $1.7$  &   $0.040$ \tabularnewline
  \hline\hline
  
  \end{tabular}
\end{center}
\end{table}

\subsection{Isolated hybrid stars}

\begin{figure}[!ht]  
  \centering
  \includegraphics[width=1.0\linewidth]{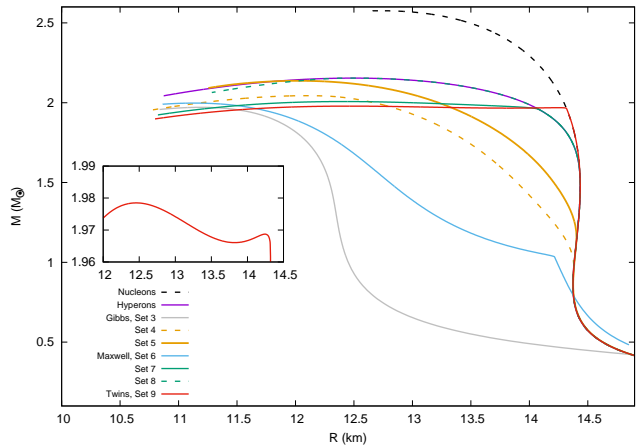}
  \caption{Mass-Radius relation for all parametrizations listed on Table \ref{Table1_sets}. The curve in red (Set 9) shows a third family of stars configuration.}
\label{Figure1}
\end{figure}

The modeling of interaction among particles with non-zero strangeness both in hadronic (hyperons) and in quark matter affects significantly phase transition features, as interaction among particles is directly related to the stiffness of the EoS. Table \ref{Table2_info} summarizes this impact from both microscopic (EoS) and macroscopic (stars properties) points of view, for all sets of parameters used in our analysis. 

First, comparing Sets 1 and 2 from Table \ref{Table2_info}, one can see that the introduction of hyperons in a hadronic model softens drastically  the EoS of hadronic matter, generaring a sequence of stars with much smaller maximum mass, as  discussed several times on literature. 
For the same reason, as hyperons start to populate the core of stars, for masses above $1.5\,\mathrm{M_{\odot}}$, the stars become more compact, i.e., present a smaller radius. Similarly, comparting hadronic stars (Sets 1 and 2) to all hybrid ones (all others), one can see a reduction in radius of $14.44\,\mathrm{km}$ to $12.32\,\mathrm{km}$ (Set 6) for the $1.4\,\mathrm{M_{\odot}}$ star, for example. 

Hybrid stars modeled under the assumption of a Gibbs construction (Sets 3-5) lead to a phase transition at low chemical potentials because of an extensive mixed phase. The critical chemical potentials range $976-1058$ MeV, depending on the 
stiffness of the EoS used. 
From the comparison of Sets 3 and 4, we see that softer hadronic EoS' lead to an earlier phase transition (lower $\mu_{n,c}$), as lower values of the $\zeta$ parameter in the MBF model generate a stiffer EoS due to weaker shielding of the scalar couplings of the model \cite{Gomes:2014aka}. 
Equivalently, Sets 4 and 5 show that a softer quark EoS also results in a lower $\mu_{n,c}$, but here this behavior is caused by the strength of the vector interaction coupling.  The critical masses for the onset of the appearance of quarks ranges between $0.36-1.15\,\mathrm{M_{\odot}}$, indicating that it is possible to obtain low mass and more compact hybrid stars for such construction. 

Although the critical baryon density at which the phase transition takes place is lower than the threshold of hyperon appearance for the parametrization of the MBF model used in the analysis, a small fraction of $\Lambda$ hyperons populate these stars because of global conservation conditions imposed in a Gibbs construction. In particular, the mixed phase covers a broad range of densities, implying that all stars present in this analysis have an incomplete phase transition to quark matter and, therefore, a core composed of a mixture of nucleons, $\Lambda$ hyperons,  leptons and quarks. 


For the corresponding stars modeled in a Maxwell construction (Set 6-8), the phase transition takes place for chemical potentials higher than the ones in a Gibbs construction, and consequently the gap between critical and maximum masses is smaller. For this case, stars are mostly hadronic, containing only a small quark core. In particular, the critical and maximum masses are the same for Set 8, indicating that all stars in this family are hadronic and correspond to Set 2, as the phase transition never takes place for stable stars. 

\begin{table}[t]
  \caption{\label{Table2_info} Information about the critical strangeness point (due to hyperons or quarks) and properties of stars for the different sets of parameters. The columns are: 
name of the set, critical chemical potential ($\mu_{c}$), critical ($M_{c}$) and maximum ($M_{max}$) mass of hybrid stars, and the radius of the $M=1.4\,\mathrm{M_{\odot}}$ star, respectively. }
\begin{center}
\begin{tabular}{ccccc}
 \hline \hline 
Set & $\mu_{n,c}$  & $M_{c}$ &  $M_{max}$ & $R_{1.4M_{\odot}}$ \\
 &  $(\mathrm{MeV})$  & $(\mathrm{M_{\odot}})$ & $(\mathrm{M_{\odot}})$ & $(\mathrm{km})$    \\
  \hline \hline
1 & n.a. & n.a. & $2.57$ & $14.44$   \tabularnewline

2 & $1122$ & $1.50$ & $2.15$ & $14.44$   \tabularnewline

3 & $976$ & $1.55$ & $1.97$ & $12.32$   \tabularnewline

4 & $1038$ & $0.98$ & $2.04$ & $14.02$   \tabularnewline

5 & $1058$ & $1.15$ & $2.14$ & $14.30$   \tabularnewline

6 & $1058$ & $1.03$ & $1.99$ & $12.87$   \tabularnewline

7 & $1260$ & $1.96$ & $2.01$ & $14.44$   \tabularnewline

8 & $1638$ & $2.15$ & $2.15$ & $14.44$   \tabularnewline

9 & $1232$ & $1.97$ & $1.97$ (N) & $14.44$
 \tabularnewline
 &  &  &  $1.98$ (Q) & \tabularnewline
  \hline\hline
  
  \end{tabular}
\end{center}
\end{table}
 
Moreover, after a scan of different parameterizations of the quark model, we identify in Set 9 a third family of stars. From Table \ref{Table2_info}, one can see that the critical chemical potential is slightly lower in comparison to other sets described by a Maxwell construction (Sets 7-8), which is associated to the lower value of the vector interaction coupling. However, the transition occurs for a baryon density of $\rho_{b,i} = 0.382\,\mathrm{fm^{-3}}$, ensuring that hyperons are present in the core of stars. For this configuration, the critical mass of $1.97\,\mathrm{M_{\odot}}$ corresponds to the maximum mass of the first branch (denoted by N, for nucleonic stars). For higher densities, a phase of instability takes place until a stable branch of hybrid stars rises for a mass of $1.968\,\mathrm{M_{\odot}}$, growing to a maximum mass of $1.978\,\mathrm{M_{\odot}}$ (denoted by Q, for stars with a quark core).  

In order to visualize the difference between these constructions, in Figure \ref{Figure1} we compare the mass-radius diagrams for all sets of parameters.
The different approaches affect directly the composition of stars and consequently their compactness.
For the $1.4\,\mathrm{M_{\odot}}$ star, from Table \ref{Table2_info} one can see  that the difference in radius can vary by more than $2\,\mathrm{km}$ when comparing hadronic and hybrid stars.

However, it is important to point out that parameterizations in Sets 3 and 6  
present a reconfinement feature at high densities  \cite{Heinimann:2016zbx}, which is removed by hand in Set 6, and ignored in Set 3 as the central densities of stars still allow for a mixed phase. Set 3 also presents a critical density slightly below saturation, but still above crust densities ($\rho_{b,c}=0.138\,\mathrm{fm^{-3}}$). In particular, Sets 3 and 6 present a lower value of the bag constant, which shifts the transition point to lower chemical potentials both for Maxwell and Gibbs constructions. The early transition leads to an increase of compactness of stars and, therefore, to lower radii. As we are going to see in the next session, this has dramatic impact on the tidal deformation of stars during mergers.


\subsection{Binary systems}

In the previous section we investigated how different features in the modeling of particle interactions affect the macroscopic properties of stars.
In the following, we explore the behavior of the tidal deformability when these stars are part of binary systems, according to the formalism presented in Session III. Note that, as Set 8 overlaps with Set 2, we do not display the results for this parametrization. 

\begin{figure}[!ht]  
  \centering
  \includegraphics[width=1.0\linewidth]{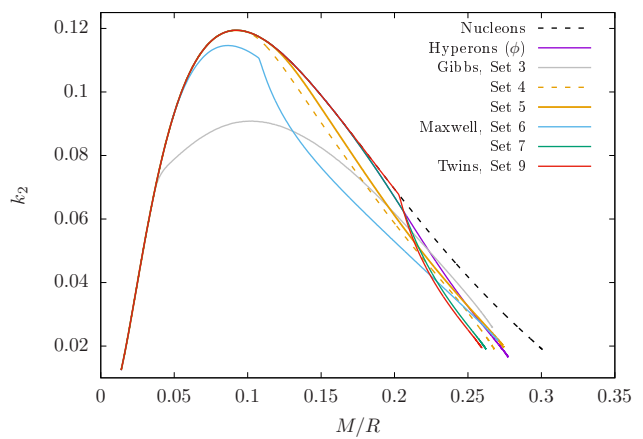}
  \caption{Love Number (vertical axis) as a function of compactness (horizontal axis) for different sets of parameters.}
\label{fig1}
\end{figure}

\begin{figure}[!ht]  
  \centering
  \includegraphics[width=1.0\linewidth]{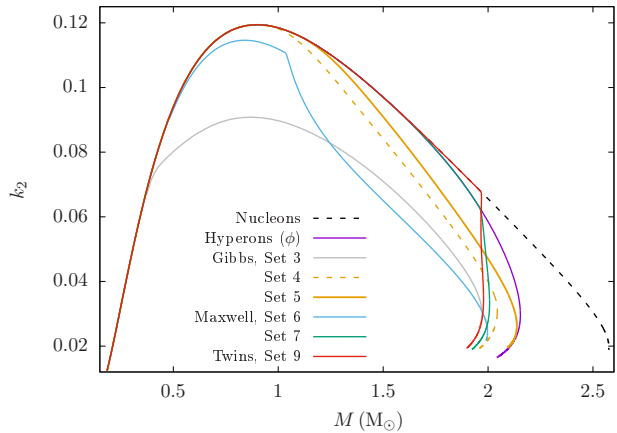}
  \caption{Same as Figure \ref{fig1}, plotted as function of the mass of the star.}
\label{fig2}
\end{figure}

Figures \ref{fig1} and \ref{fig2} show the values of the Love number $k_2$ as a function of compactness and mass, respectively. 
One can see that the spread is little for the sets considered in this analysis, when $k_2$ is shown as a function of compactness in comparison to the mass of the star. The value of $k_2$ peaks around $1\,\mathrm{M}_\odot$, while it is much lower at both higher and lower masses. This behavior comes from the fact that in both limits of mass, stars are more dense in the center and therefore more difficult to deform. 
In other words, if the mass is mostly concentrated in the center, then, regardless of how  deformed the outer region could become, it will not contribute much to the quadrupole moment. If the radius is large and the mass is also uniformly distributed, then the quadrupolar contribution is also larger.

\begin{figure}[!ht]  
  \centering
  \includegraphics[width=1.0\linewidth]{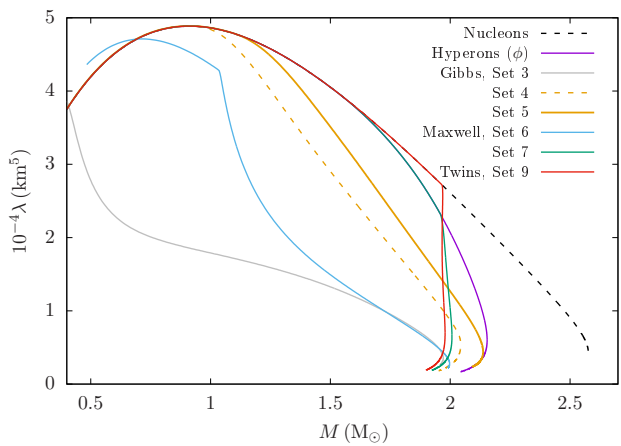}
  \caption{Tidal deformability (vertical axis) as a function of mass  for different sets of parameters.}
\label{fig3}
\end{figure}

\begin{figure}[!ht]  
  \centering
  \includegraphics[width=1.0\linewidth]{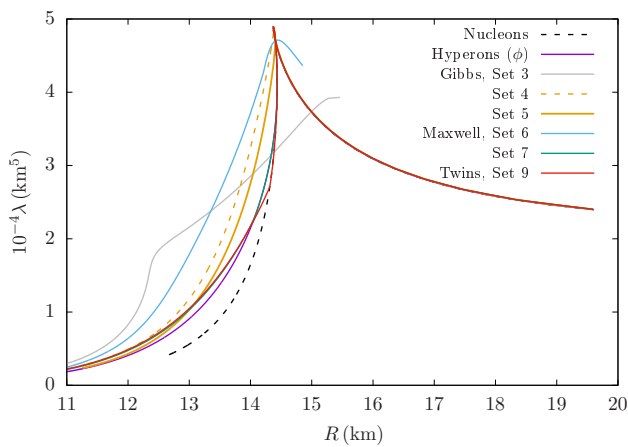}
  \caption{Same as Figure \ref{fig3}, as a function of the stellar radius.}
\label{fig4}
\end{figure}

Tidal deformability $\lambda$, which is directly measurable from GW observations, is plotted in Figures \ref{fig3} and \ref{fig4} as a function of mass and radius of the stars, respectively. 
Again, the peak in $\lambda$ occurs for stars of nearly $1$ M$_\odot$, but with a more pronounced spread than for the Love number $k_2$, which is a consequence of the dependence on $R^5$. Our results are consistent with previous studies \cite{Hinderer:2009ca,Postnikov:2010yn}, and are here reported for the first time in a broad analysis of hadron-quark transition comparing Gibbs and Maxwell constructions and their respective impact on the tidal deformability of hybrid stars. 

Regarding Gibbs and Maxwell constructions for hybrid stars, one can see that the range of values of $\lambda$ and $k_2$ lie essentially in the same region for different values of the vector interaction couplings, 
but branches of two sequences from nucleonic sequence is different, being smoother for a Gibbs construction, due to the presence of a mixed phase.
Also, as Gibbs transitions occur at lower densities than Maxwell, a considerable amount of quarks appears making the EoS softer and stars more compact, i.e. less deformed.
More specifically, from Figure \ref{fig3}, it is possible to see that the tidal deformability for a $1.6 M_{\odot}$ star, is $\lambda = 3.79127 \times 10^4\,\mathrm{km^5}$ for a nucleonic star (set 1), whereas for a hybrid star in a Gibbs construction (set 4) it is $\lambda = 2.49365 \times 10^4\,\mathrm{km^5}$, indicating a reduction of $\sim 35 \%$. For a $1.8\,\mathrm{M_{\odot}}$ star, 
the same comparison leads to an even larger relative reduction of $\sim 48 \%$. For the sets with a Maxwell construction, the value of $\lambda$ falls drastically immediately after the transition due to the sharper transition, indicating that this construction creates more compact stars in a similar mass range, in relation to sets 1 and 2.

\begin{figure}[!ht]  
  \centering
  \includegraphics[width=1.0\linewidth]{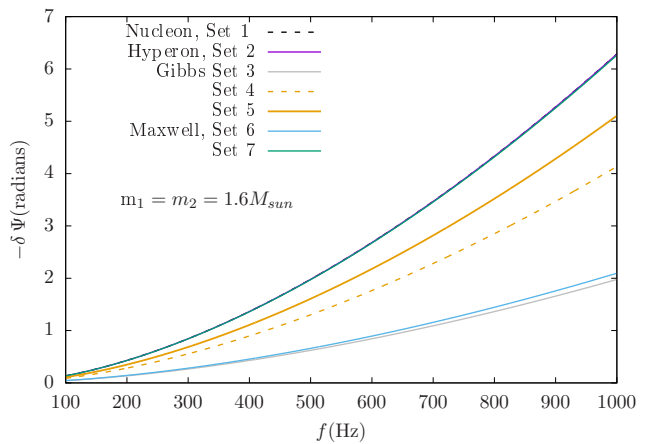}
  \caption{Total accumulated phase as a function of gravitational wave frequency for an equal mass binary system of $1.6$ $M_\odot$ stars. }
\label{fig5}
\end{figure}

\begin{figure}[!ht]  
  \centering
  \includegraphics[width=1.0\linewidth]{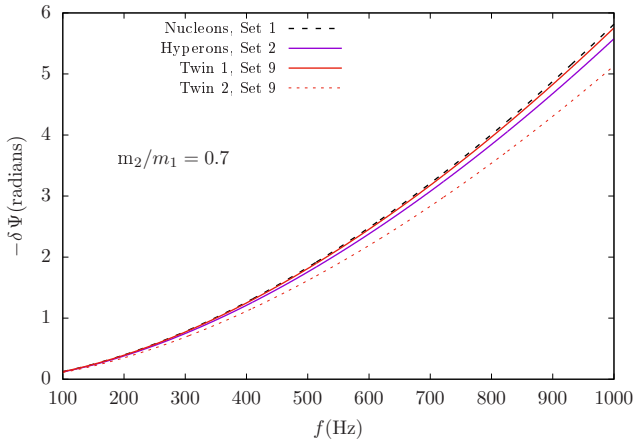}
  \caption{Same as Figure \ref{fig5}, but for a binary system of $1.97$ $M_\odot$ stars with a companion of $1.39$ $M_\odot$.}
\label{fig6}
\end{figure}

Furthermore, in order to study the phase lag, we plot the reduction of accumulated gravitational wave phase due to tidal interaction in Figures \ref{fig5} and \ref{fig6}. As the innermost stable circular orbit is supposed to correspond to a phase lag frequency between $400$ and $500$ Hz, which can be termed as the end of the early inspiral stage \cite{Hinderer:2009ca}, we fix the value at $450$ Hz, in order to analyze each set of parameters. 

For an equal mass binary where $M_1 = M_2 = 1.6\,\mathrm{M_\odot}$, from Equation \ref{eq:lambda} we have $\tilde\lambda = \lambda_1 = \lambda_2$. As shown in Figure \ref{fig5}, we find the curves for nucleonic (set 1), hyperonic (set 2) and most Maxwell constructions (sets 7,9) essentially overlap, as the mass of the system lies only slightly above the critical one for set 2 (only a few hyperons populate the star), and below the critical mass for the appearance of quarks for these parameterizations. However, for the Gibbs construction (sets 3-5) and set 6 with Maxwell, quarks are present in the core of stars starting at much lower densities. In order to quantify our results, the phase lag for a nucleonic system is $1.6614$ rad, and $1.0928$ rad for set 4, with an associated difference of $\sim 34 \%$. This means that the gravitational wave phase is more delayed when stars are more deformed (described with a stiffer EoS). 
These stars absorb more orbital energy, leading to a faster orbital decay and, hence, an earlier merger.

The weighted average of deformation $\tilde{\lambda}$ is the combined effect of both stellar deformations on the waveform, as defined in Eq. 20 of Ref. \cite{Hinderer:2009ca}. In Figure \ref{fig6}, we use this quantity for evaluating a pair of twin stars in set 9, defining Twin-1 (nucleonic) and Twin-2 (hybrid), for a fixed mass of $M$ = 1.97\,$\mathrm{M_\odot}$. We take the mass ratio of the stars in in the binary to $M_2/M_1=0.7$, for the partner star of mass, radius and deformation of $M$ = 1.39 $M_\odot$, $14.43$ km and $4.313 \times 10^4 ~\mathrm{km^5}$, respectively. The results for weighted average of deformation, phase lag and radius are shown in Table \ref{Table_197}, where we also compare with a similar mass ratio for nucleonic (set 1) and hyperonic (set 2) stars. From the results one can see once again that the more compact stars are less deformed and present also lower phase lag, as discussed before. 


\begin{table}[t]
  \caption{\label{Table_197} Properties of the $1.97\,\mathrm{M_{\odot}}$ star in a binary system described in Figure \ref{fig6}, for a particular binary where $m_1 = 1.97\,\mathrm{M_{\odot}}$, the mass ratio is $m_2/m_1=0.7$, and the values for phase correspond to the frequency $450\,\mathrm{Hz}$. The columns read: set of parameters, weighted average deformation $\tilde{\lambda}$, phase lag $-\delta\,\Psi$, compactness $M/R$  and radius, respectively.}
\begin{center}
\begin{tabular}{ccccc}
 \hline
Set & $ \tilde{\lambda}\,(\mathrm{ 10^ {4}\,km^{5}})$  & $-\delta\,\Psi\, (\mathrm{rad})$ &  $M/R$ & $R\,(\mathrm{km})$ \\
  \hline \hline
    
1          & 3.976 & 1.5365 & 0.138 & 14.32  \tabularnewline
2          & 3.814 & 1.4737 & 0.140 & 14.05  \tabularnewline
9 (Twin 1) & 3.934 & 1.5208 & 0.138 & 14.25  \tabularnewline
9 (Twin 2) & 3.510 & 1.3565 & 0.147 & 13.42  \tabularnewline
  \hline\hline  
  \end{tabular}
\end{center}
\end{table}

\begin{figure}[!ht]  
  \centering
  \includegraphics[width=1.0\linewidth]{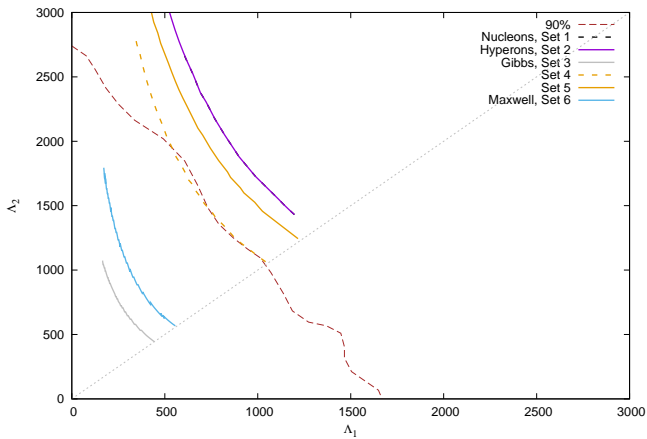}
  \caption{Tidal deformabilities associated with the individual components of the binary of GW170817.}
\label{fig8}
\end{figure}

Finally, we plot the individual tidal deformabilities ($\Lambda_1$ \& $\Lambda_2$) for each of the components of the binary associated with GW170817. The detected chirp mass [${\cal M}= (M_1 M_2)^{3/5}/(M_1+M_2)^{1/5}$] for this event is $1.188\,\mathrm{M_\odot}$. We vary the high mass component ($M_1$) from $1.36 M_\odot$ to $1.6 M_\odot$ keeping the chirp mass fixed while the low mass component ($M_2$) being varied from $1.17\,\mathrm{M_\odot}$ to $1.36\,\mathrm{M_\odot}$ and plot the corresponding $\Lambda$s in Figure \ref{fig8}. The brown dashed line signifies the $90\%$ probability contour found from the event GW170817 assuming low spinning components of the binary. This contour tells us that allowed EOS should form smaller and more compact stars. We also find the set 3 and set 6 comfortably satisfy the constraint set by the event. In both the cases, we get a very early phase transition due to the smaller value of the bag constant. Hence, a large quark fraction exists even in the low mass stars in the sequences which also have smaller radii leading to very compact object formation with smaller deformabilities. The stars on set 4  also have smaller radii due to early onset of the transition and lie very close to the contour. But, the other EOS have larger radii, therefore larger deformabilities, which place them fairly outside the constraint. 

\section{Conclusions}

The new era of multi-messenger astronomy and especially  the results from GW170817 make the tidal deformability a powerful tool to constrain the equation of nuclear and quark matter at high densities. 
In this work, we have investigated the impact of hadron-quarks phase transitions 
on the properties of isolated hybrid stars, as well as hybrid stars in binary systems. We have also for the first time compared hybrid stars in Gibbs and Maxwell construction on the light of the new data from GW170817.

For a fixed parametrization of the MBF model, we have changed the hadronic phase content, in order to also compare our results with nucleonic and hyperonic stars. The quark phase was described with a MIT Bag model with vector interaction, for Gibbs and Maxwell constructions, with the vector interaction coupling and bag constant as free parameters.  

We have shown that a Gibbs construction allows for a phase transition that happens at distinctly lower densities than the same set of parameters with a Maxwell construction. These results show a drastic impact on the composition of stars, making essentially the entire family of hybrid stars in a Gibbs construction contain only a thin lawyer of nucleonic matter and a core mostly composed of a mixture of phases, whereas for a  Maxwell construction, only massive stars are hybrid, with a large portion of their composition consisting of hadronic matter (including hyperons) and only a small quark core. We also identify a third family of stars using a Maxwell construction which allows for the description of high mass twin stars of up to $1.97\,\mathrm{M_{\odot}}$ (maximum mass for nucleonic) and $1.98\,\mathrm{M_{\odot}}$ (maximum mass for hybrid).

For binary systems, we have investigated the tidal deformability and love number for stars in the different sets (different compositions).  
We have shown that only two among the nine sets of parameters analyzed agree with all the current observational constraints. In particular, we have shown that hybrid stars described in a Gibbs construction are more compact and less deformed because of the higher fraction of quarks inside the stars due to an early phase transition in comparison to Maxwell constructions. 
We have also calculated the phase lag due to tidal interaction, showing that a higher phase lag leads to faster orbital decay, being a potential tool to distinguish twin stars.


It is important to stress that Maxwell and Gibbs constructions are to a certain extent only simplified ways of modeling the smoothness of the interface between the phases. In reality, the transition might be more complex and, therefore, effects such as surface and Coulomb interaction can and should be incorporated explicitly when describing realistic hybrid stars \cite{Wu:2017xaz}. 

Our results show that still both Gibbs and Maxwell scenarios are possible for the description of hybrid stars, as long as the transition takes place at low enough densities that allow for a high amount of quarks. However, future observations of new NS-NS binary systems can bring new insights on the distinguishing of these two possible constructions. Nucleonic and hyperonic parametrization of the MBF model used in our analysis do not fulfill the tidal deformability constraints from GW170817. In this context, further investigations on the correlation of nuclear matter at saturation and tidal deformability are already being carried out.

\begin{acknowledgements}
The authors acknowledge support from NewCompstar, COST Action MP 1304 and HIC for FAIR. R.O. Gomes would like to thank V. Dexheimer and S.O. Kepler for fruitful discussions and suggestions. P. Char acknowledges support from the Navajbai Ratan Tata Trust.

\end{acknowledgements}

\bibliography{main}

\begin{thebibliography}{153}%
\makeatletter
\providecommand \@ifxundefined [1]{%
 \@ifx{#1\undefined}
}%
\providecommand \@ifnum [1]{%
 \ifnum #1\expandafter \@firstoftwo
 \else \expandafter \@secondoftwo
 \fi
}%
\providecommand \@ifx [1]{%
 \ifx #1\expandafter \@firstoftwo
 \else \expandafter \@secondoftwo
 \fi
}%
\providecommand \natexlab [1]{#1}%
\providecommand \enquote  [1]{``#1''}%
\providecommand \bibnamefont  [1]{#1}%
\providecommand \bibfnamefont [1]{#1}%
\providecommand \citenamefont [1]{#1}%
\providecommand \href@noop [0]{\@secondoftwo}%
\providecommand \href [0]{\begingroup \@sanitize@url \@href}%
\providecommand \@href[1]{\@@startlink{#1}\@@href}%
\providecommand \@@href[1]{\endgroup#1\@@endlink}%
\providecommand \@sanitize@url [0]{\catcode `\\12\catcode `\$12\catcode
  `\&12\catcode `\#12\catcode `\^12\catcode `\_12\catcode `\%12\relax}%
\providecommand \@@startlink[1]{}%
\providecommand \@@endlink[0]{}%
\providecommand \url  [0]{\begingroup\@sanitize@url \@url }%
\providecommand \@url [1]{\endgroup\@href {#1}{\urlprefix }}%
\providecommand \urlprefix  [0]{URL }%
\providecommand \Eprint [0]{\href }%
\providecommand \doibase [0]{http://dx.doi.org/}%
\providecommand \selectlanguage [0]{\@gobble}%
\providecommand \bibinfo  [0]{\@secondoftwo}%
\providecommand \bibfield  [0]{\@secondoftwo}%
\providecommand \translation [1]{[#1]}%
\providecommand \BibitemOpen [0]{}%
\providecommand \bibitemStop [0]{}%
\providecommand \bibitemNoStop [0]{.\EOS\space}%
\providecommand \EOS [0]{\spacefactor3000\relax}%
\providecommand \BibitemShut  [1]{\csname bibitem#1\endcsname}%
\let\auto@bib@innerbib\@empty
\bibitem [{\citenamefont {Dexheimer}\ and\ \citenamefont
  {Schramm}(2008)}]{Dexheimer:2008ax}%
  \BibitemOpen
  \bibfield  {author} {\bibinfo {author} {\bibfnamefont {V.}~\bibnamefont
  {Dexheimer}}\ and\ \bibinfo {author} {\bibfnamefont {S.}~\bibnamefont
  {Schramm}},\ }\href {\doibase 10.1086/589735} {\bibfield  {journal} {\bibinfo
   {journal} {Astrophys. J.}\ }\textbf {\bibinfo {volume} {683}},\ \bibinfo
  {pages} {943} (\bibinfo {year} {2008})},\ \Eprint
  {http://arxiv.org/abs/0802.1999} {arXiv:0802.1999 [astro-ph]} \BibitemShut
  {NoStop}%
\bibitem [{\citenamefont {Gomes}\ \emph {et~al.}(2015)\citenamefont {Gomes},
  \citenamefont {Dexheimer}, \citenamefont {Schramm},\ and\ \citenamefont
  {Vasconcellos}}]{Gomes:2014aka}%
  \BibitemOpen
  \bibfield  {author} {\bibinfo {author} {\bibfnamefont {R.~O.}\ \bibnamefont
  {Gomes}}, \bibinfo {author} {\bibfnamefont {V.}~\bibnamefont {Dexheimer}},
  \bibinfo {author} {\bibfnamefont {S.}~\bibnamefont {Schramm}}, \ and\
  \bibinfo {author} {\bibfnamefont {C.~A.~Z.}\ \bibnamefont {Vasconcellos}},\
  }\href {\doibase 10.1088/0004-637X/808/1/8} {\bibfield  {journal} {\bibinfo
  {journal} {Astrophys. J.}\ }\textbf {\bibinfo {volume} {808}},\ \bibinfo
  {pages} {8} (\bibinfo {year} {2015})},\ \Eprint
  {http://arxiv.org/abs/1411.4875} {arXiv:1411.4875 [astro-ph.SR]} \BibitemShut
  {NoStop}%
\bibitem [{\citenamefont {Oertel}\ \emph {et~al.}(2015)\citenamefont {Oertel},
  \citenamefont {Providência}, \citenamefont {Gulminelli},\ and\ \citenamefont
  {Raduta}}]{Oertel:2014qza}%
  \BibitemOpen
  \bibfield  {author} {\bibinfo {author} {\bibfnamefont {M.}~\bibnamefont
  {Oertel}}, \bibinfo {author} {\bibfnamefont {C.}~\bibnamefont
  {Providência}}, \bibinfo {author} {\bibfnamefont {F.}~\bibnamefont
  {Gulminelli}}, \ and\ \bibinfo {author} {\bibfnamefont {A.~R.}\ \bibnamefont
  {Raduta}},\ }\href {\doibase 10.1088/0954-3899/42/7/075202} {\bibfield
  {journal} {\bibinfo  {journal} {J. Phys.}\ }\textbf {\bibinfo {volume}
  {G42}},\ \bibinfo {pages} {075202} (\bibinfo {year} {2015})},\ \Eprint
  {http://arxiv.org/abs/1412.4545} {arXiv:1412.4545 [nucl-th]} \BibitemShut
  {NoStop}%
\bibitem [{\citenamefont {Chatterjee}\ and\ \citenamefont
  {Vidaña}(2016)}]{Chatterjee:2015pua}%
  \BibitemOpen
  \bibfield  {author} {\bibinfo {author} {\bibfnamefont {D.}~\bibnamefont
  {Chatterjee}}\ and\ \bibinfo {author} {\bibfnamefont {I.}~\bibnamefont
  {Vidaña}},\ }\href {\doibase 10.1140/epja/i2016-16029-x} {\bibfield
  {journal} {\bibinfo  {journal} {Eur. Phys. J.}\ }\textbf {\bibinfo {volume}
  {A52}},\ \bibinfo {pages} {29} (\bibinfo {year} {2016})},\ \Eprint
  {http://arxiv.org/abs/1510.06306} {arXiv:1510.06306 [nucl-th]} \BibitemShut
  {NoStop}%
\bibitem [{\citenamefont {Vidaña}(2016)}]{Vidana:2015rsa}%
  \BibitemOpen
  \bibfield  {author} {\bibinfo {author} {\bibfnamefont {I.}~\bibnamefont
  {Vidaña}},\ }\bibfield  {booktitle} {\emph {\bibinfo {booktitle}
  {{Proceedings, 15th International Conference on Strangeness in Quark Matter
  (SQM 2015): Dubna, Moscow region, Russia, July 6-11, 2015}}},\ }\href
  {\doibase 10.1088/1742-6596/668/1/012031} {\bibfield  {journal} {\bibinfo
  {journal} {J. Phys. Conf. Ser.}\ }\textbf {\bibinfo {volume} {668}},\
  \bibinfo {pages} {012031} (\bibinfo {year} {2016})},\ \Eprint
  {http://arxiv.org/abs/1509.03587} {arXiv:1509.03587 [nucl-th]} \BibitemShut
  {NoStop}%
\bibitem [{\citenamefont {Yamamoto}\ \emph {et~al.}(2016)\citenamefont
  {Yamamoto}, \citenamefont {Furumoto}, \citenamefont {Yasutake},\ and\
  \citenamefont {Rijken}}]{Yamamoto:2015lwa}%
  \BibitemOpen
  \bibfield  {author} {\bibinfo {author} {\bibfnamefont {Y.}~\bibnamefont
  {Yamamoto}}, \bibinfo {author} {\bibfnamefont {T.}~\bibnamefont {Furumoto}},
  \bibinfo {author} {\bibfnamefont {N.}~\bibnamefont {Yasutake}}, \ and\
  \bibinfo {author} {\bibfnamefont {T.~A.}\ \bibnamefont {Rijken}},\ }\href
  {\doibase 10.1140/epja/i2016-16019-0} {\bibfield  {journal} {\bibinfo
  {journal} {Eur. Phys. J.}\ }\textbf {\bibinfo {volume} {A52}},\ \bibinfo
  {pages} {19} (\bibinfo {year} {2016})},\ \Eprint
  {http://arxiv.org/abs/1510.06099} {arXiv:1510.06099 [nucl-th]} \BibitemShut
  {NoStop}%
\bibitem [{\citenamefont {Torres}\ \emph {et~al.}(2017)\citenamefont {Torres},
  \citenamefont {Gulminelli},\ and\ \citenamefont {Menezes}}]{Torres:2016ydl}%
  \BibitemOpen
  \bibfield  {author} {\bibinfo {author} {\bibfnamefont {J.~R.}\ \bibnamefont
  {Torres}}, \bibinfo {author} {\bibfnamefont {F.}~\bibnamefont {Gulminelli}},
  \ and\ \bibinfo {author} {\bibfnamefont {D.~P.}\ \bibnamefont {Menezes}},\
  }\href {\doibase 10.1103/PhysRevC.95.025201} {\bibfield  {journal} {\bibinfo
  {journal} {Phys. Rev.}\ }\textbf {\bibinfo {volume} {C95}},\ \bibinfo {pages}
  {025201} (\bibinfo {year} {2017})},\ \Eprint
  {http://arxiv.org/abs/1608.05108} {arXiv:1608.05108 [nucl-th]} \BibitemShut
  {NoStop}%
\bibitem [{\citenamefont {Tolos}\ \emph
  {et~al.}(2017{\natexlab{a}})\citenamefont {Tolos}, \citenamefont
  {Centelles},\ and\ \citenamefont {Ramos}}]{Tolos:2016hhl}%
  \BibitemOpen
  \bibfield  {author} {\bibinfo {author} {\bibfnamefont {L.}~\bibnamefont
  {Tolos}}, \bibinfo {author} {\bibfnamefont {M.}~\bibnamefont {Centelles}}, \
  and\ \bibinfo {author} {\bibfnamefont {A.}~\bibnamefont {Ramos}},\ }\href
  {\doibase 10.3847/1538-4357/834/1/3} {\bibfield  {journal} {\bibinfo
  {journal} {Astrophys. J.}\ }\textbf {\bibinfo {volume} {834}},\ \bibinfo
  {pages} {3} (\bibinfo {year} {2017}{\natexlab{a}})},\ \Eprint
  {http://arxiv.org/abs/1610.00919} {arXiv:1610.00919 [astro-ph.HE]}
  \BibitemShut {NoStop}%
\bibitem [{\citenamefont {Mishra}\ \emph {et~al.}(2016)\citenamefont {Mishra},
  \citenamefont {Sahoo}, \citenamefont {Panda}, \citenamefont {Barik},\ and\
  \citenamefont {Frederico}}]{Mishra:2016qhw}%
  \BibitemOpen
  \bibfield  {author} {\bibinfo {author} {\bibfnamefont {R.~N.}\ \bibnamefont
  {Mishra}}, \bibinfo {author} {\bibfnamefont {H.~S.}\ \bibnamefont {Sahoo}},
  \bibinfo {author} {\bibfnamefont {P.~K.}\ \bibnamefont {Panda}}, \bibinfo
  {author} {\bibfnamefont {N.}~\bibnamefont {Barik}}, \ and\ \bibinfo {author}
  {\bibfnamefont {T.}~\bibnamefont {Frederico}},\ }\href {\doibase
  10.1103/PhysRevC.94.035805} {\bibfield  {journal} {\bibinfo  {journal} {Phys.
  Rev.}\ }\textbf {\bibinfo {volume} {C94}},\ \bibinfo {pages} {035805}
  (\bibinfo {year} {2016})},\ \Eprint {http://arxiv.org/abs/1609.02708}
  {arXiv:1609.02708 [nucl-th]} \BibitemShut {NoStop}%
\bibitem [{\citenamefont {Schurhoff}\ \emph {et~al.}(2010)\citenamefont
  {Schurhoff}, \citenamefont {Schramm},\ and\ \citenamefont
  {Dexheimer}}]{Schurhoff:2010ph}%
  \BibitemOpen
  \bibfield  {author} {\bibinfo {author} {\bibfnamefont {T.}~\bibnamefont
  {Schurhoff}}, \bibinfo {author} {\bibfnamefont {S.}~\bibnamefont {Schramm}},
  \ and\ \bibinfo {author} {\bibfnamefont {V.}~\bibnamefont {Dexheimer}},\
  }\href {\doibase 10.1088/2041-8205/724/1/L74} {\bibfield  {journal} {\bibinfo
   {journal} {Astrophys. J.}\ }\textbf {\bibinfo {volume} {724}},\ \bibinfo
  {pages} {L74} (\bibinfo {year} {2010})},\ \Eprint
  {http://arxiv.org/abs/1008.0957} {arXiv:1008.0957 [astro-ph.SR]} \BibitemShut
  {NoStop}%
\bibitem [{\citenamefont {Drago}\ \emph {et~al.}(2014)\citenamefont {Drago},
  \citenamefont {Lavagno},\ and\ \citenamefont {Pagliara}}]{Drago:2013fsa}%
  \BibitemOpen
  \bibfield  {author} {\bibinfo {author} {\bibfnamefont {A.}~\bibnamefont
  {Drago}}, \bibinfo {author} {\bibfnamefont {A.}~\bibnamefont {Lavagno}}, \
  and\ \bibinfo {author} {\bibfnamefont {G.}~\bibnamefont {Pagliara}},\ }\href
  {\doibase 10.1103/PhysRevD.89.043014} {\bibfield  {journal} {\bibinfo
  {journal} {Phys. Rev.}\ }\textbf {\bibinfo {volume} {D89}},\ \bibinfo {pages}
  {043014} (\bibinfo {year} {2014})},\ \Eprint {http://arxiv.org/abs/1309.7263}
  {arXiv:1309.7263 [nucl-th]} \BibitemShut {NoStop}%
\bibitem [{\citenamefont {Drago}\ \emph {et~al.}(2016)\citenamefont {Drago},
  \citenamefont {Lavagno}, \citenamefont {Pagliara},\ and\ \citenamefont
  {Pigato}}]{Drago:2015cea}%
  \BibitemOpen
  \bibfield  {author} {\bibinfo {author} {\bibfnamefont {A.}~\bibnamefont
  {Drago}}, \bibinfo {author} {\bibfnamefont {A.}~\bibnamefont {Lavagno}},
  \bibinfo {author} {\bibfnamefont {G.}~\bibnamefont {Pagliara}}, \ and\
  \bibinfo {author} {\bibfnamefont {D.}~\bibnamefont {Pigato}},\ }\href
  {\doibase 10.1140/epja/i2016-16040-3} {\bibfield  {journal} {\bibinfo
  {journal} {Eur. Phys. J.}\ }\textbf {\bibinfo {volume} {A52}},\ \bibinfo
  {pages} {40} (\bibinfo {year} {2016})},\ \Eprint
  {http://arxiv.org/abs/1509.02131} {arXiv:1509.02131 [astro-ph.SR]}
  \BibitemShut {NoStop}%
\bibitem [{\citenamefont {Cai}\ \emph {et~al.}(2015)\citenamefont {Cai},
  \citenamefont {Fattoyev}, \citenamefont {Li},\ and\ \citenamefont
  {Newton}}]{Cai:2015hya}%
  \BibitemOpen
  \bibfield  {author} {\bibinfo {author} {\bibfnamefont {B.-J.}\ \bibnamefont
  {Cai}}, \bibinfo {author} {\bibfnamefont {F.~J.}\ \bibnamefont {Fattoyev}},
  \bibinfo {author} {\bibfnamefont {B.-A.}\ \bibnamefont {Li}}, \ and\ \bibinfo
  {author} {\bibfnamefont {W.~G.}\ \bibnamefont {Newton}},\ }\href {\doibase
  10.1103/PhysRevC.92.015802} {\bibfield  {journal} {\bibinfo  {journal} {Phys.
  Rev.}\ }\textbf {\bibinfo {volume} {C92}},\ \bibinfo {pages} {015802}
  (\bibinfo {year} {2015})},\ \Eprint {http://arxiv.org/abs/1501.01680}
  {arXiv:1501.01680 [nucl-th]} \BibitemShut {NoStop}%
\bibitem [{\citenamefont {Zhu}\ \emph {et~al.}(2016)\citenamefont {Zhu},
  \citenamefont {Li}, \citenamefont {Hu},\ and\ \citenamefont
  {Sagawa}}]{Zhu:2016mtc}%
  \BibitemOpen
  \bibfield  {author} {\bibinfo {author} {\bibfnamefont {Z.-Y.}\ \bibnamefont
  {Zhu}}, \bibinfo {author} {\bibfnamefont {A.}~\bibnamefont {Li}}, \bibinfo
  {author} {\bibfnamefont {J.-N.}\ \bibnamefont {Hu}}, \ and\ \bibinfo {author}
  {\bibfnamefont {H.}~\bibnamefont {Sagawa}},\ }\href {\doibase
  10.1103/PhysRevC.94.045803} {\bibfield  {journal} {\bibinfo  {journal} {Phys.
  Rev.}\ }\textbf {\bibinfo {volume} {C94}},\ \bibinfo {pages} {045803}
  (\bibinfo {year} {2016})},\ \Eprint {http://arxiv.org/abs/1607.04007}
  {arXiv:1607.04007 [nucl-th]} \BibitemShut {NoStop}%
\bibitem [{\citenamefont {Menezes}\ \emph {et~al.}(2005)\citenamefont
  {Menezes}, \citenamefont {Panda},\ and\ \citenamefont
  {Providencia}}]{Menezes:2005ic}%
  \BibitemOpen
  \bibfield  {author} {\bibinfo {author} {\bibfnamefont {D.~P.}\ \bibnamefont
  {Menezes}}, \bibinfo {author} {\bibfnamefont {P.~K.}\ \bibnamefont {Panda}},
  \ and\ \bibinfo {author} {\bibfnamefont {C.}~\bibnamefont {Providencia}},\
  }\href {\doibase 10.1103/PhysRevC.72.035802} {\bibfield  {journal} {\bibinfo
  {journal} {Phys. Rev.}\ }\textbf {\bibinfo {volume} {C72}},\ \bibinfo {pages}
  {035802} (\bibinfo {year} {2005})},\ \Eprint
  {http://arxiv.org/abs/astro-ph/0506196} {arXiv:astro-ph/0506196 [astro-ph]}
  \BibitemShut {NoStop}%
\bibitem [{\citenamefont {Mishra}\ \emph {et~al.}(2010)\citenamefont {Mishra},
  \citenamefont {Kumar}, \citenamefont {Sanyal}, \citenamefont {Dexheimer},\
  and\ \citenamefont {Schramm}}]{Mishra:2009bp}%
  \BibitemOpen
  \bibfield  {author} {\bibinfo {author} {\bibfnamefont {A.}~\bibnamefont
  {Mishra}}, \bibinfo {author} {\bibfnamefont {A.}~\bibnamefont {Kumar}},
  \bibinfo {author} {\bibfnamefont {S.}~\bibnamefont {Sanyal}}, \bibinfo
  {author} {\bibfnamefont {V.}~\bibnamefont {Dexheimer}}, \ and\ \bibinfo
  {author} {\bibfnamefont {S.}~\bibnamefont {Schramm}},\ }\href {\doibase
  10.1140/epja/i2010-10986-x} {\bibfield  {journal} {\bibinfo  {journal} {Eur.
  Phys. J.}\ }\textbf {\bibinfo {volume} {A45}},\ \bibinfo {pages} {169}
  (\bibinfo {year} {2010})},\ \Eprint {http://arxiv.org/abs/0905.3518}
  {arXiv:0905.3518 [nucl-th]} \BibitemShut {NoStop}%
\bibitem [{\citenamefont {Alford}\ \emph {et~al.}(2010)\citenamefont {Alford},
  \citenamefont {Braby},\ and\ \citenamefont {Mahmoodifar}}]{Alford:2009jm}%
  \BibitemOpen
  \bibfield  {author} {\bibinfo {author} {\bibfnamefont {M.~G.}\ \bibnamefont
  {Alford}}, \bibinfo {author} {\bibfnamefont {M.}~\bibnamefont {Braby}}, \
  and\ \bibinfo {author} {\bibfnamefont {S.}~\bibnamefont {Mahmoodifar}},\
  }\href {\doibase 10.1103/PhysRevC.81.025202} {\bibfield  {journal} {\bibinfo
  {journal} {Phys. Rev.}\ }\textbf {\bibinfo {volume} {C81}},\ \bibinfo {pages}
  {025202} (\bibinfo {year} {2010})},\ \Eprint {http://arxiv.org/abs/0910.2180}
  {arXiv:0910.2180 [nucl-th]} \BibitemShut {NoStop}%
\bibitem [{\citenamefont {Fernandez}\ \emph {et~al.}(2010)\citenamefont
  {Fernandez}, \citenamefont {Mesquita}, \citenamefont {Razeira},\ and\
  \citenamefont {Vasconcellos}}]{Fernandez:2010zzc}%
  \BibitemOpen
  \bibfield  {author} {\bibinfo {author} {\bibfnamefont {F.}~\bibnamefont
  {Fernandez}}, \bibinfo {author} {\bibfnamefont {A.}~\bibnamefont {Mesquita}},
  \bibinfo {author} {\bibfnamefont {M.}~\bibnamefont {Razeira}}, \ and\
  \bibinfo {author} {\bibfnamefont {C.~A.~Z.}\ \bibnamefont {Vasconcellos}},\
  }\bibfield  {booktitle} {\emph {\bibinfo {booktitle} {{Astronomy and
  relativistic astrophysics. Proceedings, 4th International Workshop, IWARA
  2009, Maresias, Sao Paulo, Brazil, October 4-8, 2009}}},\ }\href {\doibase
  10.1142/S0218271810017299} {\bibfield  {journal} {\bibinfo  {journal} {Int.
  J. Mod. Phys.}\ }\textbf {\bibinfo {volume} {D19}},\ \bibinfo {pages} {1545}
  (\bibinfo {year} {2010})}\BibitemShut {NoStop}%
\bibitem [{\citenamefont {Mesquita}\ \emph {et~al.}(2010)\citenamefont
  {Mesquita}, \citenamefont {Razeira}, \citenamefont {Vasconcellos},\ and\
  \citenamefont {Fernandez}}]{Mesquita:2010zzb}%
  \BibitemOpen
  \bibfield  {author} {\bibinfo {author} {\bibfnamefont {A.}~\bibnamefont
  {Mesquita}}, \bibinfo {author} {\bibfnamefont {M.}~\bibnamefont {Razeira}},
  \bibinfo {author} {\bibfnamefont {C.~A.~Z.}\ \bibnamefont {Vasconcellos}}, \
  and\ \bibinfo {author} {\bibfnamefont {F.}~\bibnamefont {Fernandez}},\
  }\bibfield  {booktitle} {\emph {\bibinfo {booktitle} {{Astronomy and
  relativistic astrophysics. Proceedings, 4th International Workshop, IWARA
  2009, Maresias, Sao Paulo, Brazil, October 4-8, 2009}}},\ }\href {\doibase
  10.1142/S0218271810017366} {\bibfield  {journal} {\bibinfo  {journal} {Int.
  J. Mod. Phys.}\ }\textbf {\bibinfo {volume} {D19}},\ \bibinfo {pages} {1549}
  (\bibinfo {year} {2010})}\BibitemShut {NoStop}%
\bibitem [{\citenamefont {Lim}\ \emph {et~al.}(2014)\citenamefont {Lim},
  \citenamefont {Kwak}, \citenamefont {Hyun},\ and\ \citenamefont
  {Lee}}]{Lim:2013tqa}%
  \BibitemOpen
  \bibfield  {author} {\bibinfo {author} {\bibfnamefont {Y.}~\bibnamefont
  {Lim}}, \bibinfo {author} {\bibfnamefont {K.}~\bibnamefont {Kwak}}, \bibinfo
  {author} {\bibfnamefont {C.~H.}\ \bibnamefont {Hyun}}, \ and\ \bibinfo
  {author} {\bibfnamefont {C.-H.}\ \bibnamefont {Lee}},\ }\href {\doibase
  10.1103/PhysRevC.89.055804} {\bibfield  {journal} {\bibinfo  {journal} {Phys.
  Rev.}\ }\textbf {\bibinfo {volume} {C89}},\ \bibinfo {pages} {055804}
  (\bibinfo {year} {2014})},\ \Eprint {http://arxiv.org/abs/1312.2640}
  {arXiv:1312.2640 [nucl-th]} \BibitemShut {NoStop}%
\bibitem [{\citenamefont {Muto}\ \emph {et~al.}(2015)\citenamefont {Muto},
  \citenamefont {Maruyama},\ and\ \citenamefont {Tatsumi}}]{Muto:2015sgx}%
  \BibitemOpen
  \bibfield  {author} {\bibinfo {author} {\bibfnamefont {T.}~\bibnamefont
  {Muto}}, \bibinfo {author} {\bibfnamefont {T.}~\bibnamefont {Maruyama}}, \
  and\ \bibinfo {author} {\bibfnamefont {T.}~\bibnamefont {Tatsumi}},\
  }\href@noop {} {\bibfield  {journal} {\bibinfo  {journal} {Acta Astron.
  Sin.}\ }\textbf {\bibinfo {volume} {56}},\ \bibinfo {pages} {43} (\bibinfo
  {year} {2015})},\ \Eprint {http://arxiv.org/abs/1512.05137} {arXiv:1512.05137
  [nucl-th]} \BibitemShut {NoStop}%
\bibitem [{\citenamefont {Buballa}\ \emph {et~al.}(2004)\citenamefont
  {Buballa}, \citenamefont {Neumann}, \citenamefont {Oertel},\ and\
  \citenamefont {Shovkovy}}]{Buballa:2003et}%
  \BibitemOpen
  \bibfield  {author} {\bibinfo {author} {\bibfnamefont {M.}~\bibnamefont
  {Buballa}}, \bibinfo {author} {\bibfnamefont {F.}~\bibnamefont {Neumann}},
  \bibinfo {author} {\bibfnamefont {M.}~\bibnamefont {Oertel}}, \ and\ \bibinfo
  {author} {\bibfnamefont {I.}~\bibnamefont {Shovkovy}},\ }\href {\doibase
  10.1016/j.physletb.2004.05.064} {\bibfield  {journal} {\bibinfo  {journal}
  {Phys. Lett.}\ }\textbf {\bibinfo {volume} {B595}},\ \bibinfo {pages} {36}
  (\bibinfo {year} {2004})},\ \Eprint {http://arxiv.org/abs/nucl-th/0312078}
  {arXiv:nucl-th/0312078 [nucl-th]} \BibitemShut {NoStop}%
\bibitem [{\citenamefont {Alford}\ \emph {et~al.}(2005)\citenamefont {Alford},
  \citenamefont {Braby}, \citenamefont {Paris},\ and\ \citenamefont
  {Reddy}}]{Alford:2004pf}%
  \BibitemOpen
  \bibfield  {author} {\bibinfo {author} {\bibfnamefont {M.}~\bibnamefont
  {Alford}}, \bibinfo {author} {\bibfnamefont {M.}~\bibnamefont {Braby}},
  \bibinfo {author} {\bibfnamefont {M.~W.}\ \bibnamefont {Paris}}, \ and\
  \bibinfo {author} {\bibfnamefont {S.}~\bibnamefont {Reddy}},\ }\href
  {\doibase 10.1086/430902} {\bibfield  {journal} {\bibinfo  {journal}
  {Astrophys. J.}\ }\textbf {\bibinfo {volume} {629}},\ \bibinfo {pages} {969}
  (\bibinfo {year} {2005})},\ \Eprint {http://arxiv.org/abs/nucl-th/0411016}
  {arXiv:nucl-th/0411016 [nucl-th]} \BibitemShut {NoStop}%
\bibitem [{\citenamefont {Bombaci}\ \emph {et~al.}(2007)\citenamefont
  {Bombaci}, \citenamefont {Lugones},\ and\ \citenamefont
  {Vidana}}]{Bombaci:2006cs}%
  \BibitemOpen
  \bibfield  {author} {\bibinfo {author} {\bibfnamefont {I.}~\bibnamefont
  {Bombaci}}, \bibinfo {author} {\bibfnamefont {G.}~\bibnamefont {Lugones}}, \
  and\ \bibinfo {author} {\bibfnamefont {I.}~\bibnamefont {Vidana}},\ }\href
  {\doibase 10.1051/0004-6361:20065259} {\bibfield  {journal} {\bibinfo
  {journal} {Astron. Astrophys.}\ }\textbf {\bibinfo {volume} {462}},\ \bibinfo
  {pages} {1017} (\bibinfo {year} {2007})},\ \Eprint
  {http://arxiv.org/abs/astro-ph/0603644} {arXiv:astro-ph/0603644 [astro-ph]}
  \BibitemShut {NoStop}%
\bibitem [{\citenamefont {Bonanno}\ and\ \citenamefont
  {Sedrakian}(2012)}]{Bonanno:2011ch}%
  \BibitemOpen
  \bibfield  {author} {\bibinfo {author} {\bibfnamefont {L.}~\bibnamefont
  {Bonanno}}\ and\ \bibinfo {author} {\bibfnamefont {A.}~\bibnamefont
  {Sedrakian}},\ }\href {\doibase 10.1051/0004-6361/201117832} {\bibfield
  {journal} {\bibinfo  {journal} {Astron. Astrophys.}\ }\textbf {\bibinfo
  {volume} {539}},\ \bibinfo {pages} {A16} (\bibinfo {year} {2012})},\ \Eprint
  {http://arxiv.org/abs/1108.0559} {arXiv:1108.0559 [astro-ph.SR]} \BibitemShut
  {NoStop}%
\bibitem [{\citenamefont {Alford}\ and\ \citenamefont
  {Sedrakian}(2017)}]{Alford:2017qgh}%
  \BibitemOpen
  \bibfield  {author} {\bibinfo {author} {\bibfnamefont {M.~G.}\ \bibnamefont
  {Alford}}\ and\ \bibinfo {author} {\bibfnamefont {A.}~\bibnamefont
  {Sedrakian}},\ }\href {\doibase 10.1103/PhysRevLett.119.161104} {\bibfield
  {journal} {\bibinfo  {journal} {Phys. Rev. Lett.}\ }\textbf {\bibinfo
  {volume} {119}},\ \bibinfo {pages} {161104} (\bibinfo {year} {2017})},\
  \Eprint {http://arxiv.org/abs/1706.01592} {arXiv:1706.01592 [astro-ph.HE]}
  \BibitemShut {NoStop}%
\bibitem [{\citenamefont {Hempel}\ \emph {et~al.}(2009)\citenamefont {Hempel},
  \citenamefont {Pagliara},\ and\ \citenamefont
  {Schaffner-Bielich}}]{Hempel:2009vp}%
  \BibitemOpen
  \bibfield  {author} {\bibinfo {author} {\bibfnamefont {M.}~\bibnamefont
  {Hempel}}, \bibinfo {author} {\bibfnamefont {G.}~\bibnamefont {Pagliara}}, \
  and\ \bibinfo {author} {\bibfnamefont {J.}~\bibnamefont
  {Schaffner-Bielich}},\ }\href {\doibase 10.1103/PhysRevD.80.125014}
  {\bibfield  {journal} {\bibinfo  {journal} {Phys. Rev.}\ }\textbf {\bibinfo
  {volume} {D80}},\ \bibinfo {pages} {125014} (\bibinfo {year} {2009})},\
  \Eprint {http://arxiv.org/abs/0907.2680} {arXiv:0907.2680 [astro-ph.HE]}
  \BibitemShut {NoStop}%
\bibitem [{\citenamefont {Klähn}\ \emph {et~al.}(2013)\citenamefont {Klähn},
  \citenamefont {Łastowiecki},\ and\ \citenamefont
  {Blaschke}}]{Klahn:2013kga}%
  \BibitemOpen
  \bibfield  {author} {\bibinfo {author} {\bibfnamefont {T.}~\bibnamefont
  {Klähn}}, \bibinfo {author} {\bibfnamefont {R.}~\bibnamefont
  {Łastowiecki}}, \ and\ \bibinfo {author} {\bibfnamefont {D.~B.}\
  \bibnamefont {Blaschke}},\ }\href {\doibase 10.1103/PhysRevD.88.085001}
  {\bibfield  {journal} {\bibinfo  {journal} {Phys. Rev.}\ }\textbf {\bibinfo
  {volume} {D88}},\ \bibinfo {pages} {085001} (\bibinfo {year} {2013})},\
  \Eprint {http://arxiv.org/abs/1307.6996} {arXiv:1307.6996 [nucl-th]}
  \BibitemShut {NoStop}%
\bibitem [{\citenamefont {Schaffner}\ and\ \citenamefont
  {Mishustin}(1996)}]{Schaffner:1995th}%
  \BibitemOpen
  \bibfield  {author} {\bibinfo {author} {\bibfnamefont {J.}~\bibnamefont
  {Schaffner}}\ and\ \bibinfo {author} {\bibfnamefont {I.~N.}\ \bibnamefont
  {Mishustin}},\ }\href {\doibase 10.1103/PhysRevC.53.1416} {\bibfield
  {journal} {\bibinfo  {journal} {Phys. Rev.}\ }\textbf {\bibinfo {volume}
  {C53}},\ \bibinfo {pages} {1416} (\bibinfo {year} {1996})},\ \Eprint
  {http://arxiv.org/abs/nucl-th/9506011} {arXiv:nucl-th/9506011 [nucl-th]}
  \BibitemShut {NoStop}%
\bibitem [{\citenamefont {Bednarek}\ \emph {et~al.}(2012)\citenamefont
  {Bednarek}, \citenamefont {Haensel}, \citenamefont {Zdunik}, \citenamefont
  {Bejger},\ and\ \citenamefont {Manka}}]{Bednarek:2011gd}%
  \BibitemOpen
  \bibfield  {author} {\bibinfo {author} {\bibfnamefont {I.}~\bibnamefont
  {Bednarek}}, \bibinfo {author} {\bibfnamefont {P.}~\bibnamefont {Haensel}},
  \bibinfo {author} {\bibfnamefont {J.~L.}\ \bibnamefont {Zdunik}}, \bibinfo
  {author} {\bibfnamefont {M.}~\bibnamefont {Bejger}}, \ and\ \bibinfo {author}
  {\bibfnamefont {R.}~\bibnamefont {Manka}},\ }\href {\doibase
  10.1051/0004-6361/201118560} {\bibfield  {journal} {\bibinfo  {journal}
  {Astron. Astrophys.}\ }\textbf {\bibinfo {volume} {543}},\ \bibinfo {pages}
  {A157} (\bibinfo {year} {2012})},\ \Eprint {http://arxiv.org/abs/1111.6942}
  {arXiv:1111.6942 [astro-ph.SR]} \BibitemShut {NoStop}%
\bibitem [{\citenamefont {Zhao}(2015)}]{Zhao:2015ncr}%
  \BibitemOpen
  \bibfield  {author} {\bibinfo {author} {\bibfnamefont {X.-F.}\ \bibnamefont
  {Zhao}},\ }\href {\doibase 10.1103/PhysRevC.92.055802} {\bibfield  {journal}
  {\bibinfo  {journal} {Phys. Rev.}\ }\textbf {\bibinfo {volume} {C92}},\
  \bibinfo {pages} {055802} (\bibinfo {year} {2015})},\ \Eprint
  {http://arxiv.org/abs/1712.08856} {arXiv:1712.08856 [nucl-th]} \BibitemShut
  {NoStop}%
\bibitem [{\citenamefont {Haidenbauer}\ \emph {et~al.}(2017)\citenamefont
  {Haidenbauer}, \citenamefont {Meißner}, \citenamefont {Kaiser},\ and\
  \citenamefont {Weise}}]{Haidenbauer:2016vfq}%
  \BibitemOpen
  \bibfield  {author} {\bibinfo {author} {\bibfnamefont {J.}~\bibnamefont
  {Haidenbauer}}, \bibinfo {author} {\bibfnamefont {U.~G.}\ \bibnamefont
  {Meißner}}, \bibinfo {author} {\bibfnamefont {N.}~\bibnamefont {Kaiser}}, \
  and\ \bibinfo {author} {\bibfnamefont {W.}~\bibnamefont {Weise}},\ }\href
  {\doibase 10.1140/epja/i2017-12316-4} {\bibfield  {journal} {\bibinfo
  {journal} {Eur. Phys. J.}\ }\textbf {\bibinfo {volume} {A53}},\ \bibinfo
  {pages} {121} (\bibinfo {year} {2017})},\ \Eprint
  {http://arxiv.org/abs/1612.03758} {arXiv:1612.03758 [nucl-th]} \BibitemShut
  {NoStop}%
\bibitem [{\citenamefont {Bombaci}(2017)}]{Bombaci:2016xzl}%
  \BibitemOpen
  \bibfield  {author} {\bibinfo {author} {\bibfnamefont {I.}~\bibnamefont
  {Bombaci}},\ }\bibfield  {booktitle} {\emph {\bibinfo {booktitle}
  {{Proceedings, 12th International Conference on Hypernuclear and Strange
  Particle Physics (HYP 2015): Sendai, Japan, September 7-12, 2015}}},\ }\href
  {\doibase 10.7566/JPSCP.17.101002} {\bibfield  {journal} {\bibinfo  {journal}
  {JPS Conf. Proc.}\ }\textbf {\bibinfo {volume} {17}},\ \bibinfo {pages}
  {101002} (\bibinfo {year} {2017})},\ \Eprint
  {http://arxiv.org/abs/1601.05339} {arXiv:1601.05339 [nucl-th]} \BibitemShut
  {NoStop}%
\bibitem [{\citenamefont {Fortin}\ \emph {et~al.}(2016)\citenamefont {Fortin},
  \citenamefont {Providencia}, \citenamefont {Raduta}, \citenamefont
  {Gulminelli}, \citenamefont {Zdunik}, \citenamefont {Haensel},\ and\
  \citenamefont {Bejger}}]{Fortin:2016hny}%
  \BibitemOpen
  \bibfield  {author} {\bibinfo {author} {\bibfnamefont {M.}~\bibnamefont
  {Fortin}}, \bibinfo {author} {\bibfnamefont {C.}~\bibnamefont {Providencia}},
  \bibinfo {author} {\bibfnamefont {A.~R.}\ \bibnamefont {Raduta}}, \bibinfo
  {author} {\bibfnamefont {F.}~\bibnamefont {Gulminelli}}, \bibinfo {author}
  {\bibfnamefont {J.~L.}\ \bibnamefont {Zdunik}}, \bibinfo {author}
  {\bibfnamefont {P.}~\bibnamefont {Haensel}}, \ and\ \bibinfo {author}
  {\bibfnamefont {M.}~\bibnamefont {Bejger}},\ }\href {\doibase
  10.1103/PhysRevC.94.035804} {\bibfield  {journal} {\bibinfo  {journal} {Phys.
  Rev.}\ }\textbf {\bibinfo {volume} {C94}},\ \bibinfo {pages} {035804}
  (\bibinfo {year} {2016})},\ \Eprint {http://arxiv.org/abs/1604.01944}
  {arXiv:1604.01944 [astro-ph.SR]} \BibitemShut {NoStop}%
\bibitem [{\citenamefont {Tolos}\ \emph
  {et~al.}(2017{\natexlab{b}})\citenamefont {Tolos}, \citenamefont
  {Centelles},\ and\ \citenamefont {Ramos}}]{Tolos:2017lgv}%
  \BibitemOpen
  \bibfield  {author} {\bibinfo {author} {\bibfnamefont {L.}~\bibnamefont
  {Tolos}}, \bibinfo {author} {\bibfnamefont {M.}~\bibnamefont {Centelles}}, \
  and\ \bibinfo {author} {\bibfnamefont {A.}~\bibnamefont {Ramos}},\ }\href
  {\doibase 10.1017/pasa.2017.60} {\bibfield  {journal} {\bibinfo  {journal}
  {Publ. Astron. Soc. Austral.}\ }\textbf {\bibinfo {volume} {34}},\ \bibinfo
  {pages} {e065} (\bibinfo {year} {2017}{\natexlab{b}})},\ \Eprint
  {http://arxiv.org/abs/1708.08681} {arXiv:1708.08681 [astro-ph.HE]}
  \BibitemShut {NoStop}%
\bibitem [{\citenamefont {de~Oliveira}\ \emph {et~al.}(2018)\citenamefont
  {de~Oliveira}, \citenamefont {Menezes}, \citenamefont {Pinto},\ and\
  \citenamefont {Gulminelli}}]{deOliveira:2017gty}%
  \BibitemOpen
  \bibfield  {author} {\bibinfo {author} {\bibfnamefont {T.}~\bibnamefont
  {de~Oliveira}}, \bibinfo {author} {\bibfnamefont {D.~P.}\ \bibnamefont
  {Menezes}}, \bibinfo {author} {\bibfnamefont {M.~B.}\ \bibnamefont {Pinto}},
  \ and\ \bibinfo {author} {\bibfnamefont {F.}~\bibnamefont {Gulminelli}},\
  }\href {\doibase 10.1103/PhysRevC.97.055205} {\bibfield  {journal} {\bibinfo
  {journal} {Phys. Rev.}\ }\textbf {\bibinfo {volume} {C97}},\ \bibinfo {pages}
  {055205} (\bibinfo {year} {2018})},\ \Eprint
  {http://arxiv.org/abs/1712.04526} {arXiv:1712.04526 [nucl-th]} \BibitemShut
  {NoStop}%
\bibitem [{\citenamefont {Weissenborn}\ \emph {et~al.}(2012)\citenamefont
  {Weissenborn}, \citenamefont {Chatterjee},\ and\ \citenamefont
  {Schaffner-Bielich}}]{Weissenborn:2011kb}%
  \BibitemOpen
  \bibfield  {author} {\bibinfo {author} {\bibfnamefont {S.}~\bibnamefont
  {Weissenborn}}, \bibinfo {author} {\bibfnamefont {D.}~\bibnamefont
  {Chatterjee}}, \ and\ \bibinfo {author} {\bibfnamefont {J.}~\bibnamefont
  {Schaffner-Bielich}},\ }\bibfield  {booktitle} {\emph {\bibinfo {booktitle}
  {{Progress in strangeness nuclear physics. Proceedings, ECT Workshop on
  Strange Hadronic Matter, Trento, Italy, September 26-30, 2011}}},\ }\href
  {\doibase 10.1016/j.nuclphysa.2012.02.012} {\bibfield  {journal} {\bibinfo
  {journal} {Nucl. Phys.}\ }\textbf {\bibinfo {volume} {A881}},\ \bibinfo
  {pages} {62} (\bibinfo {year} {2012})},\ \Eprint
  {http://arxiv.org/abs/1111.6049} {arXiv:1111.6049 [astro-ph.HE]} \BibitemShut
  {NoStop}%
\bibitem [{\citenamefont {Sun}\ \emph {et~al.}(2018)\citenamefont {Sun},
  \citenamefont {Xia}, \citenamefont {Zhang},\ and\ \citenamefont
  {Smith}}]{Sun:2017fnf}%
  \BibitemOpen
  \bibfield  {author} {\bibinfo {author} {\bibfnamefont {T.-T.}\ \bibnamefont
  {Sun}}, \bibinfo {author} {\bibfnamefont {C.-J.}\ \bibnamefont {Xia}},
  \bibinfo {author} {\bibfnamefont {S.-S.}\ \bibnamefont {Zhang}}, \ and\
  \bibinfo {author} {\bibfnamefont {M.~S.}\ \bibnamefont {Smith}},\ }\href
  {\doibase 10.1088/1674-1137/42/2/025101} {\bibfield  {journal} {\bibinfo
  {journal} {Chin. Phys.}\ }\textbf {\bibinfo {volume} {C42}},\ \bibinfo
  {pages} {025101} (\bibinfo {year} {2018})},\ \Eprint
  {http://arxiv.org/abs/1712.05569} {arXiv:1712.05569 [nucl-th]} \BibitemShut
  {NoStop}%
\bibitem [{\citenamefont {Fortin}\ \emph {et~al.}(2017)\citenamefont {Fortin},
  \citenamefont {Avancini}, \citenamefont {Providência},\ and\ \citenamefont
  {Vidaña}}]{Fortin:2017cvt}%
  \BibitemOpen
  \bibfield  {author} {\bibinfo {author} {\bibfnamefont {M.}~\bibnamefont
  {Fortin}}, \bibinfo {author} {\bibfnamefont {S.~S.}\ \bibnamefont
  {Avancini}}, \bibinfo {author} {\bibfnamefont {C.}~\bibnamefont
  {Providência}}, \ and\ \bibinfo {author} {\bibfnamefont {I.}~\bibnamefont
  {Vidaña}},\ }\href {\doibase 10.1103/PhysRevC.95.065803} {\bibfield
  {journal} {\bibinfo  {journal} {Phys. Rev.}\ }\textbf {\bibinfo {volume}
  {C95}},\ \bibinfo {pages} {065803} (\bibinfo {year} {2017})},\ \Eprint
  {http://arxiv.org/abs/1701.06373} {arXiv:1701.06373 [nucl-th]} \BibitemShut
  {NoStop}%
\bibitem [{\citenamefont {Shuryak}(2009)}]{Shuryak:2008eq}%
  \BibitemOpen
  \bibfield  {author} {\bibinfo {author} {\bibfnamefont {E.}~\bibnamefont
  {Shuryak}},\ }\href {\doibase 10.1016/j.ppnp.2008.09.001} {\bibfield
  {journal} {\bibinfo  {journal} {Prog. Part. Nucl. Phys.}\ }\textbf {\bibinfo
  {volume} {62}},\ \bibinfo {pages} {48} (\bibinfo {year} {2009})},\ \Eprint
  {http://arxiv.org/abs/0807.3033} {arXiv:0807.3033 [hep-ph]} \BibitemShut
  {NoStop}%
\bibitem [{\citenamefont {Bombaci}\ \emph {et~al.}(2009)\citenamefont
  {Bombaci}, \citenamefont {Logoteta}, \citenamefont {Panda}, \citenamefont
  {Providencia},\ and\ \citenamefont {Vidana}}]{Bombaci:2009jt}%
  \BibitemOpen
  \bibfield  {author} {\bibinfo {author} {\bibfnamefont {I.}~\bibnamefont
  {Bombaci}}, \bibinfo {author} {\bibfnamefont {D.}~\bibnamefont {Logoteta}},
  \bibinfo {author} {\bibfnamefont {P.~K.}\ \bibnamefont {Panda}}, \bibinfo
  {author} {\bibfnamefont {C.}~\bibnamefont {Providencia}}, \ and\ \bibinfo
  {author} {\bibfnamefont {I.}~\bibnamefont {Vidana}},\ }\href {\doibase
  10.1016/j.physletb.2009.09.039} {\bibfield  {journal} {\bibinfo  {journal}
  {Phys. Lett.}\ }\textbf {\bibinfo {volume} {B680}},\ \bibinfo {pages} {448}
  (\bibinfo {year} {2009})},\ \Eprint {http://arxiv.org/abs/0910.4109}
  {arXiv:0910.4109 [astro-ph.SR]} \BibitemShut {NoStop}%
\bibitem [{\citenamefont {Bombaci}\ \emph {et~al.}(2016)\citenamefont
  {Bombaci}, \citenamefont {Logoteta}, \citenamefont {Vidaña},\ and\
  \citenamefont {Providência}}]{Bombaci:2016xuj}%
  \BibitemOpen
  \bibfield  {author} {\bibinfo {author} {\bibfnamefont {I.}~\bibnamefont
  {Bombaci}}, \bibinfo {author} {\bibfnamefont {D.}~\bibnamefont {Logoteta}},
  \bibinfo {author} {\bibfnamefont {I.}~\bibnamefont {Vidaña}}, \ and\
  \bibinfo {author} {\bibfnamefont {C.}~\bibnamefont {Providência}},\ }\href
  {\doibase 10.1140/epja/i2016-16058-5} {\bibfield  {journal} {\bibinfo
  {journal} {Eur. Phys. J.}\ }\textbf {\bibinfo {volume} {A52}},\ \bibinfo
  {pages} {58} (\bibinfo {year} {2016})},\ \Eprint
  {http://arxiv.org/abs/1601.04559} {arXiv:1601.04559 [astro-ph.HE]}
  \BibitemShut {NoStop}%
\bibitem [{\citenamefont {Ippolito}\ \emph {et~al.}(2008)\citenamefont
  {Ippolito}, \citenamefont {Ruggieri}, \citenamefont {Rischke}, \citenamefont
  {Sedrakian},\ and\ \citenamefont {Weber}}]{Ippolito:2007hn}%
  \BibitemOpen
  \bibfield  {author} {\bibinfo {author} {\bibfnamefont {N.}~\bibnamefont
  {Ippolito}}, \bibinfo {author} {\bibfnamefont {M.}~\bibnamefont {Ruggieri}},
  \bibinfo {author} {\bibfnamefont {D.}~\bibnamefont {Rischke}}, \bibinfo
  {author} {\bibfnamefont {A.}~\bibnamefont {Sedrakian}}, \ and\ \bibinfo
  {author} {\bibfnamefont {F.}~\bibnamefont {Weber}},\ }\href {\doibase
  10.1103/PhysRevD.77.023004} {\bibfield  {journal} {\bibinfo  {journal} {Phys.
  Rev.}\ }\textbf {\bibinfo {volume} {D77}},\ \bibinfo {pages} {023004}
  (\bibinfo {year} {2008})},\ \Eprint {http://arxiv.org/abs/0710.3874}
  {arXiv:0710.3874 [astro-ph]} \BibitemShut {NoStop}%
\bibitem [{\citenamefont {Pereira}\ \emph {et~al.}(2017)\citenamefont
  {Pereira}, \citenamefont {Flores},\ and\ \citenamefont
  {Lugones}}]{Pereira:2017rmp}%
  \BibitemOpen
  \bibfield  {author} {\bibinfo {author} {\bibfnamefont {J.~P.}\ \bibnamefont
  {Pereira}}, \bibinfo {author} {\bibfnamefont {C.~V.}\ \bibnamefont {Flores}},
  \ and\ \bibinfo {author} {\bibfnamefont {G.}~\bibnamefont {Lugones}},\
  }\href@noop {} {\  (\bibinfo {year} {2017})},\ \Eprint
  {http://arxiv.org/abs/1706.09371} {arXiv:1706.09371 [gr-qc]} \BibitemShut
  {NoStop}%
\bibitem [{\citenamefont {Glendenning}\ \emph {et~al.}(1997)\citenamefont
  {Glendenning}, \citenamefont {Pei},\ and\ \citenamefont
  {Weber}}]{Glendenning:1997fy}%
  \BibitemOpen
  \bibfield  {author} {\bibinfo {author} {\bibfnamefont {N.~K.}\ \bibnamefont
  {Glendenning}}, \bibinfo {author} {\bibfnamefont {S.}~\bibnamefont {Pei}}, \
  and\ \bibinfo {author} {\bibfnamefont {F.}~\bibnamefont {Weber}},\ }\href
  {\doibase 10.1103/PhysRevLett.79.1603} {\bibfield  {journal} {\bibinfo
  {journal} {Phys. Rev. Lett.}\ }\textbf {\bibinfo {volume} {79}},\ \bibinfo
  {pages} {1603} (\bibinfo {year} {1997})},\ \Eprint
  {http://arxiv.org/abs/astro-ph/9705235} {arXiv:astro-ph/9705235 [astro-ph]}
  \BibitemShut {NoStop}%
\bibitem [{\citenamefont {Chubarian}\ \emph {et~al.}(2000)\citenamefont
  {Chubarian}, \citenamefont {Grigorian}, \citenamefont {Poghosyan},\ and\
  \citenamefont {Blaschke}}]{Chubarian:1999yn}%
  \BibitemOpen
  \bibfield  {author} {\bibinfo {author} {\bibfnamefont {E.}~\bibnamefont
  {Chubarian}}, \bibinfo {author} {\bibfnamefont {H.}~\bibnamefont
  {Grigorian}}, \bibinfo {author} {\bibfnamefont {G.~S.}\ \bibnamefont
  {Poghosyan}}, \ and\ \bibinfo {author} {\bibfnamefont {D.}~\bibnamefont
  {Blaschke}},\ }\href@noop {} {\bibfield  {journal} {\bibinfo  {journal}
  {Astron. Astrophys.}\ }\textbf {\bibinfo {volume} {357}},\ \bibinfo {pages}
  {968} (\bibinfo {year} {2000})},\ \Eprint
  {http://arxiv.org/abs/astro-ph/9903489} {arXiv:astro-ph/9903489 [astro-ph]}
  \BibitemShut {NoStop}%
\bibitem [{\citenamefont {Glendenning}\ and\ \citenamefont
  {Weber}(2001)}]{Glendenning:2000zz}%
  \BibitemOpen
  \bibfield  {author} {\bibinfo {author} {\bibfnamefont {N.~K.}\ \bibnamefont
  {Glendenning}}\ and\ \bibinfo {author} {\bibfnamefont {F.}~\bibnamefont
  {Weber}},\ }\href {\doibase 10.1086/323972} {\bibfield  {journal} {\bibinfo
  {journal} {Astrophys. J.}\ }\textbf {\bibinfo {volume} {559}},\ \bibinfo
  {pages} {L119} (\bibinfo {year} {2001})},\ \Eprint
  {http://arxiv.org/abs/astro-ph/0003426} {arXiv:astro-ph/0003426 [astro-ph]}
  \BibitemShut {NoStop}%
\bibitem [{\citenamefont {Zdunik}\ \emph {et~al.}(2006)\citenamefont {Zdunik},
  \citenamefont {Bejger}, \citenamefont {Haensel},\ and\ \citenamefont
  {Gourgoulhon}}]{Zdunik:2005kh}%
  \BibitemOpen
  \bibfield  {author} {\bibinfo {author} {\bibfnamefont {J.~L.}\ \bibnamefont
  {Zdunik}}, \bibinfo {author} {\bibfnamefont {M.}~\bibnamefont {Bejger}},
  \bibinfo {author} {\bibfnamefont {P.}~\bibnamefont {Haensel}}, \ and\
  \bibinfo {author} {\bibfnamefont {E.}~\bibnamefont {Gourgoulhon}},\ }\href
  {\doibase 10.1051/0004-6361:20054260} {\bibfield  {journal} {\bibinfo
  {journal} {Astron. Astrophys.}\ }\textbf {\bibinfo {volume} {450}},\ \bibinfo
  {pages} {747} (\bibinfo {year} {2006})},\ \Eprint
  {http://arxiv.org/abs/astro-ph/0509806} {arXiv:astro-ph/0509806 [astro-ph]}
  \BibitemShut {NoStop}%
\bibitem [{\citenamefont {Dimmelmeier}\ \emph {et~al.}(2009)\citenamefont
  {Dimmelmeier}, \citenamefont {Bejger}, \citenamefont {Haensel},\ and\
  \citenamefont {Zdunik}}]{Dimmelmeier:2009vw}%
  \BibitemOpen
  \bibfield  {author} {\bibinfo {author} {\bibfnamefont {H.}~\bibnamefont
  {Dimmelmeier}}, \bibinfo {author} {\bibfnamefont {M.}~\bibnamefont {Bejger}},
  \bibinfo {author} {\bibfnamefont {P.}~\bibnamefont {Haensel}}, \ and\
  \bibinfo {author} {\bibfnamefont {J.~L.}\ \bibnamefont {Zdunik}},\ }\href
  {\doibase 10.1111/j.1365-2966.2009.14891.x} {\bibfield  {journal} {\bibinfo
  {journal} {Mon. Not. Roy. Astron. Soc.}\ }\textbf {\bibinfo {volume} {396}},\
  \bibinfo {pages} {2269} (\bibinfo {year} {2009})},\ \Eprint
  {http://arxiv.org/abs/0901.3819} {arXiv:0901.3819 [astro-ph.SR]} \BibitemShut
  {NoStop}%
\bibitem [{\citenamefont {Ayvazyan}\ \emph {et~al.}(2013)\citenamefont
  {Ayvazyan}, \citenamefont {Colucci}, \citenamefont {Rischke},\ and\
  \citenamefont {Sedrakian}}]{Ayvazyan:2013cva}%
  \BibitemOpen
  \bibfield  {author} {\bibinfo {author} {\bibfnamefont {N.~S.}\ \bibnamefont
  {Ayvazyan}}, \bibinfo {author} {\bibfnamefont {G.}~\bibnamefont {Colucci}},
  \bibinfo {author} {\bibfnamefont {D.~H.}\ \bibnamefont {Rischke}}, \ and\
  \bibinfo {author} {\bibfnamefont {A.}~\bibnamefont {Sedrakian}},\ }\href
  {\doibase 10.1051/0004-6361/201322484} {\bibfield  {journal} {\bibinfo
  {journal} {Astron. Astrophys.}\ }\textbf {\bibinfo {volume} {559}},\ \bibinfo
  {pages} {A118} (\bibinfo {year} {2013})},\ \Eprint
  {http://arxiv.org/abs/1308.3053} {arXiv:1308.3053 [astro-ph.SR]} \BibitemShut
  {NoStop}%
\bibitem [{\citenamefont {Bejger}\ \emph {et~al.}(2017)\citenamefont {Bejger},
  \citenamefont {Blaschke}, \citenamefont {Haensel}, \citenamefont {Zdunik},\
  and\ \citenamefont {Fortin}}]{Bejger:2016emu}%
  \BibitemOpen
  \bibfield  {author} {\bibinfo {author} {\bibfnamefont {M.}~\bibnamefont
  {Bejger}}, \bibinfo {author} {\bibfnamefont {D.}~\bibnamefont {Blaschke}},
  \bibinfo {author} {\bibfnamefont {P.}~\bibnamefont {Haensel}}, \bibinfo
  {author} {\bibfnamefont {J.~L.}\ \bibnamefont {Zdunik}}, \ and\ \bibinfo
  {author} {\bibfnamefont {M.}~\bibnamefont {Fortin}},\ }\href {\doibase
  10.1051/0004-6361/201629580} {\bibfield  {journal} {\bibinfo  {journal}
  {Astron. Astrophys.}\ }\textbf {\bibinfo {volume} {600}},\ \bibinfo {pages}
  {A39} (\bibinfo {year} {2017})},\ \Eprint {http://arxiv.org/abs/1608.07049}
  {arXiv:1608.07049 [astro-ph.HE]} \BibitemShut {NoStop}%
\bibitem [{\citenamefont {Rabhi}\ \emph {et~al.}(2009)\citenamefont {Rabhi},
  \citenamefont {Pais}, \citenamefont {Panda},\ and\ \citenamefont
  {Providencia}}]{Rabhi:2009ih}%
  \BibitemOpen
  \bibfield  {author} {\bibinfo {author} {\bibfnamefont {A.}~\bibnamefont
  {Rabhi}}, \bibinfo {author} {\bibfnamefont {H.}~\bibnamefont {Pais}},
  \bibinfo {author} {\bibfnamefont {P.~K.}\ \bibnamefont {Panda}}, \ and\
  \bibinfo {author} {\bibfnamefont {C.}~\bibnamefont {Providencia}},\ }\href
  {\doibase 10.1088/0954-3899/36/11/115204} {\bibfield  {journal} {\bibinfo
  {journal} {J. Phys.}\ }\textbf {\bibinfo {volume} {G36}},\ \bibinfo {pages}
  {115204} (\bibinfo {year} {2009})},\ \Eprint {http://arxiv.org/abs/0909.1114}
  {arXiv:0909.1114 [nucl-th]} \BibitemShut {NoStop}%
\bibitem [{\citenamefont {Sotani}\ and\ \citenamefont
  {Tatsumi}(2015)}]{Sotani:2014rva}%
  \BibitemOpen
  \bibfield  {author} {\bibinfo {author} {\bibfnamefont {H.}~\bibnamefont
  {Sotani}}\ and\ \bibinfo {author} {\bibfnamefont {T.}~\bibnamefont
  {Tatsumi}},\ }\href {\doibase 10.1093/mnras/stu2677} {\bibfield  {journal}
  {\bibinfo  {journal} {Mon. Not. Roy. Astron. Soc.}\ }\textbf {\bibinfo
  {volume} {447}},\ \bibinfo {pages} {3155} (\bibinfo {year} {2015})},\ \Eprint
  {http://arxiv.org/abs/1412.4610} {arXiv:1412.4610 [astro-ph.HE]} \BibitemShut
  {NoStop}%
\bibitem [{\citenamefont {Franzon}\ \emph {et~al.}(2015)\citenamefont
  {Franzon}, \citenamefont {Dexheimer},\ and\ \citenamefont
  {Schramm}}]{Franzon:2015sya}%
  \BibitemOpen
  \bibfield  {author} {\bibinfo {author} {\bibfnamefont {B.}~\bibnamefont
  {Franzon}}, \bibinfo {author} {\bibfnamefont {V.}~\bibnamefont {Dexheimer}},
  \ and\ \bibinfo {author} {\bibfnamefont {S.}~\bibnamefont {Schramm}},\ }\href
  {\doibase 10.1093/mnras/stv2606} {\bibfield  {journal} {\bibinfo  {journal}
  {Mon. Not. Roy. Astron. Soc.}\ }\textbf {\bibinfo {volume} {456}},\ \bibinfo
  {pages} {2937} (\bibinfo {year} {2015})},\ \Eprint
  {http://arxiv.org/abs/1508.04431} {arXiv:1508.04431 [astro-ph.HE]}
  \BibitemShut {NoStop}%
\bibitem [{\citenamefont {Franzon}\ \emph {et~al.}(2016)\citenamefont
  {Franzon}, \citenamefont {Gomes},\ and\ \citenamefont
  {Schramm}}]{Franzon:2016urz}%
  \BibitemOpen
  \bibfield  {author} {\bibinfo {author} {\bibfnamefont {B.}~\bibnamefont
  {Franzon}}, \bibinfo {author} {\bibfnamefont {R.~O.}\ \bibnamefont {Gomes}},
  \ and\ \bibinfo {author} {\bibfnamefont {S.}~\bibnamefont {Schramm}},\ }\href
  {\doibase 10.1093/mnras/stw1967} {\bibfield  {journal} {\bibinfo  {journal}
  {Mon. Not. Roy. Astron. Soc.}\ }\textbf {\bibinfo {volume} {463}},\ \bibinfo
  {pages} {571} (\bibinfo {year} {2016})},\ \Eprint
  {http://arxiv.org/abs/1608.02845} {arXiv:1608.02845 [astro-ph.HE]}
  \BibitemShut {NoStop}%
\bibitem [{\citenamefont {Page}\ \emph {et~al.}(2006)\citenamefont {Page},
  \citenamefont {Geppert},\ and\ \citenamefont {Weber}}]{Page:2005fq}%
  \BibitemOpen
  \bibfield  {author} {\bibinfo {author} {\bibfnamefont {D.}~\bibnamefont
  {Page}}, \bibinfo {author} {\bibfnamefont {U.}~\bibnamefont {Geppert}}, \
  and\ \bibinfo {author} {\bibfnamefont {F.}~\bibnamefont {Weber}},\ }\href
  {\doibase 10.1016/j.nuclphysa.2005.09.019} {\bibfield  {journal} {\bibinfo
  {journal} {Nucl. Phys.}\ }\textbf {\bibinfo {volume} {A777}},\ \bibinfo
  {pages} {497} (\bibinfo {year} {2006})},\ \Eprint
  {http://arxiv.org/abs/astro-ph/0508056} {arXiv:astro-ph/0508056 [astro-ph]}
  \BibitemShut {NoStop}%
\bibitem [{\citenamefont {Dexheimer}\ \emph {et~al.}(2013)\citenamefont
  {Dexheimer}, \citenamefont {Steinheimer}, \citenamefont {Negreiros},\ and\
  \citenamefont {Schramm}}]{Dexheimer:2012eu}%
  \BibitemOpen
  \bibfield  {author} {\bibinfo {author} {\bibfnamefont {V.}~\bibnamefont
  {Dexheimer}}, \bibinfo {author} {\bibfnamefont {J.}~\bibnamefont
  {Steinheimer}}, \bibinfo {author} {\bibfnamefont {R.}~\bibnamefont
  {Negreiros}}, \ and\ \bibinfo {author} {\bibfnamefont {S.}~\bibnamefont
  {Schramm}},\ }\href {\doibase 10.1103/PhysRevC.87.015804} {\bibfield
  {journal} {\bibinfo  {journal} {Phys. Rev.}\ }\textbf {\bibinfo {volume}
  {C87}},\ \bibinfo {pages} {015804} (\bibinfo {year} {2013})},\ \Eprint
  {http://arxiv.org/abs/1206.3086} {arXiv:1206.3086 [astro-ph.HE]} \BibitemShut
  {NoStop}%
\bibitem [{\citenamefont {Dexheimer}\ \emph {et~al.}(2015)\citenamefont
  {Dexheimer}, \citenamefont {Negreiros},\ and\ \citenamefont
  {Schramm}}]{Dexheimer:2014pea}%
  \BibitemOpen
  \bibfield  {author} {\bibinfo {author} {\bibfnamefont {V.}~\bibnamefont
  {Dexheimer}}, \bibinfo {author} {\bibfnamefont {R.}~\bibnamefont
  {Negreiros}}, \ and\ \bibinfo {author} {\bibfnamefont {S.}~\bibnamefont
  {Schramm}},\ }\href {\doibase 10.1103/PhysRevC.91.055808} {\bibfield
  {journal} {\bibinfo  {journal} {Phys. Rev.}\ }\textbf {\bibinfo {volume}
  {C91}},\ \bibinfo {pages} {055808} (\bibinfo {year} {2015})},\ \Eprint
  {http://arxiv.org/abs/1411.4623} {arXiv:1411.4623 [astro-ph.HE]} \BibitemShut
  {NoStop}%
\bibitem [{\citenamefont {de~Carvalho}\ \emph {et~al.}(2015)\citenamefont
  {de~Carvalho}, \citenamefont {Negreiros}, \citenamefont {Orsaria},
  \citenamefont {Contrera}, \citenamefont {Weber},\ and\ \citenamefont
  {Spinella}}]{deCarvalho:2015lpa}%
  \BibitemOpen
  \bibfield  {author} {\bibinfo {author} {\bibfnamefont {S.~M.}\ \bibnamefont
  {de~Carvalho}}, \bibinfo {author} {\bibfnamefont {R.}~\bibnamefont
  {Negreiros}}, \bibinfo {author} {\bibfnamefont {M.}~\bibnamefont {Orsaria}},
  \bibinfo {author} {\bibfnamefont {G.~A.}\ \bibnamefont {Contrera}}, \bibinfo
  {author} {\bibfnamefont {F.}~\bibnamefont {Weber}}, \ and\ \bibinfo {author}
  {\bibfnamefont {W.}~\bibnamefont {Spinella}},\ }\href {\doibase
  10.1103/PhysRevC.92.035810} {\bibfield  {journal} {\bibinfo  {journal} {Phys.
  Rev.}\ }\textbf {\bibinfo {volume} {C92}},\ \bibinfo {pages} {035810}
  (\bibinfo {year} {2015})},\ \Eprint {http://arxiv.org/abs/1601.02938}
  {arXiv:1601.02938 [nucl-th]} \BibitemShut {NoStop}%
\bibitem [{\citenamefont {Pons}\ \emph {et~al.}(2001)\citenamefont {Pons},
  \citenamefont {Steiner}, \citenamefont {Prakash},\ and\ \citenamefont
  {Lattimer}}]{Pons:2001ar}%
  \BibitemOpen
  \bibfield  {author} {\bibinfo {author} {\bibfnamefont {J.~A.}\ \bibnamefont
  {Pons}}, \bibinfo {author} {\bibfnamefont {A.~W.}\ \bibnamefont {Steiner}},
  \bibinfo {author} {\bibfnamefont {M.}~\bibnamefont {Prakash}}, \ and\
  \bibinfo {author} {\bibfnamefont {J.~M.}\ \bibnamefont {Lattimer}},\ }\href
  {\doibase 10.1103/PhysRevLett.86.5223} {\bibfield  {journal} {\bibinfo
  {journal} {Phys. Rev. Lett.}\ }\textbf {\bibinfo {volume} {86}},\ \bibinfo
  {pages} {5223} (\bibinfo {year} {2001})},\ \Eprint
  {http://arxiv.org/abs/astro-ph/0102015} {arXiv:astro-ph/0102015 [astro-ph]}
  \BibitemShut {NoStop}%
\bibitem [{\citenamefont {Yasutake}\ and\ \citenamefont
  {Kashiwa}(2009)}]{Yasutake:2009kj}%
  \BibitemOpen
  \bibfield  {author} {\bibinfo {author} {\bibfnamefont {N.}~\bibnamefont
  {Yasutake}}\ and\ \bibinfo {author} {\bibfnamefont {K.}~\bibnamefont
  {Kashiwa}},\ }\href {\doibase 10.1103/PhysRevD.79.043012} {\bibfield
  {journal} {\bibinfo  {journal} {Phys. Rev.}\ }\textbf {\bibinfo {volume}
  {D79}},\ \bibinfo {pages} {043012} (\bibinfo {year} {2009})},\ \Eprint
  {http://arxiv.org/abs/0902.0111} {arXiv:0902.0111 [astro-ph.HE]} \BibitemShut
  {NoStop}%
\bibitem [{\citenamefont {Mariani}\ \emph {et~al.}(2017)\citenamefont
  {Mariani}, \citenamefont {Orsaria},\ and\ \citenamefont
  {Vucetich}}]{Mariani:2017fqr}%
  \BibitemOpen
  \bibfield  {author} {\bibinfo {author} {\bibfnamefont {M.}~\bibnamefont
  {Mariani}}, \bibinfo {author} {\bibfnamefont {M.}~\bibnamefont {Orsaria}}, \
  and\ \bibinfo {author} {\bibfnamefont {H.}~\bibnamefont {Vucetich}},\
  }\bibfield  {booktitle} {\emph {\bibinfo {booktitle} {{Proceedings, 7th
  International Workshop on Astronomy and Relativistic Astrophysics (IWARA
  2016): Gramado, Brazil, October 9-13, 2016}}},\ }\href {\doibase
  10.1142/S2010194517600412} {\bibfield  {journal} {\bibinfo  {journal} {Int.
  J. Mod. Phys. Conf. Ser.}\ }\textbf {\bibinfo {volume} {45}},\ \bibinfo
  {pages} {1760041} (\bibinfo {year} {2017})},\ \Eprint
  {http://arxiv.org/abs/1704.07732} {arXiv:1704.07732 [nucl-th]} \BibitemShut
  {NoStop}%
\bibitem [{\citenamefont {Heinimann}\ \emph {et~al.}(2016)\citenamefont
  {Heinimann}, \citenamefont {Hempel},\ and\ \citenamefont
  {Thielemann}}]{Heinimann:2016zbx}%
  \BibitemOpen
  \bibfield  {author} {\bibinfo {author} {\bibfnamefont {O.}~\bibnamefont
  {Heinimann}}, \bibinfo {author} {\bibfnamefont {M.}~\bibnamefont {Hempel}}, \
  and\ \bibinfo {author} {\bibfnamefont {F.-K.}\ \bibnamefont {Thielemann}},\
  }\href {\doibase 10.1103/PhysRevD.94.103008} {\bibfield  {journal} {\bibinfo
  {journal} {Phys. Rev.}\ }\textbf {\bibinfo {volume} {D94}},\ \bibinfo {pages}
  {103008} (\bibinfo {year} {2016})},\ \Eprint
  {http://arxiv.org/abs/1608.08862} {arXiv:1608.08862 [astro-ph.SR]}
  \BibitemShut {NoStop}%
\bibitem [{\citenamefont {Brillante}\ and\ \citenamefont
  {Mishustin}(2014)}]{Brillante:2014lwa}%
  \BibitemOpen
  \bibfield  {author} {\bibinfo {author} {\bibfnamefont {A.}~\bibnamefont
  {Brillante}}\ and\ \bibinfo {author} {\bibfnamefont {I.~N.}\ \bibnamefont
  {Mishustin}},\ }\href {\doibase 10.1209/0295-5075/105/39001} {\bibfield
  {journal} {\bibinfo  {journal} {Europhys. Lett.}\ }\textbf {\bibinfo {volume}
  {105}},\ \bibinfo {pages} {39001} (\bibinfo {year} {2014})},\ \Eprint
  {http://arxiv.org/abs/1401.7915} {arXiv:1401.7915 [astro-ph.SR]} \BibitemShut
  {NoStop}%
\bibitem [{\citenamefont {Alford}\ and\ \citenamefont
  {Han}(2016)}]{Alford:2015gna}%
  \BibitemOpen
  \bibfield  {author} {\bibinfo {author} {\bibfnamefont {M.~G.}\ \bibnamefont
  {Alford}}\ and\ \bibinfo {author} {\bibfnamefont {S.}~\bibnamefont {Han}},\
  }\href {\doibase 10.1140/epja/i2016-16062-9} {\bibfield  {journal} {\bibinfo
  {journal} {Eur. Phys. J.}\ }\textbf {\bibinfo {volume} {A52}},\ \bibinfo
  {pages} {62} (\bibinfo {year} {2016})},\ \Eprint
  {http://arxiv.org/abs/1508.01261} {arXiv:1508.01261 [nucl-th]} \BibitemShut
  {NoStop}%
\bibitem [{\citenamefont {Lenzi}\ and\ \citenamefont
  {Lugones}(2012)}]{Lenzi:2012xz}%
  \BibitemOpen
  \bibfield  {author} {\bibinfo {author} {\bibfnamefont {C.~H.}\ \bibnamefont
  {Lenzi}}\ and\ \bibinfo {author} {\bibfnamefont {G.}~\bibnamefont
  {Lugones}},\ }\href {\doibase 10.1088/0004-637X/759/1/57} {\bibfield
  {journal} {\bibinfo  {journal} {Astrophys. J.}\ }\textbf {\bibinfo {volume}
  {759}},\ \bibinfo {pages} {57} (\bibinfo {year} {2012})},\ \Eprint
  {http://arxiv.org/abs/1206.4108} {arXiv:1206.4108 [astro-ph.SR]} \BibitemShut
  {NoStop}%
\bibitem [{\citenamefont {Shao}\ \emph {et~al.}(2013)\citenamefont {Shao},
  \citenamefont {Colonna}, \citenamefont {Di~Toro}, \citenamefont {Liu},\ and\
  \citenamefont {Liu}}]{Shao:2013toa}%
  \BibitemOpen
  \bibfield  {author} {\bibinfo {author} {\bibfnamefont {G.~Y.}\ \bibnamefont
  {Shao}}, \bibinfo {author} {\bibfnamefont {M.}~\bibnamefont {Colonna}},
  \bibinfo {author} {\bibfnamefont {M.}~\bibnamefont {Di~Toro}}, \bibinfo
  {author} {\bibfnamefont {Y.~X.}\ \bibnamefont {Liu}}, \ and\ \bibinfo
  {author} {\bibfnamefont {B.}~\bibnamefont {Liu}},\ }\href {\doibase
  10.1103/PhysRevD.87.096012} {\bibfield  {journal} {\bibinfo  {journal} {Phys.
  Rev.}\ }\textbf {\bibinfo {volume} {D87}},\ \bibinfo {pages} {096012}
  (\bibinfo {year} {2013})},\ \Eprint {http://arxiv.org/abs/1305.1176}
  {arXiv:1305.1176 [nucl-th]} \BibitemShut {NoStop}%
\bibitem [{\citenamefont {Benic}\ \emph {et~al.}(2015)\citenamefont {Benic},
  \citenamefont {Blaschke}, \citenamefont {Alvarez-Castillo}, \citenamefont
  {Fischer},\ and\ \citenamefont {Typel}}]{Benic:2014jia}%
  \BibitemOpen
  \bibfield  {author} {\bibinfo {author} {\bibfnamefont {S.}~\bibnamefont
  {Benic}}, \bibinfo {author} {\bibfnamefont {D.}~\bibnamefont {Blaschke}},
  \bibinfo {author} {\bibfnamefont {D.~E.}\ \bibnamefont {Alvarez-Castillo}},
  \bibinfo {author} {\bibfnamefont {T.}~\bibnamefont {Fischer}}, \ and\
  \bibinfo {author} {\bibfnamefont {S.}~\bibnamefont {Typel}},\ }\href
  {\doibase 10.1051/0004-6361/201425318} {\bibfield  {journal} {\bibinfo
  {journal} {Astron. Astrophys.}\ }\textbf {\bibinfo {volume} {577}},\ \bibinfo
  {pages} {A40} (\bibinfo {year} {2015})},\ \Eprint
  {http://arxiv.org/abs/1411.2856} {arXiv:1411.2856 [astro-ph.HE]} \BibitemShut
  {NoStop}%
\bibitem [{\citenamefont {Ranea-Sandoval}\ \emph {et~al.}(2016)\citenamefont
  {Ranea-Sandoval}, \citenamefont {Han}, \citenamefont {Orsaria}, \citenamefont
  {Contrera}, \citenamefont {Weber},\ and\ \citenamefont
  {Alford}}]{Ranea-Sandoval:2015ldr}%
  \BibitemOpen
  \bibfield  {author} {\bibinfo {author} {\bibfnamefont {I.~F.}\ \bibnamefont
  {Ranea-Sandoval}}, \bibinfo {author} {\bibfnamefont {S.}~\bibnamefont {Han}},
  \bibinfo {author} {\bibfnamefont {M.~G.}\ \bibnamefont {Orsaria}}, \bibinfo
  {author} {\bibfnamefont {G.~A.}\ \bibnamefont {Contrera}}, \bibinfo {author}
  {\bibfnamefont {F.}~\bibnamefont {Weber}}, \ and\ \bibinfo {author}
  {\bibfnamefont {M.~G.}\ \bibnamefont {Alford}},\ }\href {\doibase
  10.1103/PhysRevC.93.045812} {\bibfield  {journal} {\bibinfo  {journal} {Phys.
  Rev.}\ }\textbf {\bibinfo {volume} {C93}},\ \bibinfo {pages} {045812}
  (\bibinfo {year} {2016})},\ \Eprint {http://arxiv.org/abs/1512.09183}
  {arXiv:1512.09183 [nucl-th]} \BibitemShut {NoStop}%
\bibitem [{\citenamefont {Alvarez-Castillo}\ \emph
  {et~al.}(2016{\natexlab{a}})\citenamefont {Alvarez-Castillo}, \citenamefont
  {Benic}, \citenamefont {Blaschke}, \citenamefont {Han},\ and\ \citenamefont
  {Typel}}]{Alvarez-Castillo:2016wqj}%
  \BibitemOpen
  \bibfield  {author} {\bibinfo {author} {\bibfnamefont {D.}~\bibnamefont
  {Alvarez-Castillo}}, \bibinfo {author} {\bibfnamefont {S.}~\bibnamefont
  {Benic}}, \bibinfo {author} {\bibfnamefont {D.}~\bibnamefont {Blaschke}},
  \bibinfo {author} {\bibfnamefont {S.}~\bibnamefont {Han}}, \ and\ \bibinfo
  {author} {\bibfnamefont {S.}~\bibnamefont {Typel}},\ }\href {\doibase
  10.1140/epja/i2016-16232-9} {\bibfield  {journal} {\bibinfo  {journal} {Eur.
  Phys. J.}\ }\textbf {\bibinfo {volume} {A52}},\ \bibinfo {pages} {232}
  (\bibinfo {year} {2016}{\natexlab{a}})},\ \Eprint
  {http://arxiv.org/abs/1608.02425} {arXiv:1608.02425 [nucl-th]} \BibitemShut
  {NoStop}%
\bibitem [{\citenamefont {Pereira}\ \emph {et~al.}(2016)\citenamefont
  {Pereira}, \citenamefont {Costa},\ and\ \citenamefont
  {Providência}}]{Pereira:2016dfg}%
  \BibitemOpen
  \bibfield  {author} {\bibinfo {author} {\bibfnamefont {R.~C.}\ \bibnamefont
  {Pereira}}, \bibinfo {author} {\bibfnamefont {P.}~\bibnamefont {Costa}}, \
  and\ \bibinfo {author} {\bibfnamefont {C.}~\bibnamefont {Providência}},\
  }\href {\doibase 10.1103/PhysRevD.94.094001} {\bibfield  {journal} {\bibinfo
  {journal} {Phys. Rev.}\ }\textbf {\bibinfo {volume} {D94}},\ \bibinfo {pages}
  {094001} (\bibinfo {year} {2016})},\ \Eprint
  {http://arxiv.org/abs/1610.06435} {arXiv:1610.06435 [nucl-th]} \BibitemShut
  {NoStop}%
\bibitem [{\citenamefont {Miyatsu}\ \emph {et~al.}(2017)\citenamefont
  {Miyatsu}, \citenamefont {Kambe},\ and\ \citenamefont
  {Saito}}]{Miyatsu:2017teh}%
  \BibitemOpen
  \bibfield  {author} {\bibinfo {author} {\bibfnamefont {T.}~\bibnamefont
  {Miyatsu}}, \bibinfo {author} {\bibfnamefont {T.}~\bibnamefont {Kambe}}, \
  and\ \bibinfo {author} {\bibfnamefont {K.}~\bibnamefont {Saito}},\ }\bibfield
   {booktitle} {\emph {\bibinfo {booktitle} {{Proceedings, 26th International
  Nuclear Physics Conference (INPC2016): Adelaide, Australia, September 11-16,
  2016}}},\ }\href@noop {} {\bibfield  {journal} {\bibinfo  {journal} {PoS}\
  }\textbf {\bibinfo {volume} {INPC2016}},\ \bibinfo {pages} {135} (\bibinfo
  {year} {2017})},\ \Eprint {http://arxiv.org/abs/1702.05871} {arXiv:1702.05871
  [nucl-th]} \BibitemShut {NoStop}%
\bibitem [{\citenamefont {Burgio}\ \emph {et~al.}(2002)\citenamefont {Burgio},
  \citenamefont {Baldo}, \citenamefont {Sahu},\ and\ \citenamefont
  {Schulze}}]{Burgio:2002sn}%
  \BibitemOpen
  \bibfield  {author} {\bibinfo {author} {\bibfnamefont {G.~F.}\ \bibnamefont
  {Burgio}}, \bibinfo {author} {\bibfnamefont {M.}~\bibnamefont {Baldo}},
  \bibinfo {author} {\bibfnamefont {P.~K.}\ \bibnamefont {Sahu}}, \ and\
  \bibinfo {author} {\bibfnamefont {H.~J.}\ \bibnamefont {Schulze}},\ }\href
  {\doibase 10.1103/PhysRevC.66.025802} {\bibfield  {journal} {\bibinfo
  {journal} {Phys. Rev.}\ }\textbf {\bibinfo {volume} {C66}},\ \bibinfo {pages}
  {025802} (\bibinfo {year} {2002})},\ \Eprint
  {http://arxiv.org/abs/nucl-th/0206009} {arXiv:nucl-th/0206009 [nucl-th]}
  \BibitemShut {NoStop}%
\bibitem [{\citenamefont {Bhattacharyya}\ \emph {et~al.}(2010)\citenamefont
  {Bhattacharyya}, \citenamefont {Mishustin},\ and\ \citenamefont
  {Greiner}}]{Bhattacharyya:2009fg}%
  \BibitemOpen
  \bibfield  {author} {\bibinfo {author} {\bibfnamefont {A.}~\bibnamefont
  {Bhattacharyya}}, \bibinfo {author} {\bibfnamefont {I.~N.}\ \bibnamefont
  {Mishustin}}, \ and\ \bibinfo {author} {\bibfnamefont {W.}~\bibnamefont
  {Greiner}},\ }\href {\doibase 10.1088/0954-3899/37/2/025201} {\bibfield
  {journal} {\bibinfo  {journal} {J. Phys.}\ }\textbf {\bibinfo {volume}
  {G37}},\ \bibinfo {pages} {025201} (\bibinfo {year} {2010})},\ \Eprint
  {http://arxiv.org/abs/0905.0352} {arXiv:0905.0352 [nucl-th]} \BibitemShut
  {NoStop}%
\bibitem [{\citenamefont {Yasutake}\ \emph {et~al.}(2011)\citenamefont
  {Yasutake}, \citenamefont {Burgio},\ and\ \citenamefont
  {Schulze}}]{Yasutake:2010eq}%
  \BibitemOpen
  \bibfield  {author} {\bibinfo {author} {\bibfnamefont {N.}~\bibnamefont
  {Yasutake}}, \bibinfo {author} {\bibfnamefont {G.~F.}\ \bibnamefont
  {Burgio}}, \ and\ \bibinfo {author} {\bibfnamefont {H.~J.}\ \bibnamefont
  {Schulze}},\ }\href {\doibase 10.1134/S1063778811100073} {\bibfield
  {journal} {\bibinfo  {journal} {Phys. Atom. Nucl.}\ }\textbf {\bibinfo
  {volume} {74}},\ \bibinfo {pages} {1502} (\bibinfo {year} {2011})},\ \Eprint
  {http://arxiv.org/abs/1012.1773} {arXiv:1012.1773 [astro-ph.SR]} \BibitemShut
  {NoStop}%
\bibitem [{\citenamefont {Shao}\ \emph {et~al.}(2012)\citenamefont {Shao},
  \citenamefont {Colonna}, \citenamefont {Di~Toro}, \citenamefont {Liu},\ and\
  \citenamefont {Matera}}]{Shao:2012tu}%
  \BibitemOpen
  \bibfield  {author} {\bibinfo {author} {\bibfnamefont {G.~Y.}\ \bibnamefont
  {Shao}}, \bibinfo {author} {\bibfnamefont {M.}~\bibnamefont {Colonna}},
  \bibinfo {author} {\bibfnamefont {M.}~\bibnamefont {Di~Toro}}, \bibinfo
  {author} {\bibfnamefont {B.}~\bibnamefont {Liu}}, \ and\ \bibinfo {author}
  {\bibfnamefont {F.}~\bibnamefont {Matera}},\ }\href {\doibase
  10.1103/PhysRevD.85.114017} {\bibfield  {journal} {\bibinfo  {journal} {Phys.
  Rev.}\ }\textbf {\bibinfo {volume} {D85}},\ \bibinfo {pages} {114017}
  (\bibinfo {year} {2012})},\ \Eprint {http://arxiv.org/abs/1202.6476}
  {arXiv:1202.6476 [nucl-th]} \BibitemShut {NoStop}%
\bibitem [{\citenamefont {Whittenbury}\ \emph {et~al.}(2016)\citenamefont
  {Whittenbury}, \citenamefont {Matevosyan},\ and\ \citenamefont
  {Thomas}}]{Whittenbury:2015ziz}%
  \BibitemOpen
  \bibfield  {author} {\bibinfo {author} {\bibfnamefont {D.~L.}\ \bibnamefont
  {Whittenbury}}, \bibinfo {author} {\bibfnamefont {H.~H.}\ \bibnamefont
  {Matevosyan}}, \ and\ \bibinfo {author} {\bibfnamefont {A.~W.}\ \bibnamefont
  {Thomas}},\ }\href {\doibase 10.1103/PhysRevC.93.035807} {\bibfield
  {journal} {\bibinfo  {journal} {Phys. Rev.}\ }\textbf {\bibinfo {volume}
  {C93}},\ \bibinfo {pages} {035807} (\bibinfo {year} {2016})},\ \Eprint
  {http://arxiv.org/abs/1511.08561} {arXiv:1511.08561 [nucl-th]} \BibitemShut
  {NoStop}%
\bibitem [{\citenamefont {Zacchi}\ \emph {et~al.}(2016)\citenamefont {Zacchi},
  \citenamefont {Hanauske},\ and\ \citenamefont
  {Schaffner-Bielich}}]{Zacchi:2015oma}%
  \BibitemOpen
  \bibfield  {author} {\bibinfo {author} {\bibfnamefont {A.}~\bibnamefont
  {Zacchi}}, \bibinfo {author} {\bibfnamefont {M.}~\bibnamefont {Hanauske}}, \
  and\ \bibinfo {author} {\bibfnamefont {J.}~\bibnamefont
  {Schaffner-Bielich}},\ }\href {\doibase 10.1103/PhysRevD.93.065011}
  {\bibfield  {journal} {\bibinfo  {journal} {Phys. Rev.}\ }\textbf {\bibinfo
  {volume} {D93}},\ \bibinfo {pages} {065011} (\bibinfo {year} {2016})},\
  \Eprint {http://arxiv.org/abs/1510.00180} {arXiv:1510.00180 [nucl-th]}
  \BibitemShut {NoStop}%
\bibitem [{\citenamefont {Dexheimer}\ and\ \citenamefont
  {Schramm}(2010)}]{Dexheimer:2009hi}%
  \BibitemOpen
  \bibfield  {author} {\bibinfo {author} {\bibfnamefont {V.~A.}\ \bibnamefont
  {Dexheimer}}\ and\ \bibinfo {author} {\bibfnamefont {S.}~\bibnamefont
  {Schramm}},\ }\href {\doibase 10.1103/PhysRevC.81.045201} {\bibfield
  {journal} {\bibinfo  {journal} {Phys. Rev.}\ }\textbf {\bibinfo {volume}
  {C81}},\ \bibinfo {pages} {045201} (\bibinfo {year} {2010})},\ \Eprint
  {http://arxiv.org/abs/0901.1748} {arXiv:0901.1748 [astro-ph.SR]} \BibitemShut
  {NoStop}%
\bibitem [{\citenamefont {Chen}\ \emph {et~al.}(2011)\citenamefont {Chen},
  \citenamefont {Baldo}, \citenamefont {Burgio},\ and\ \citenamefont
  {Schulze}}]{Chen:2011my}%
  \BibitemOpen
  \bibfield  {author} {\bibinfo {author} {\bibfnamefont {H.}~\bibnamefont
  {Chen}}, \bibinfo {author} {\bibfnamefont {M.}~\bibnamefont {Baldo}},
  \bibinfo {author} {\bibfnamefont {G.~F.}\ \bibnamefont {Burgio}}, \ and\
  \bibinfo {author} {\bibfnamefont {H.~J.}\ \bibnamefont {Schulze}},\ }\href
  {\doibase 10.1103/PhysRevD.84.105023} {\bibfield  {journal} {\bibinfo
  {journal} {Phys. Rev.}\ }\textbf {\bibinfo {volume} {D84}},\ \bibinfo {pages}
  {105023} (\bibinfo {year} {2011})},\ \Eprint {http://arxiv.org/abs/1107.2497}
  {arXiv:1107.2497 [nucl-th]} \BibitemShut {NoStop}%
\bibitem [{\citenamefont {Li}\ \emph {et~al.}(2015)\citenamefont {Li},
  \citenamefont {Zuo},\ and\ \citenamefont {Peng}}]{Li:2015ida}%
  \BibitemOpen
  \bibfield  {author} {\bibinfo {author} {\bibfnamefont {A.}~\bibnamefont
  {Li}}, \bibinfo {author} {\bibfnamefont {W.}~\bibnamefont {Zuo}}, \ and\
  \bibinfo {author} {\bibfnamefont {G.~X.}\ \bibnamefont {Peng}},\ }\href
  {\doibase 10.1103/PhysRevC.91.035803} {\bibfield  {journal} {\bibinfo
  {journal} {Phys. Rev.}\ }\textbf {\bibinfo {volume} {C91}},\ \bibinfo {pages}
  {035803} (\bibinfo {year} {2015})},\ \Eprint
  {http://arxiv.org/abs/1503.02739} {arXiv:1503.02739 [astro-ph.SR]}
  \BibitemShut {NoStop}%
\bibitem [{\citenamefont {Schramm}\ \emph {et~al.}(2016)\citenamefont
  {Schramm}, \citenamefont {Dexheimer},\ and\ \citenamefont
  {Negreiros}}]{Schramm:2015hba}%
  \BibitemOpen
  \bibfield  {author} {\bibinfo {author} {\bibfnamefont {S.}~\bibnamefont
  {Schramm}}, \bibinfo {author} {\bibfnamefont {V.}~\bibnamefont {Dexheimer}},
  \ and\ \bibinfo {author} {\bibfnamefont {R.}~\bibnamefont {Negreiros}},\
  }\href {\doibase 10.1140/epja/i2016-16014-5} {\bibfield  {journal} {\bibinfo
  {journal} {Eur. Phys. J.}\ }\textbf {\bibinfo {volume} {A52}},\ \bibinfo
  {pages} {14} (\bibinfo {year} {2016})},\ \Eprint
  {http://arxiv.org/abs/1508.04699} {arXiv:1508.04699 [nucl-th]} \BibitemShut
  {NoStop}%
\bibitem [{\citenamefont {Burgio}\ and\ \citenamefont
  {Zappalà}(2016)}]{Burgio:2015zka}%
  \BibitemOpen
  \bibfield  {author} {\bibinfo {author} {\bibfnamefont {G.~F.}\ \bibnamefont
  {Burgio}}\ and\ \bibinfo {author} {\bibfnamefont {D.}~\bibnamefont
  {Zappalà}},\ }\href {\doibase 10.1140/epja/i2016-16060-y} {\bibfield
  {journal} {\bibinfo  {journal} {Eur. Phys. J.}\ }\textbf {\bibinfo {volume}
  {A52}},\ \bibinfo {pages} {60} (\bibinfo {year} {2016})},\ \Eprint
  {http://arxiv.org/abs/1509.00841} {arXiv:1509.00841 [nucl-th]} \BibitemShut
  {NoStop}%
\bibitem [{\citenamefont {Alvarez-Castillo}\ \emph
  {et~al.}(2016{\natexlab{b}})\citenamefont {Alvarez-Castillo}, \citenamefont
  {Ayriyan}, \citenamefont {Benic}, \citenamefont {Blaschke}, \citenamefont
  {Grigorian},\ and\ \citenamefont {Typel}}]{Alvarez-Castillo:2016oln}%
  \BibitemOpen
  \bibfield  {author} {\bibinfo {author} {\bibfnamefont {D.}~\bibnamefont
  {Alvarez-Castillo}}, \bibinfo {author} {\bibfnamefont {A.}~\bibnamefont
  {Ayriyan}}, \bibinfo {author} {\bibfnamefont {S.}~\bibnamefont {Benic}},
  \bibinfo {author} {\bibfnamefont {D.}~\bibnamefont {Blaschke}}, \bibinfo
  {author} {\bibfnamefont {H.}~\bibnamefont {Grigorian}}, \ and\ \bibinfo
  {author} {\bibfnamefont {S.}~\bibnamefont {Typel}},\ }\href {\doibase
  10.1140/epja/i2016-16069-2} {\bibfield  {journal} {\bibinfo  {journal} {Eur.
  Phys. J.}\ }\textbf {\bibinfo {volume} {A52}},\ \bibinfo {pages} {69}
  (\bibinfo {year} {2016}{\natexlab{b}})},\ \Eprint
  {http://arxiv.org/abs/1603.03457} {arXiv:1603.03457 [nucl-th]} \BibitemShut
  {NoStop}%
\bibitem [{\citenamefont {Glendenning}(1992)}]{Glendenning:1992vb}%
  \BibitemOpen
  \bibfield  {author} {\bibinfo {author} {\bibfnamefont {N.~K.}\ \bibnamefont
  {Glendenning}},\ }\href {\doibase 10.1103/PhysRevD.46.1274} {\bibfield
  {journal} {\bibinfo  {journal} {Phys. Rev.}\ }\textbf {\bibinfo {volume}
  {D46}},\ \bibinfo {pages} {1274} (\bibinfo {year} {1992})}\BibitemShut
  {NoStop}%
\bibitem [{\citenamefont {Heiselberg}\ \emph {et~al.}(1993)\citenamefont
  {Heiselberg}, \citenamefont {Pethick},\ and\ \citenamefont
  {Staubo}}]{Heiselberg:1992dx}%
  \BibitemOpen
  \bibfield  {author} {\bibinfo {author} {\bibfnamefont {H.}~\bibnamefont
  {Heiselberg}}, \bibinfo {author} {\bibfnamefont {C.~J.}\ \bibnamefont
  {Pethick}}, \ and\ \bibinfo {author} {\bibfnamefont {E.~F.}\ \bibnamefont
  {Staubo}},\ }\href {\doibase 10.1103/PhysRevLett.70.1355} {\bibfield
  {journal} {\bibinfo  {journal} {Phys. Rev. Lett.}\ }\textbf {\bibinfo
  {volume} {70}},\ \bibinfo {pages} {1355} (\bibinfo {year}
  {1993})}\BibitemShut {NoStop}%
\bibitem [{\citenamefont {Voskresensky}\ \emph {et~al.}(2003)\citenamefont
  {Voskresensky}, \citenamefont {Yasuhira},\ and\ \citenamefont
  {Tatsumi}}]{Voskresensky:2002hu}%
  \BibitemOpen
  \bibfield  {author} {\bibinfo {author} {\bibfnamefont {D.~N.}\ \bibnamefont
  {Voskresensky}}, \bibinfo {author} {\bibfnamefont {M.}~\bibnamefont
  {Yasuhira}}, \ and\ \bibinfo {author} {\bibfnamefont {T.}~\bibnamefont
  {Tatsumi}},\ }\href {\doibase 10.1016/S0375-9474(03)01313-7} {\bibfield
  {journal} {\bibinfo  {journal} {Nucl. Phys.}\ }\textbf {\bibinfo {volume}
  {A723}},\ \bibinfo {pages} {291} (\bibinfo {year} {2003})},\ \Eprint
  {http://arxiv.org/abs/nucl-th/0208067} {arXiv:nucl-th/0208067 [nucl-th]}
  \BibitemShut {NoStop}%
\bibitem [{\citenamefont {Alford}\ \emph {et~al.}(2001)\citenamefont {Alford},
  \citenamefont {Rajagopal}, \citenamefont {Reddy},\ and\ \citenamefont
  {Wilczek}}]{Alford:2001zr}%
  \BibitemOpen
  \bibfield  {author} {\bibinfo {author} {\bibfnamefont {M.~G.}\ \bibnamefont
  {Alford}}, \bibinfo {author} {\bibfnamefont {K.}~\bibnamefont {Rajagopal}},
  \bibinfo {author} {\bibfnamefont {S.}~\bibnamefont {Reddy}}, \ and\ \bibinfo
  {author} {\bibfnamefont {F.}~\bibnamefont {Wilczek}},\ }\href {\doibase
  10.1103/PhysRevD.64.074017} {\bibfield  {journal} {\bibinfo  {journal} {Phys.
  Rev.}\ }\textbf {\bibinfo {volume} {D64}},\ \bibinfo {pages} {074017}
  (\bibinfo {year} {2001})},\ \Eprint {http://arxiv.org/abs/hep-ph/0105009}
  {arXiv:hep-ph/0105009 [hep-ph]} \BibitemShut {NoStop}%
\bibitem [{\citenamefont {Pinto}\ \emph {et~al.}(2012)\citenamefont {Pinto},
  \citenamefont {Koch},\ and\ \citenamefont {Randrup}}]{Pinto:2012aq}%
  \BibitemOpen
  \bibfield  {author} {\bibinfo {author} {\bibfnamefont {M.~B.}\ \bibnamefont
  {Pinto}}, \bibinfo {author} {\bibfnamefont {V.}~\bibnamefont {Koch}}, \ and\
  \bibinfo {author} {\bibfnamefont {J.}~\bibnamefont {Randrup}},\ }\href
  {\doibase 10.1103/PhysRevC.86.025203} {\bibfield  {journal} {\bibinfo
  {journal} {Phys. Rev.}\ }\textbf {\bibinfo {volume} {C86}},\ \bibinfo {pages}
  {025203} (\bibinfo {year} {2012})},\ \Eprint {http://arxiv.org/abs/1207.5186}
  {arXiv:1207.5186 [hep-ph]} \BibitemShut {NoStop}%
\bibitem [{\citenamefont {Stiele}\ and\ \citenamefont
  {Schaffner-Bielich}(2016)}]{Stiele:2016cfs}%
  \BibitemOpen
  \bibfield  {author} {\bibinfo {author} {\bibfnamefont {R.}~\bibnamefont
  {Stiele}}\ and\ \bibinfo {author} {\bibfnamefont {J.}~\bibnamefont
  {Schaffner-Bielich}},\ }\href {\doibase 10.1103/PhysRevD.93.094014}
  {\bibfield  {journal} {\bibinfo  {journal} {Phys. Rev.}\ }\textbf {\bibinfo
  {volume} {D93}},\ \bibinfo {pages} {094014} (\bibinfo {year} {2016})},\
  \Eprint {http://arxiv.org/abs/1601.05731} {arXiv:1601.05731 [hep-ph]}
  \BibitemShut {NoStop}%
\bibitem [{\citenamefont {Alaverdyan}\ \emph {et~al.}(2010)\citenamefont
  {Alaverdyan}, \citenamefont {Alaverdyan},\ and\ \citenamefont
  {Chiladze}}]{Alaverdyan:2010zz}%
  \BibitemOpen
  \bibfield  {author} {\bibinfo {author} {\bibfnamefont {A.~G.}\ \bibnamefont
  {Alaverdyan}}, \bibinfo {author} {\bibfnamefont {G.~B.}\ \bibnamefont
  {Alaverdyan}}, \ and\ \bibinfo {author} {\bibfnamefont {A.~O.}\ \bibnamefont
  {Chiladze}},\ }\bibfield  {booktitle} {\emph {\bibinfo {booktitle}
  {{Astronomy and relativistic astrophysics. Proceedings, 4th International
  Workshop, IWARA 2009, Maresias, Sao Paulo, Brazil, October 4-8, 2009}}},\
  }\href {\doibase 10.1142/S0218271810017408} {\bibfield  {journal} {\bibinfo
  {journal} {Int. J. Mod. Phys.}\ }\textbf {\bibinfo {volume} {D19}},\ \bibinfo
  {pages} {1557} (\bibinfo {year} {2010})},\ \Eprint
  {http://arxiv.org/abs/1207.3549} {arXiv:1207.3549 [astro-ph.SR]} \BibitemShut
  {NoStop}%
\bibitem [{\citenamefont {Contrera}\ \emph {et~al.}(2017)\citenamefont
  {Contrera}, \citenamefont {Orsaria}, \citenamefont {Ranea-Sandoval},\ and\
  \citenamefont {Weber}}]{Contrera:2016phj}%
  \BibitemOpen
  \bibfield  {author} {\bibinfo {author} {\bibfnamefont {G.~A.}\ \bibnamefont
  {Contrera}}, \bibinfo {author} {\bibfnamefont {M.}~\bibnamefont {Orsaria}},
  \bibinfo {author} {\bibfnamefont {I.~F.}\ \bibnamefont {Ranea-Sandoval}}, \
  and\ \bibinfo {author} {\bibfnamefont {F.}~\bibnamefont {Weber}},\ }\bibfield
   {booktitle} {\emph {\bibinfo {booktitle} {{Proceedings, 7th International
  Workshop on Astronomy and Relativistic Astrophysics (IWARA 2016): Gramado,
  Brazil, October 9-13, 2016}}},\ }\href {\doibase 10.1142/S2010194517600266}
  {\bibfield  {journal} {\bibinfo  {journal} {Int. J. Mod. Phys. Conf. Ser.}\
  }\textbf {\bibinfo {volume} {45}},\ \bibinfo {pages} {1760026} (\bibinfo
  {year} {2017})},\ \Eprint {http://arxiv.org/abs/1612.09485} {arXiv:1612.09485
  [nucl-th]} \BibitemShut {NoStop}%
\bibitem [{\citenamefont {Ayriyan}\ \emph {et~al.}(2017)\citenamefont
  {Ayriyan}, \citenamefont {Alvarez-Castillo}, \citenamefont {Benic},
  \citenamefont {Blaschke}, \citenamefont {Grigorian},\ and\ \citenamefont
  {Typel}}]{Ayriyan:2017nhp}%
  \BibitemOpen
  \bibfield  {author} {\bibinfo {author} {\bibfnamefont {A.}~\bibnamefont
  {Ayriyan}}, \bibinfo {author} {\bibfnamefont {D.~E.}\ \bibnamefont
  {Alvarez-Castillo}}, \bibinfo {author} {\bibfnamefont {S.}~\bibnamefont
  {Benic}}, \bibinfo {author} {\bibfnamefont {D.}~\bibnamefont {Blaschke}},
  \bibinfo {author} {\bibfnamefont {H.}~\bibnamefont {Grigorian}}, \ and\
  \bibinfo {author} {\bibfnamefont {S.}~\bibnamefont {Typel}},\ }in\ \href
  {https://inspirehep.net/record/1512076/files/arXiv:1702.00801.pdf} {\emph
  {\bibinfo {booktitle} {{10th International Workshop on Critical Point and
  Onset of Deconfinement (CPOD 2016) Wrocław, Poland, May 30-June 4, 2016}}}}\
  (\bibinfo {year} {2017})\ \Eprint {http://arxiv.org/abs/1702.00801}
  {arXiv:1702.00801 [astro-ph.HE]} \BibitemShut {NoStop}%
\bibitem [{\citenamefont {Demorest}\ \emph {et~al.}(2010)\citenamefont
  {Demorest}, \citenamefont {Pennucci}, \citenamefont {Ransom}, \citenamefont
  {Roberts},\ and\ \citenamefont {Hessels}}]{Demorest:2010bx}%
  \BibitemOpen
  \bibfield  {author} {\bibinfo {author} {\bibfnamefont {P.}~\bibnamefont
  {Demorest}}, \bibinfo {author} {\bibfnamefont {T.}~\bibnamefont {Pennucci}},
  \bibinfo {author} {\bibfnamefont {S.}~\bibnamefont {Ransom}}, \bibinfo
  {author} {\bibfnamefont {M.}~\bibnamefont {Roberts}}, \ and\ \bibinfo
  {author} {\bibfnamefont {J.}~\bibnamefont {Hessels}},\ }\href {\doibase
  10.1038/nature09466} {\bibfield  {journal} {\bibinfo  {journal} {Nature}\
  }\textbf {\bibinfo {volume} {467}},\ \bibinfo {pages} {1081} (\bibinfo {year}
  {2010})},\ \Eprint {http://arxiv.org/abs/1010.5788} {arXiv:1010.5788
  [astro-ph.HE]} \BibitemShut {NoStop}%
\bibitem [{\citenamefont {Antoniadis}\ \emph {et~al.}(2013)\citenamefont
  {Antoniadis} \emph {et~al.}}]{Antoniadis:2013pzd}%
  \BibitemOpen
  \bibfield  {author} {\bibinfo {author} {\bibfnamefont {J.}~\bibnamefont
  {Antoniadis}} \emph {et~al.},\ }\href {\doibase 10.1126/science.1233232}
  {\bibfield  {journal} {\bibinfo  {journal} {Science}\ }\textbf {\bibinfo
  {volume} {340}},\ \bibinfo {pages} {6131} (\bibinfo {year} {2013})},\ \Eprint
  {http://arxiv.org/abs/1304.6875} {arXiv:1304.6875 [astro-ph.HE]} \BibitemShut
  {NoStop}%
\bibitem [{Note1()}]{Note1}%
  \BibitemOpen
  \bibinfo {note}
  {Http://heasarc.gsfc.nasa.gov/docs/nicer/index.html}\BibitemShut {NoStop}%
\bibitem [{Note2()}]{Note2}%
  \BibitemOpen
  \bibinfo {note} {Https://www.nustar.caltech.edu}\BibitemShut {NoStop}%
\bibitem [{Note3()}]{Note3}%
  \BibitemOpen
  \bibinfo {note} {Http://www.skatelescope.org}\BibitemShut {NoStop}%
\bibitem [{\citenamefont {Abbott}\ \emph {et~al.}(2017)\citenamefont {Abbott}
  \emph {et~al.}}]{TheLIGOScientific:2017qsa}%
  \BibitemOpen
  \bibfield  {author} {\bibinfo {author} {\bibfnamefont {B.~P.}\ \bibnamefont
  {Abbott}} \emph {et~al.} (\bibinfo {collaboration} {Virgo, LIGO
  Scientific}),\ }\href {\doibase 10.1103/PhysRevLett.119.161101} {\bibfield
  {journal} {\bibinfo  {journal} {Phys. Rev. Lett.}\ }\textbf {\bibinfo
  {volume} {119}},\ \bibinfo {pages} {161101} (\bibinfo {year} {2017})},\
  \Eprint {http://arxiv.org/abs/1710.05832} {arXiv:1710.05832 [gr-qc]}
  \BibitemShut {NoStop}%
\bibitem [{\citenamefont {Banik}\ and\ \citenamefont
  {Bandyopadhyay}(2017)}]{Banik:2017zia}%
  \BibitemOpen
  \bibfield  {author} {\bibinfo {author} {\bibfnamefont {S.}~\bibnamefont
  {Banik}}\ and\ \bibinfo {author} {\bibfnamefont {D.}~\bibnamefont
  {Bandyopadhyay}}\ }(\bibinfo {year} {2017})\ \Eprint
  {http://arxiv.org/abs/1712.09760} {arXiv:1712.09760 [astro-ph.HE]}
  \BibitemShut {NoStop}%
\bibitem [{Note4()}]{Note4}%
  \BibitemOpen
  \bibinfo {note} {LIGO: http://www.ligo.caltech.edu.}\BibitemShut {Stop}%
\bibitem [{Note5()}]{Note5}%
  \BibitemOpen
  \bibinfo {note} {VIRGO: http://virgo.infn.it}\BibitemShut {NoStop}%
\bibitem [{Note6()}]{Note6}%
  \BibitemOpen
  \bibinfo {note} {Kagra: http://gwcenter.icrr.u-tokyo.ac.jp/en/}\BibitemShut
  {NoStop}%
\bibitem [{Note7()}]{Note7}%
  \BibitemOpen
  \bibinfo {note} {ET: http://www.et-gw.eu/}\BibitemShut {NoStop}%
\bibitem [{\citenamefont {De}\ \emph {et~al.}(2018)\citenamefont {De},
  \citenamefont {Finstad}, \citenamefont {Lattimer}, \citenamefont {Brown},
  \citenamefont {Berger},\ and\ \citenamefont {Biwer}}]{De:2018uhw}%
  \BibitemOpen
  \bibfield  {author} {\bibinfo {author} {\bibfnamefont {S.}~\bibnamefont
  {De}}, \bibinfo {author} {\bibfnamefont {D.}~\bibnamefont {Finstad}},
  \bibinfo {author} {\bibfnamefont {J.~M.}\ \bibnamefont {Lattimer}}, \bibinfo
  {author} {\bibfnamefont {D.~A.}\ \bibnamefont {Brown}}, \bibinfo {author}
  {\bibfnamefont {E.}~\bibnamefont {Berger}}, \ and\ \bibinfo {author}
  {\bibfnamefont {C.~M.}\ \bibnamefont {Biwer}},\ }\href@noop {} {\  (\bibinfo
  {year} {2018})},\ \Eprint {http://arxiv.org/abs/1804.08583} {arXiv:1804.08583
  [astro-ph.HE]} \BibitemShut {NoStop}%
\bibitem [{\citenamefont {Radice}\ \emph {et~al.}(2018)\citenamefont {Radice},
  \citenamefont {Perego}, \citenamefont {Zappa},\ and\ \citenamefont
  {Bernuzzi}}]{Radice:2017lry}%
  \BibitemOpen
  \bibfield  {author} {\bibinfo {author} {\bibfnamefont {D.}~\bibnamefont
  {Radice}}, \bibinfo {author} {\bibfnamefont {A.}~\bibnamefont {Perego}},
  \bibinfo {author} {\bibfnamefont {F.}~\bibnamefont {Zappa}}, \ and\ \bibinfo
  {author} {\bibfnamefont {S.}~\bibnamefont {Bernuzzi}},\ }\href {\doibase
  10.3847/2041-8213/aaa402} {\bibfield  {journal} {\bibinfo  {journal}
  {Astrophys. J.}\ }\textbf {\bibinfo {volume} {852}},\ \bibinfo {pages} {L29}
  (\bibinfo {year} {2018})},\ \Eprint {http://arxiv.org/abs/1711.03647}
  {arXiv:1711.03647 [astro-ph.HE]} \BibitemShut {NoStop}%
\bibitem [{\citenamefont {{Love}}(1909)}]{love}%
  \BibitemOpen
  \bibfield  {author} {\bibinfo {author} {\bibfnamefont {A.~E.~H.}\
  \bibnamefont {{Love}}},\ }\href {\doibase 10.1098/rspa.1909.0008} {\bibfield
  {journal} {\bibinfo  {journal} {Proceedings of the Royal Society of London
  Series A}\ }\textbf {\bibinfo {volume} {82}},\ \bibinfo {pages} {73}
  (\bibinfo {year} {1909})}\BibitemShut {NoStop}%
\bibitem [{\citenamefont {Hinderer}\ \emph {et~al.}(2010)\citenamefont
  {Hinderer}, \citenamefont {Lackey}, \citenamefont {Lang},\ and\ \citenamefont
  {Read}}]{Hinderer:2009ca}%
  \BibitemOpen
  \bibfield  {author} {\bibinfo {author} {\bibfnamefont {T.}~\bibnamefont
  {Hinderer}}, \bibinfo {author} {\bibfnamefont {B.~D.}\ \bibnamefont
  {Lackey}}, \bibinfo {author} {\bibfnamefont {R.~N.}\ \bibnamefont {Lang}}, \
  and\ \bibinfo {author} {\bibfnamefont {J.~S.}\ \bibnamefont {Read}},\ }\href
  {\doibase 10.1103/PhysRevD.81.123016} {\bibfield  {journal} {\bibinfo
  {journal} {Phys. Rev.}\ }\textbf {\bibinfo {volume} {D81}},\ \bibinfo {pages}
  {123016} (\bibinfo {year} {2010})},\ \Eprint {http://arxiv.org/abs/0911.3535}
  {arXiv:0911.3535 [astro-ph.HE]} \BibitemShut {NoStop}%
\bibitem [{\citenamefont {Damour}\ and\ \citenamefont
  {Nagar}(2009)}]{Damour:2009vw}%
  \BibitemOpen
  \bibfield  {author} {\bibinfo {author} {\bibfnamefont {T.}~\bibnamefont
  {Damour}}\ and\ \bibinfo {author} {\bibfnamefont {A.}~\bibnamefont {Nagar}},\
  }\href {\doibase 10.1103/PhysRevD.80.084035} {\bibfield  {journal} {\bibinfo
  {journal} {Phys. Rev.}\ }\textbf {\bibinfo {volume} {D80}},\ \bibinfo {pages}
  {084035} (\bibinfo {year} {2009})},\ \Eprint {http://arxiv.org/abs/0906.0096}
  {arXiv:0906.0096 [gr-qc]} \BibitemShut {NoStop}%
\bibitem [{\citenamefont {Postnikov}\ \emph {et~al.}(2010)\citenamefont
  {Postnikov}, \citenamefont {Prakash},\ and\ \citenamefont
  {Lattimer}}]{Postnikov:2010yn}%
  \BibitemOpen
  \bibfield  {author} {\bibinfo {author} {\bibfnamefont {S.}~\bibnamefont
  {Postnikov}}, \bibinfo {author} {\bibfnamefont {M.}~\bibnamefont {Prakash}},
  \ and\ \bibinfo {author} {\bibfnamefont {J.~M.}\ \bibnamefont {Lattimer}},\
  }\href {\doibase 10.1103/PhysRevD.82.024016} {\bibfield  {journal} {\bibinfo
  {journal} {Phys. Rev.}\ }\textbf {\bibinfo {volume} {D82}},\ \bibinfo {pages}
  {024016} (\bibinfo {year} {2010})},\ \Eprint {http://arxiv.org/abs/1004.5098}
  {arXiv:1004.5098 [astro-ph.SR]} \BibitemShut {NoStop}%
\bibitem [{\citenamefont {Drago}\ and\ \citenamefont
  {Pagliara}(2018)}]{Drago:2017bnf}%
  \BibitemOpen
  \bibfield  {author} {\bibinfo {author} {\bibfnamefont {A.}~\bibnamefont
  {Drago}}\ and\ \bibinfo {author} {\bibfnamefont {G.}~\bibnamefont
  {Pagliara}},\ }\href {\doibase 10.3847/2041-8213/aaa40a} {\bibfield
  {journal} {\bibinfo  {journal} {Astrophys. J.}\ }\textbf {\bibinfo {volume}
  {852}},\ \bibinfo {pages} {L32} (\bibinfo {year} {2018})},\ \Eprint
  {http://arxiv.org/abs/1710.02003} {arXiv:1710.02003 [astro-ph.HE]}
  \BibitemShut {NoStop}%
\bibitem [{\citenamefont {Nandi}\ and\ \citenamefont
  {Char}(2018)}]{Nandi:2017rhy}%
  \BibitemOpen
  \bibfield  {author} {\bibinfo {author} {\bibfnamefont {R.}~\bibnamefont
  {Nandi}}\ and\ \bibinfo {author} {\bibfnamefont {P.}~\bibnamefont {Char}},\
  }\href {\doibase 10.3847/1538-4357/aab78c} {\bibfield  {journal} {\bibinfo
  {journal} {Astrophys. J.}\ }\textbf {\bibinfo {volume} {857}},\ \bibinfo
  {pages} {12} (\bibinfo {year} {2018})},\ \Eprint
  {http://arxiv.org/abs/1712.08094} {arXiv:1712.08094 [astro-ph.HE]}
  \BibitemShut {NoStop}%
\bibitem [{\citenamefont {Kumar}\ \emph {et~al.}(2017)\citenamefont {Kumar},
  \citenamefont {Biswal},\ and\ \citenamefont {Patra}}]{Kumar:2016dks}%
  \BibitemOpen
  \bibfield  {author} {\bibinfo {author} {\bibfnamefont {B.}~\bibnamefont
  {Kumar}}, \bibinfo {author} {\bibfnamefont {S.~K.}\ \bibnamefont {Biswal}}, \
  and\ \bibinfo {author} {\bibfnamefont {S.~K.}\ \bibnamefont {Patra}},\ }\href
  {\doibase 10.1103/PhysRevC.95.015801} {\bibfield  {journal} {\bibinfo
  {journal} {Phys. Rev.}\ }\textbf {\bibinfo {volume} {C95}},\ \bibinfo {pages}
  {015801} (\bibinfo {year} {2017})},\ \Eprint
  {http://arxiv.org/abs/1609.08863} {arXiv:1609.08863 [nucl-th]} \BibitemShut
  {NoStop}%
\bibitem [{\citenamefont {Marques}\ \emph {et~al.}(2017)\citenamefont
  {Marques}, \citenamefont {Oertel}, \citenamefont {Hempel},\ and\
  \citenamefont {Novak}}]{Marques:2017zju}%
  \BibitemOpen
  \bibfield  {author} {\bibinfo {author} {\bibfnamefont {M.}~\bibnamefont
  {Marques}}, \bibinfo {author} {\bibfnamefont {M.}~\bibnamefont {Oertel}},
  \bibinfo {author} {\bibfnamefont {M.}~\bibnamefont {Hempel}}, \ and\ \bibinfo
  {author} {\bibfnamefont {J.}~\bibnamefont {Novak}},\ }\href {\doibase
  10.1103/PhysRevC.96.045806} {\bibfield  {journal} {\bibinfo  {journal} {Phys.
  Rev.}\ }\textbf {\bibinfo {volume} {C96}},\ \bibinfo {pages} {045806}
  (\bibinfo {year} {2017})},\ \Eprint {http://arxiv.org/abs/1706.02913}
  {arXiv:1706.02913 [nucl-th]} \BibitemShut {NoStop}%
\bibitem [{\citenamefont {Ayriyan}\ \emph {et~al.}(2018)\citenamefont
  {Ayriyan}, \citenamefont {Bastian}, \citenamefont {Blaschke}, \citenamefont
  {Grigorian}, \citenamefont {Maslov},\ and\ \citenamefont
  {Voskresensky}}]{Ayriyan:2017nby}%
  \BibitemOpen
  \bibfield  {author} {\bibinfo {author} {\bibfnamefont {A.}~\bibnamefont
  {Ayriyan}}, \bibinfo {author} {\bibfnamefont {N.~U.}\ \bibnamefont
  {Bastian}}, \bibinfo {author} {\bibfnamefont {D.}~\bibnamefont {Blaschke}},
  \bibinfo {author} {\bibfnamefont {H.}~\bibnamefont {Grigorian}}, \bibinfo
  {author} {\bibfnamefont {K.}~\bibnamefont {Maslov}}, \ and\ \bibinfo {author}
  {\bibfnamefont {D.~N.}\ \bibnamefont {Voskresensky}},\ }\href {\doibase
  10.1103/PhysRevC.97.045802} {\bibfield  {journal} {\bibinfo  {journal} {Phys.
  Rev.}\ }\textbf {\bibinfo {volume} {C97}},\ \bibinfo {pages} {045802}
  (\bibinfo {year} {2018})},\ \Eprint {http://arxiv.org/abs/1711.03926}
  {arXiv:1711.03926 [nucl-th]} \BibitemShut {NoStop}%
\bibitem [{\citenamefont {Paschalidis}\ \emph {et~al.}(2018)\citenamefont
  {Paschalidis}, \citenamefont {Yagi}, \citenamefont {Alvarez-Castillo},
  \citenamefont {Blaschke},\ and\ \citenamefont
  {Sedrakian}}]{Paschalidis:2017qmb}%
  \BibitemOpen
  \bibfield  {author} {\bibinfo {author} {\bibfnamefont {V.}~\bibnamefont
  {Paschalidis}}, \bibinfo {author} {\bibfnamefont {K.}~\bibnamefont {Yagi}},
  \bibinfo {author} {\bibfnamefont {D.}~\bibnamefont {Alvarez-Castillo}},
  \bibinfo {author} {\bibfnamefont {D.~B.}\ \bibnamefont {Blaschke}}, \ and\
  \bibinfo {author} {\bibfnamefont {A.}~\bibnamefont {Sedrakian}},\ }\href
  {\doibase 10.1103/PhysRevD.97.084038} {\bibfield  {journal} {\bibinfo
  {journal} {Phys. Rev.}\ }\textbf {\bibinfo {volume} {D97}},\ \bibinfo {pages}
  {084038} (\bibinfo {year} {2018})},\ \Eprint
  {http://arxiv.org/abs/1712.00451} {arXiv:1712.00451 [astro-ph.HE]}
  \BibitemShut {NoStop}%
\bibitem [{\citenamefont {Most}\ \emph {et~al.}(2018)\citenamefont {Most},
  \citenamefont {Weih}, \citenamefont {Rezzolla},\ and\ \citenamefont
  {Schaffner-Bielich}}]{Most:2018hfd}%
  \BibitemOpen
  \bibfield  {author} {\bibinfo {author} {\bibfnamefont {E.~R.}\ \bibnamefont
  {Most}}, \bibinfo {author} {\bibfnamefont {L.~R.}\ \bibnamefont {Weih}},
  \bibinfo {author} {\bibfnamefont {L.}~\bibnamefont {Rezzolla}}, \ and\
  \bibinfo {author} {\bibfnamefont {J.}~\bibnamefont {Schaffner-Bielich}},\
  }\href@noop {} {\  (\bibinfo {year} {2018})},\ \Eprint
  {http://arxiv.org/abs/1803.00549} {arXiv:1803.00549 [gr-qc]} \BibitemShut
  {NoStop}%
\bibitem [{\citenamefont {Burgio}\ \emph {et~al.}(2018)\citenamefont {Burgio},
  \citenamefont {Drago}, \citenamefont {Pagliara}, \citenamefont {Schulze},\
  and\ \citenamefont {Wei}}]{Burgio:2018yix}%
  \BibitemOpen
  \bibfield  {author} {\bibinfo {author} {\bibfnamefont {G.~F.}\ \bibnamefont
  {Burgio}}, \bibinfo {author} {\bibfnamefont {A.}~\bibnamefont {Drago}},
  \bibinfo {author} {\bibfnamefont {G.}~\bibnamefont {Pagliara}}, \bibinfo
  {author} {\bibfnamefont {H.~J.}\ \bibnamefont {Schulze}}, \ and\ \bibinfo
  {author} {\bibfnamefont {J.~B.}\ \bibnamefont {Wei}},\ }\href@noop {} {\
  (\bibinfo {year} {2018})},\ \Eprint {http://arxiv.org/abs/1803.09696}
  {arXiv:1803.09696 [astro-ph.HE]} \BibitemShut {NoStop}%
\bibitem [{\citenamefont {Alvarez-Castillo}\ \emph {et~al.}(2018)\citenamefont
  {Alvarez-Castillo}, \citenamefont {Blaschke}, \citenamefont {Grunfeld},\ and\
  \citenamefont {Pagura}}]{Alvarez-Castillo:2018pve}%
  \BibitemOpen
  \bibfield  {author} {\bibinfo {author} {\bibfnamefont {D.~E.}\ \bibnamefont
  {Alvarez-Castillo}}, \bibinfo {author} {\bibfnamefont {D.~B.}\ \bibnamefont
  {Blaschke}}, \bibinfo {author} {\bibfnamefont {A.~G.}\ \bibnamefont
  {Grunfeld}}, \ and\ \bibinfo {author} {\bibfnamefont {V.~P.}\ \bibnamefont
  {Pagura}},\ }\href@noop {} {\  (\bibinfo {year} {2018})},\ \Eprint
  {http://arxiv.org/abs/1805.04105} {arXiv:1805.04105 [hep-ph]} \BibitemShut
  {NoStop}%
\bibitem [{\citenamefont {Rueda}\ \emph {et~al.}(2018)\citenamefont {Rueda}
  \emph {et~al.}}]{Rueda:2018fky}%
  \BibitemOpen
  \bibfield  {author} {\bibinfo {author} {\bibfnamefont {J.~A.}\ \bibnamefont
  {Rueda}} \emph {et~al.},\ }\href@noop {} {\  (\bibinfo {year} {2018})},\
  \Eprint {http://arxiv.org/abs/1802.10027} {arXiv:1802.10027 [astro-ph.HE]}
  \BibitemShut {NoStop}%
\bibitem [{\citenamefont {Annala}\ \emph {et~al.}(2018)\citenamefont {Annala},
  \citenamefont {Gorda}, \citenamefont {Kurkela},\ and\ \citenamefont
  {Vuorinen}}]{Annala:2017llu}%
  \BibitemOpen
  \bibfield  {author} {\bibinfo {author} {\bibfnamefont {E.}~\bibnamefont
  {Annala}}, \bibinfo {author} {\bibfnamefont {T.}~\bibnamefont {Gorda}},
  \bibinfo {author} {\bibfnamefont {A.}~\bibnamefont {Kurkela}}, \ and\
  \bibinfo {author} {\bibfnamefont {A.}~\bibnamefont {Vuorinen}},\ }\href
  {\doibase 10.1103/PhysRevLett.120.172703} {\bibfield  {journal} {\bibinfo
  {journal} {Phys. Rev. Lett.}\ }\textbf {\bibinfo {volume} {120}},\ \bibinfo
  {pages} {172703} (\bibinfo {year} {2018})},\ \Eprint
  {http://arxiv.org/abs/1711.02644} {arXiv:1711.02644 [astro-ph.HE]}
  \BibitemShut {NoStop}%
\bibitem [{\citenamefont {Zhou}\ \emph {et~al.}(2018)\citenamefont {Zhou},
  \citenamefont {Zhou},\ and\ \citenamefont {Li}}]{Zhou:2017pha}%
  \BibitemOpen
  \bibfield  {author} {\bibinfo {author} {\bibfnamefont {E.-P.}\ \bibnamefont
  {Zhou}}, \bibinfo {author} {\bibfnamefont {X.}~\bibnamefont {Zhou}}, \ and\
  \bibinfo {author} {\bibfnamefont {A.}~\bibnamefont {Li}},\ }\href {\doibase
  10.1103/PhysRevD.97.083015} {\bibfield  {journal} {\bibinfo  {journal} {Phys.
  Rev.}\ }\textbf {\bibinfo {volume} {D97}},\ \bibinfo {pages} {083015}
  (\bibinfo {year} {2018})},\ \Eprint {http://arxiv.org/abs/1711.04312}
  {arXiv:1711.04312 [astro-ph.HE]} \BibitemShut {NoStop}%
\bibitem [{\citenamefont {Drago}\ \emph {et~al.}(2018)\citenamefont {Drago},
  \citenamefont {Pagliara}, \citenamefont {Popov}, \citenamefont {Traversi},\
  and\ \citenamefont {Wiktorowicz}}]{Drago:2018nzf}%
  \BibitemOpen
  \bibfield  {author} {\bibinfo {author} {\bibfnamefont {A.}~\bibnamefont
  {Drago}}, \bibinfo {author} {\bibfnamefont {G.}~\bibnamefont {Pagliara}},
  \bibinfo {author} {\bibfnamefont {S.~B.}\ \bibnamefont {Popov}}, \bibinfo
  {author} {\bibfnamefont {S.}~\bibnamefont {Traversi}}, \ and\ \bibinfo
  {author} {\bibfnamefont {G.}~\bibnamefont {Wiktorowicz}},\ }\bibfield
  {booktitle} {\emph {\bibinfo {booktitle} {{Proceedings, Compact Stars in the
  QCD Phase Diagram VI (CSQCD VI): Dubna, Russia, September 26-29, 2017}}},\
  }\href {\doibase 10.3390/universe4030050} {\bibfield  {journal} {\bibinfo
  {journal} {Universe}\ }\textbf {\bibinfo {volume} {4}},\ \bibinfo {pages}
  {50} (\bibinfo {year} {2018})},\ \Eprint {http://arxiv.org/abs/1802.02495}
  {arXiv:1802.02495 [astro-ph.HE]} \BibitemShut {NoStop}%
\bibitem [{\citenamefont {Krastev}\ and\ \citenamefont
  {Li}(2018)}]{Krastev:2018nwr}%
  \BibitemOpen
  \bibfield  {author} {\bibinfo {author} {\bibfnamefont {P.~G.}\ \bibnamefont
  {Krastev}}\ and\ \bibinfo {author} {\bibfnamefont {B.-A.}\ \bibnamefont
  {Li}},\ }\href@noop {} {\  (\bibinfo {year} {2018})},\ \Eprint
  {http://arxiv.org/abs/1801.04620} {arXiv:1801.04620 [nucl-th]} \BibitemShut
  {NoStop}%
\bibitem [{\citenamefont {Zhu}\ \emph {et~al.}(2018)\citenamefont {Zhu},
  \citenamefont {Zhou},\ and\ \citenamefont {Li}}]{Zhu:2018ona}%
  \BibitemOpen
  \bibfield  {author} {\bibinfo {author} {\bibfnamefont {Z.-Y.}\ \bibnamefont
  {Zhu}}, \bibinfo {author} {\bibfnamefont {E.-P.}\ \bibnamefont {Zhou}}, \
  and\ \bibinfo {author} {\bibfnamefont {A.}~\bibnamefont {Li}},\ }\href@noop
  {} {\  (\bibinfo {year} {2018})},\ \Eprint {http://arxiv.org/abs/7} {arXiv:7
  [nucl-th]} \BibitemShut {NoStop}%
\bibitem [{\citenamefont {Raithel}\ \emph {et~al.}(2018)\citenamefont
  {Raithel}, \citenamefont {Özel},\ and\ \citenamefont
  {Psaltis}}]{Raithel:2018ncd}%
  \BibitemOpen
  \bibfield  {author} {\bibinfo {author} {\bibfnamefont {C.}~\bibnamefont
  {Raithel}}, \bibinfo {author} {\bibfnamefont {F.}~\bibnamefont {Özel}}, \
  and\ \bibinfo {author} {\bibfnamefont {D.}~\bibnamefont {Psaltis}},\ }\href
  {\doibase 10.3847/2041-8213/aabcbf} {\bibfield  {journal} {\bibinfo
  {journal} {Astrophys. J.}\ }\textbf {\bibinfo {volume} {857}},\ \bibinfo
  {pages} {L23} (\bibinfo {year} {2018})},\ \Eprint
  {http://arxiv.org/abs/1803.07687} {arXiv:1803.07687 [astro-ph.HE]}
  \BibitemShut {NoStop}%
\bibitem [{\citenamefont {Malik}\ \emph {et~al.}(2018)\citenamefont {Malik},
  \citenamefont {Alam}, \citenamefont {Fortin}, \citenamefont {Providência},
  \citenamefont {Agrawal}, \citenamefont {Jha}, \citenamefont {Kumar},\ and\
  \citenamefont {Patra}}]{Malik:2018zcf}%
  \BibitemOpen
  \bibfield  {author} {\bibinfo {author} {\bibfnamefont {T.}~\bibnamefont
  {Malik}}, \bibinfo {author} {\bibfnamefont {N.}~\bibnamefont {Alam}},
  \bibinfo {author} {\bibfnamefont {M.}~\bibnamefont {Fortin}}, \bibinfo
  {author} {\bibfnamefont {C.}~\bibnamefont {Providência}}, \bibinfo {author}
  {\bibfnamefont {B.~K.}\ \bibnamefont {Agrawal}}, \bibinfo {author}
  {\bibfnamefont {T.~K.}\ \bibnamefont {Jha}}, \bibinfo {author} {\bibfnamefont
  {B.}~\bibnamefont {Kumar}}, \ and\ \bibinfo {author} {\bibfnamefont {S.~K.}\
  \bibnamefont {Patra}},\ }\href@noop {} {\  (\bibinfo {year} {2018})},\
  \Eprint {http://arxiv.org/abs/1805.11963} {arXiv:1805.11963 [nucl-th]}
  \BibitemShut {NoStop}%
\bibitem [{\citenamefont {Flanagan}\ and\ \citenamefont
  {Hinderer}(2008)}]{Flanagan:2007ix}%
  \BibitemOpen
  \bibfield  {author} {\bibinfo {author} {\bibfnamefont {E.~E.}\ \bibnamefont
  {Flanagan}}\ and\ \bibinfo {author} {\bibfnamefont {T.}~\bibnamefont
  {Hinderer}},\ }\href {\doibase 10.1103/PhysRevD.77.021502} {\bibfield
  {journal} {\bibinfo  {journal} {Phys. Rev.}\ }\textbf {\bibinfo {volume}
  {D77}},\ \bibinfo {pages} {021502} (\bibinfo {year} {2008})},\ \Eprint
  {http://arxiv.org/abs/0709.1915} {arXiv:0709.1915 [astro-ph]} \BibitemShut
  {NoStop}%
\bibitem [{\citenamefont {Dover}\ and\ \citenamefont
  {Gal}(1985)}]{Dover:1985ba}%
  \BibitemOpen
  \bibfield  {author} {\bibinfo {author} {\bibfnamefont {C.~B.}\ \bibnamefont
  {Dover}}\ and\ \bibinfo {author} {\bibfnamefont {A.}~\bibnamefont {Gal}},\
  }\href {\doibase 10.1016/0146-6410(84)90004-8} {\bibfield  {journal}
  {\bibinfo  {journal} {Prog. Part. Nucl. Phys.}\ }\textbf {\bibinfo {volume}
  {12}},\ \bibinfo {pages} {171} (\bibinfo {year} {1985})}\BibitemShut
  {NoStop}%
\bibitem [{\citenamefont {Schaffner}\ \emph {et~al.}(1994)\citenamefont
  {Schaffner}, \citenamefont {Dover}, \citenamefont {Gal}, \citenamefont
  {Greiner}, \citenamefont {Millener},\ and\ \citenamefont
  {Stoecker}}]{Schaffner:1993qj}%
  \BibitemOpen
  \bibfield  {author} {\bibinfo {author} {\bibfnamefont {J.}~\bibnamefont
  {Schaffner}}, \bibinfo {author} {\bibfnamefont {C.~B.}\ \bibnamefont
  {Dover}}, \bibinfo {author} {\bibfnamefont {A.}~\bibnamefont {Gal}}, \bibinfo
  {author} {\bibfnamefont {C.}~\bibnamefont {Greiner}}, \bibinfo {author}
  {\bibfnamefont {D.~J.}\ \bibnamefont {Millener}}, \ and\ \bibinfo {author}
  {\bibfnamefont {H.}~\bibnamefont {Stoecker}},\ }\href {\doibase
  10.1006/aphy.1994.1090} {\bibfield  {journal} {\bibinfo  {journal} {Annals
  Phys.}\ }\textbf {\bibinfo {volume} {235}},\ \bibinfo {pages} {35} (\bibinfo
  {year} {1994})}\BibitemShut {NoStop}%
\bibitem [{\citenamefont {Dexheimer}\ \emph {et~al.}(2017)\citenamefont
  {Dexheimer}, \citenamefont {Franzon}, \citenamefont {Gomes}, \citenamefont
  {Farias}, \citenamefont {Avancini},\ and\ \citenamefont
  {Schramm}}]{Dexheimer:2017fhy}%
  \BibitemOpen
  \bibfield  {author} {\bibinfo {author} {\bibfnamefont {V.}~\bibnamefont
  {Dexheimer}}, \bibinfo {author} {\bibfnamefont {B.}~\bibnamefont {Franzon}},
  \bibinfo {author} {\bibfnamefont {R.~O.}\ \bibnamefont {Gomes}}, \bibinfo
  {author} {\bibfnamefont {R.~L.~S.}\ \bibnamefont {Farias}}, \bibinfo {author}
  {\bibfnamefont {S.~S.}\ \bibnamefont {Avancini}}, \ and\ \bibinfo {author}
  {\bibfnamefont {S.}~\bibnamefont {Schramm}},\ }in\ \href
  {https://inspirehep.net/record/1621798/files/arXiv:1709.01914.pdf} {\emph
  {\bibinfo {booktitle} {{4th Caribbean Symposium on Cosmology, Gravitation,
  Nuclear and Astroparticle Physics (STARS2017) Havana, Cuba, May 7-13,
  2017}}}}\ (\bibinfo {year} {2017})\ \Eprint {http://arxiv.org/abs/1709.01914}
  {arXiv:1709.01914 [astro-ph.HE]} \BibitemShut {NoStop}%
\bibitem [{\citenamefont {Farhi}\ and\ \citenamefont
  {Jaffe}(1984)}]{Farhi:1984qu}%
  \BibitemOpen
  \bibfield  {author} {\bibinfo {author} {\bibfnamefont {E.}~\bibnamefont
  {Farhi}}\ and\ \bibinfo {author} {\bibfnamefont {R.~L.}\ \bibnamefont
  {Jaffe}},\ }\href {\doibase 10.1103/PhysRevD.30.2379} {\bibfield  {journal}
  {\bibinfo  {journal} {Phys. Rev.}\ }\textbf {\bibinfo {volume} {D30}},\
  \bibinfo {pages} {2379} (\bibinfo {year} {1984})}\BibitemShut {NoStop}%
\bibitem [{\citenamefont {Masuda}\ \emph {et~al.}(2013)\citenamefont {Masuda},
  \citenamefont {Hatsuda},\ and\ \citenamefont {Takatsuka}}]{Masuda:2012kf}%
  \BibitemOpen
  \bibfield  {author} {\bibinfo {author} {\bibfnamefont {K.}~\bibnamefont
  {Masuda}}, \bibinfo {author} {\bibfnamefont {T.}~\bibnamefont {Hatsuda}}, \
  and\ \bibinfo {author} {\bibfnamefont {T.}~\bibnamefont {Takatsuka}},\ }\href
  {\doibase 10.1088/0004-637X/764/1/12} {\bibfield  {journal} {\bibinfo
  {journal} {Astrophys. J.}\ }\textbf {\bibinfo {volume} {764}},\ \bibinfo
  {pages} {12} (\bibinfo {year} {2013})},\ \Eprint
  {http://arxiv.org/abs/1205.3621} {arXiv:1205.3621 [nucl-th]} \BibitemShut
  {NoStop}%
\bibitem [{\citenamefont {Denke}\ and\ \citenamefont
  {Pinto}(2013)}]{denke2013influence}%
  \BibitemOpen
  \bibfield  {author} {\bibinfo {author} {\bibfnamefont {R.~Z.}\ \bibnamefont
  {Denke}}\ and\ \bibinfo {author} {\bibfnamefont {M.~B.}\ \bibnamefont
  {Pinto}},\ }\href@noop {} {\bibfield  {journal} {\bibinfo  {journal}
  {Physical Review D}\ }\textbf {\bibinfo {volume} {88}},\ \bibinfo {pages}
  {056008} (\bibinfo {year} {2013})}\BibitemShut {NoStop}%
\bibitem [{\citenamefont {Contrera}\ \emph {et~al.}(2014)\citenamefont
  {Contrera}, \citenamefont {Spinella}, \citenamefont {Orsaria},\ and\
  \citenamefont {Weber}}]{contrera2014hadron}%
  \BibitemOpen
  \bibfield  {author} {\bibinfo {author} {\bibfnamefont {G.~A.}\ \bibnamefont
  {Contrera}}, \bibinfo {author} {\bibfnamefont {W.}~\bibnamefont {Spinella}},
  \bibinfo {author} {\bibfnamefont {M.}~\bibnamefont {Orsaria}}, \ and\
  \bibinfo {author} {\bibfnamefont {F.}~\bibnamefont {Weber}},\ }\href@noop {}
  {\bibfield  {journal} {\bibinfo  {journal} {arXiv preprint arXiv:1403.7415}\
  } (\bibinfo {year} {2014})}\BibitemShut {NoStop}%
\bibitem [{\citenamefont {Menezes}\ \emph {et~al.}(2014)\citenamefont
  {Menezes}, \citenamefont {Pinto}, \citenamefont {Castro}, \citenamefont
  {Costa},\ and\ \citenamefont {Provid{\^e}ncia}}]{menezes2014repulsive}%
  \BibitemOpen
  \bibfield  {author} {\bibinfo {author} {\bibfnamefont {D.~P.}\ \bibnamefont
  {Menezes}}, \bibinfo {author} {\bibfnamefont {M.~B.}\ \bibnamefont {Pinto}},
  \bibinfo {author} {\bibfnamefont {L.~B.}\ \bibnamefont {Castro}}, \bibinfo
  {author} {\bibfnamefont {P.}~\bibnamefont {Costa}}, \ and\ \bibinfo {author}
  {\bibfnamefont {C.}~\bibnamefont {Provid{\^e}ncia}},\ }\href@noop {}
  {\bibfield  {journal} {\bibinfo  {journal} {Physical Review C}\ }\textbf
  {\bibinfo {volume} {89}},\ \bibinfo {pages} {055207} (\bibinfo {year}
  {2014})}\BibitemShut {NoStop}%
\bibitem [{\citenamefont {Kl{\"a}hn}\ and\ \citenamefont
  {Fischer}(2015)}]{klahn2015vector}%
  \BibitemOpen
  \bibfield  {author} {\bibinfo {author} {\bibfnamefont {T.}~\bibnamefont
  {Kl{\"a}hn}}\ and\ \bibinfo {author} {\bibfnamefont {T.}~\bibnamefont
  {Fischer}},\ }\href@noop {} {\bibfield  {journal} {\bibinfo  {journal} {The
  Astrophysical Journal}\ }\textbf {\bibinfo {volume} {810}},\ \bibinfo {pages}
  {134} (\bibinfo {year} {2015})}\BibitemShut {NoStop}%
\bibitem [{\citenamefont {Ranea-Sandoval}\ \emph {et~al.}(2015)\citenamefont
  {Ranea-Sandoval}, \citenamefont {Han}, \citenamefont {Orsaria}, \citenamefont
  {Contrera}, \citenamefont {Weber},\ and\ \citenamefont
  {Alford}}]{ranea2015constant}%
  \BibitemOpen
  \bibfield  {author} {\bibinfo {author} {\bibfnamefont {I.~F.}\ \bibnamefont
  {Ranea-Sandoval}}, \bibinfo {author} {\bibfnamefont {S.}~\bibnamefont {Han}},
  \bibinfo {author} {\bibfnamefont {M.~G.}\ \bibnamefont {Orsaria}}, \bibinfo
  {author} {\bibfnamefont {G.~A.}\ \bibnamefont {Contrera}}, \bibinfo {author}
  {\bibfnamefont {F.}~\bibnamefont {Weber}}, \ and\ \bibinfo {author}
  {\bibfnamefont {M.~G.}\ \bibnamefont {Alford}},\ }\href@noop {} {\bibfield
  {journal} {\bibinfo  {journal} {arXiv preprint arXiv:1512.09183}\ } (\bibinfo
  {year} {2015})}\BibitemShut {NoStop}%
\bibitem [{\citenamefont {Fraga}\ \emph {et~al.}(2001)\citenamefont {Fraga},
  \citenamefont {Pisarski},\ and\ \citenamefont
  {Schaffner-Bielich}}]{Fraga:2001id}%
  \BibitemOpen
  \bibfield  {author} {\bibinfo {author} {\bibfnamefont {E.~S.}\ \bibnamefont
  {Fraga}}, \bibinfo {author} {\bibfnamefont {R.~D.}\ \bibnamefont {Pisarski}},
  \ and\ \bibinfo {author} {\bibfnamefont {J.}~\bibnamefont
  {Schaffner-Bielich}},\ }\href {\doibase 10.1103/PhysRevD.63.121702}
  {\bibfield  {journal} {\bibinfo  {journal} {Phys. Rev.}\ }\textbf {\bibinfo
  {volume} {D63}},\ \bibinfo {pages} {121702} (\bibinfo {year} {2001})},\
  \Eprint {http://arxiv.org/abs/hep-ph/0101143} {arXiv:hep-ph/0101143 [hep-ph]}
  \BibitemShut {NoStop}%
\bibitem [{\citenamefont {Fraga}\ \emph {et~al.}(2014)\citenamefont {Fraga},
  \citenamefont {Kurkela},\ and\ \citenamefont {Vuorinen}}]{Fraga:2013qra}%
  \BibitemOpen
  \bibfield  {author} {\bibinfo {author} {\bibfnamefont {E.~S.}\ \bibnamefont
  {Fraga}}, \bibinfo {author} {\bibfnamefont {A.}~\bibnamefont {Kurkela}}, \
  and\ \bibinfo {author} {\bibfnamefont {A.}~\bibnamefont {Vuorinen}},\ }\href
  {\doibase 10.1088/2041-8205/781/2/L25} {\bibfield  {journal} {\bibinfo
  {journal} {Astrophys. J.}\ }\textbf {\bibinfo {volume} {781}},\ \bibinfo
  {pages} {L25} (\bibinfo {year} {2014})},\ \Eprint
  {http://arxiv.org/abs/1311.5154} {arXiv:1311.5154 [nucl-th]} \BibitemShut
  {NoStop}%
\bibitem [{\citenamefont {Restrepo}\ \emph {et~al.}(2015)\citenamefont
  {Restrepo}, \citenamefont {Macias}, \citenamefont {Pinto},\ and\
  \citenamefont {Ferrari}}]{Restrepo:2014fna}%
  \BibitemOpen
  \bibfield  {author} {\bibinfo {author} {\bibfnamefont {T.~E.}\ \bibnamefont
  {Restrepo}}, \bibinfo {author} {\bibfnamefont {J.~C.}\ \bibnamefont
  {Macias}}, \bibinfo {author} {\bibfnamefont {M.~B.}\ \bibnamefont {Pinto}}, \
  and\ \bibinfo {author} {\bibfnamefont {G.~N.}\ \bibnamefont {Ferrari}},\
  }\href {\doibase 10.1103/PhysRevD.91.065017} {\bibfield  {journal} {\bibinfo
  {journal} {Phys. Rev.}\ }\textbf {\bibinfo {volume} {D91}},\ \bibinfo {pages}
  {065017} (\bibinfo {year} {2015})},\ \Eprint {http://arxiv.org/abs/1412.3074}
  {arXiv:1412.3074 [hep-ph]} \BibitemShut {NoStop}%
\bibitem [{\citenamefont {Maruyama}\ \emph {et~al.}(2007)\citenamefont
  {Maruyama}, \citenamefont {Chiba}, \citenamefont {Schulze},\ and\
  \citenamefont {Tatsumi}}]{Maruyama:2007ey}%
  \BibitemOpen
  \bibfield  {author} {\bibinfo {author} {\bibfnamefont {T.}~\bibnamefont
  {Maruyama}}, \bibinfo {author} {\bibfnamefont {S.}~\bibnamefont {Chiba}},
  \bibinfo {author} {\bibfnamefont {H.-J.}\ \bibnamefont {Schulze}}, \ and\
  \bibinfo {author} {\bibfnamefont {T.}~\bibnamefont {Tatsumi}},\ }\href
  {\doibase 10.1103/PhysRevD.76.123015} {\bibfield  {journal} {\bibinfo
  {journal} {Phys. Rev.}\ }\textbf {\bibinfo {volume} {D76}},\ \bibinfo {pages}
  {123015} (\bibinfo {year} {2007})},\ \Eprint {http://arxiv.org/abs/0708.3277}
  {arXiv:0708.3277 [nucl-th]} \BibitemShut {NoStop}%
\bibitem [{\citenamefont {Maruyama}\ \emph {et~al.}(2008)\citenamefont
  {Maruyama}, \citenamefont {Chiba}, \citenamefont {Schulze},\ and\
  \citenamefont {Tatsumi}}]{Maruyama:2007ss}%
  \BibitemOpen
  \bibfield  {author} {\bibinfo {author} {\bibfnamefont {T.}~\bibnamefont
  {Maruyama}}, \bibinfo {author} {\bibfnamefont {S.}~\bibnamefont {Chiba}},
  \bibinfo {author} {\bibfnamefont {H.-J.}\ \bibnamefont {Schulze}}, \ and\
  \bibinfo {author} {\bibfnamefont {T.}~\bibnamefont {Tatsumi}},\ }\href
  {\doibase 10.1016/j.physletb.2007.10.056} {\bibfield  {journal} {\bibinfo
  {journal} {Phys. Lett.}\ }\textbf {\bibinfo {volume} {B659}},\ \bibinfo
  {pages} {192} (\bibinfo {year} {2008})},\ \Eprint
  {http://arxiv.org/abs/nucl-th/0702088} {arXiv:nucl-th/0702088 [NUCL-TH]}
  \BibitemShut {NoStop}%
\bibitem [{\citenamefont {Palhares}\ and\ \citenamefont
  {Fraga}(2010)}]{Palhares:2010be}%
  \BibitemOpen
  \bibfield  {author} {\bibinfo {author} {\bibfnamefont {L.~F.}\ \bibnamefont
  {Palhares}}\ and\ \bibinfo {author} {\bibfnamefont {E.~S.}\ \bibnamefont
  {Fraga}},\ }\href {\doibase 10.1103/PhysRevD.82.125018} {\bibfield  {journal}
  {\bibinfo  {journal} {Phys. Rev.}\ }\textbf {\bibinfo {volume} {D82}},\
  \bibinfo {pages} {125018} (\bibinfo {year} {2010})},\ \Eprint
  {http://arxiv.org/abs/1006.2357} {arXiv:1006.2357 [hep-ph]} \BibitemShut
  {NoStop}%
\bibitem [{\citenamefont {Yasutake}\ \emph {et~al.}(2013)\citenamefont
  {Yasutake}, \citenamefont {Chen}, \citenamefont {Maruyama},\ and\
  \citenamefont {Tatsumi}}]{Yasutake:2013sza}%
  \BibitemOpen
  \bibfield  {author} {\bibinfo {author} {\bibfnamefont {N.}~\bibnamefont
  {Yasutake}}, \bibinfo {author} {\bibfnamefont {H.}~\bibnamefont {Chen}},
  \bibinfo {author} {\bibfnamefont {T.}~\bibnamefont {Maruyama}}, \ and\
  \bibinfo {author} {\bibfnamefont {T.}~\bibnamefont {Tatsumi}}\ }(\bibinfo
  {year} {2013})\ \Eprint {http://arxiv.org/abs/1309.1954} {arXiv:1309.1954
  [astro-ph.HE]} \BibitemShut {NoStop}%
\bibitem [{\citenamefont {Lugones}\ \emph {et~al.}(2013)\citenamefont
  {Lugones}, \citenamefont {Grunfeld},\ and\ \citenamefont
  {Al~Ajmi}}]{Lugones:2013ema}%
  \BibitemOpen
  \bibfield  {author} {\bibinfo {author} {\bibfnamefont {G.}~\bibnamefont
  {Lugones}}, \bibinfo {author} {\bibfnamefont {A.~G.}\ \bibnamefont
  {Grunfeld}}, \ and\ \bibinfo {author} {\bibfnamefont {M.}~\bibnamefont
  {Al~Ajmi}},\ }\href {\doibase 10.1103/PhysRevC.88.045803} {\bibfield
  {journal} {\bibinfo  {journal} {Phys. Rev.}\ }\textbf {\bibinfo {volume}
  {C88}},\ \bibinfo {pages} {045803} (\bibinfo {year} {2013})},\ \Eprint
  {http://arxiv.org/abs/1308.1452} {arXiv:1308.1452 [hep-ph]} \BibitemShut
  {NoStop}%
\bibitem [{\citenamefont {Garcia}\ and\ \citenamefont
  {Pinto}(2013)}]{Garcia:2013eaa}%
  \BibitemOpen
  \bibfield  {author} {\bibinfo {author} {\bibfnamefont {A.~F.}\ \bibnamefont
  {Garcia}}\ and\ \bibinfo {author} {\bibfnamefont {M.~B.}\ \bibnamefont
  {Pinto}},\ }\href {\doibase 10.1103/PhysRevC.88.025207} {\bibfield  {journal}
  {\bibinfo  {journal} {Phys. Rev.}\ }\textbf {\bibinfo {volume} {C88}},\
  \bibinfo {pages} {025207} (\bibinfo {year} {2013})},\ \Eprint
  {http://arxiv.org/abs/1306.3090} {arXiv:1306.3090 [hep-ph]} \BibitemShut
  {NoStop}%
\bibitem [{\citenamefont {Lugones}\ and\ \citenamefont
  {Grunfeld}(2017)}]{Lugones:2016ytl}%
  \BibitemOpen
  \bibfield  {author} {\bibinfo {author} {\bibfnamefont {G.}~\bibnamefont
  {Lugones}}\ and\ \bibinfo {author} {\bibfnamefont {A.~G.}\ \bibnamefont
  {Grunfeld}},\ }\href {\doibase 10.1103/PhysRevC.95.015804} {\bibfield
  {journal} {\bibinfo  {journal} {Phys. Rev.}\ }\textbf {\bibinfo {volume}
  {C95}},\ \bibinfo {pages} {015804} (\bibinfo {year} {2017})},\ \Eprint
  {http://arxiv.org/abs/1610.05875} {arXiv:1610.05875 [nucl-th]} \BibitemShut
  {NoStop}%
\bibitem [{\citenamefont {Endo}\ \emph {et~al.}(2006)\citenamefont {Endo},
  \citenamefont {Maruyama}, \citenamefont {Chiba},\ and\ \citenamefont
  {Tatsumi}}]{Endo:2005zt}%
  \BibitemOpen
  \bibfield  {author} {\bibinfo {author} {\bibfnamefont {T.}~\bibnamefont
  {Endo}}, \bibinfo {author} {\bibfnamefont {T.}~\bibnamefont {Maruyama}},
  \bibinfo {author} {\bibfnamefont {S.}~\bibnamefont {Chiba}}, \ and\ \bibinfo
  {author} {\bibfnamefont {T.}~\bibnamefont {Tatsumi}},\ }\href {\doibase
  10.1143/PTP.115.337} {\bibfield  {journal} {\bibinfo  {journal} {Prog. Theor.
  Phys.}\ }\textbf {\bibinfo {volume} {115}},\ \bibinfo {pages} {337} (\bibinfo
  {year} {2006})},\ \Eprint {http://arxiv.org/abs/hep-ph/0510279}
  {arXiv:hep-ph/0510279 [hep-ph]} \BibitemShut {NoStop}%
\bibitem [{\citenamefont {Yasutake}\ \emph {et~al.}(2014)\citenamefont
  {Yasutake}, \citenamefont {Lastowiecki}, \citenamefont {Benic}, \citenamefont
  {Blaschke}, \citenamefont {Maruyama},\ and\ \citenamefont
  {Tatsumi}}]{Yasutake:2014oxa}%
  \BibitemOpen
  \bibfield  {author} {\bibinfo {author} {\bibfnamefont {N.}~\bibnamefont
  {Yasutake}}, \bibinfo {author} {\bibfnamefont {R.}~\bibnamefont
  {Lastowiecki}}, \bibinfo {author} {\bibfnamefont {S.}~\bibnamefont {Benic}},
  \bibinfo {author} {\bibfnamefont {D.}~\bibnamefont {Blaschke}}, \bibinfo
  {author} {\bibfnamefont {T.}~\bibnamefont {Maruyama}}, \ and\ \bibinfo
  {author} {\bibfnamefont {T.}~\bibnamefont {Tatsumi}},\ }\href {\doibase
  10.1103/PhysRevC.89.065803} {\bibfield  {journal} {\bibinfo  {journal} {Phys.
  Rev.}\ }\textbf {\bibinfo {volume} {C89}},\ \bibinfo {pages} {065803}
  (\bibinfo {year} {2014})},\ \Eprint {http://arxiv.org/abs/1403.7492}
  {arXiv:1403.7492 [astro-ph.HE]} \BibitemShut {NoStop}%
\bibitem [{\citenamefont {Hinderer}(2008)}]{Hinderer:2007mb}%
  \BibitemOpen
  \bibfield  {author} {\bibinfo {author} {\bibfnamefont {T.}~\bibnamefont
  {Hinderer}},\ }\href {\doibase 10.1086/533487} {\bibfield  {journal}
  {\bibinfo  {journal} {Astrophys. J.}\ }\textbf {\bibinfo {volume} {677}},\
  \bibinfo {pages} {1216} (\bibinfo {year} {2008})},\ \Eprint
  {http://arxiv.org/abs/0711.2420} {arXiv:0711.2420 [astro-ph]} \BibitemShut
  {NoStop}%
\bibitem [{\citenamefont {Regge}\ and\ \citenamefont
  {Wheeler}(1957)}]{Regge:1957td}%
  \BibitemOpen
  \bibfield  {author} {\bibinfo {author} {\bibfnamefont {T.}~\bibnamefont
  {Regge}}\ and\ \bibinfo {author} {\bibfnamefont {J.~A.}\ \bibnamefont
  {Wheeler}},\ }\href {\doibase 10.1103/PhysRev.108.1063} {\bibfield  {journal}
  {\bibinfo  {journal} {Phys. Rev.}\ }\textbf {\bibinfo {volume} {108}},\
  \bibinfo {pages} {1063} (\bibinfo {year} {1957})}\BibitemShut {NoStop}%
\bibitem [{\citenamefont {Wu}\ and\ \citenamefont {Shen}(2017)}]{Wu:2017xaz}%
  \BibitemOpen
  \bibfield  {author} {\bibinfo {author} {\bibfnamefont {X.}~\bibnamefont
  {Wu}}\ and\ \bibinfo {author} {\bibfnamefont {H.}~\bibnamefont {Shen}},\
  }\href {\doibase 10.1103/PhysRevC.96.025802} {\bibfield  {journal} {\bibinfo
  {journal} {Phys. Rev.}\ }\textbf {\bibinfo {volume} {C96}},\ \bibinfo {pages}
  {025802} (\bibinfo {year} {2017})},\ \Eprint
  {http://arxiv.org/abs/1708.01878} {arXiv:1708.01878 [nucl-th]} \BibitemShut
  {NoStop}%
\end{thebibliography}%

\end{document}